\documentclass[conference]{IEEEtran}
\usepackage{extarrows}
\usepackage{amsmath}
\usepackage{amssymb}
\usepackage{proof}
\usepackage{subfig}
\usepackage[all]{xy}
\usepackage{setspace}
\usepackage{url}

\newtheorem{theorem}{\bf Theorem}
\newtheorem{lemma}{\bf Lemma}
\newtheorem{proposition}{\bf Proposition}
\newtheorem{example}{\bf Example}
\newtheorem{definition}{\bf Definition}

\ifCLASSOPTIONcompsoc
  \usepackage[nocompress]{cite}
\else
  \usepackage{cite}
\fi

\ifCLASSINFOpdf
\else
\fi

\begin{document}
%
\title{Value-passing CCS for Trees: A Theory for Concurrent Systems}

\author{\IEEEauthorblockN{Shichao Liu}
\IEEEauthorblockA{State Key Laboratory of Computer Science,\\ Institute of Software,
 Chinese Academy of Sciences,\\ Beijing, China\\
University of Chinese Academy of Sciences,
Beijing, China\\
Email: liusc@ios.ac.cn}
\and
\IEEEauthorblockN{Ying Jiang}
\IEEEauthorblockA{State Key Laboratory of Computer Science,\\ Institute of Software,
 Chinese Academy of Sciences,\\ Beijing, China\\
Email: jy@ios.ac.cn}
}


%

\maketitle
\begin{abstract}
In this paper, we extend the theory  CCS for trees (CCTS) to value-passing CCTS (VCCTS), whose symbols have the capacity for receiving and sending data values, and a non-sequential semantics is proposed in an operational approach. In this concurrent model, a weak barbed congruence and a localized early weak bisimilarity are defined, and the latter relation is proved to be sufficient to justify the former.
As an illustration of potential applications of VCCTS, a semantics based on VCCTS
is given to a toy multi-threaded programming language featuring a core of C/C++ concurrency;
and a formalization based on the operational semantics of VCCTS is proposed for some relaxed memory models,
and a DRF-guarantee property with respect to VCCTS is proved.
\end{abstract}

\IEEEpeerreviewmaketitle

\section{Introduction}\label{Intro}
Nowadays, multiprocessor machines are ubiquitous
and are designed to allow their primitive instructions to be performed simultaneously.
The most common way to exploit the power of multiprocessors is to design multi-threaded programs,
in which threads could communicate either through operating on shared variables or through sending messages to each other.
The simplest way to model multi-threaded programs is based on interleaving semantics,
which simulates concurrency by a nondeterministic choice of one available thread to execute at each step.
For example, the program
$${\sf x:=1}\parallel {\sf y:= 2}\eqno{(1)}$$
concurrently writing number ${\sf 1}$ to variable ${\sf x}$ and ${\sf 2}$ to ${\sf y}$, is regarded as
$$({\sf x:=1;y:=2}) + ({\sf y:=2;x:=1}) \eqno{(2)}$$
In the program $(1)$, ${\sf x:=1}$ and ${\sf y:= 2}$ are independent and concurrent, providing that ${\sf x}$ and ${\sf y}$ are different,
while in the program $(2)$ the two instructions have a causal relationship, i.e. an instruction cannot execute until the other one finishes.
Therefore, the interleaving approach loses the concurrent feature of the original program.
This raises the necessity of a new approach from the non-sequential perspective.

The most well studied non-sequential models are Petri nets \cite{reisig1985petri} and Event Structures \cite{winskel1989introduction},
which take the concurrency, causality and conflict relations as primitives on events.
In these models, processes are represented in terms of events, and independent events are allowed to occur simultaneously.
However, as G. Winskel pointed out in \cite{winskel2009events}, these models lack a systematical support to give structural operational semantics for process languages and programming languages.
A typical structural operational semantics is
G. Plotkin's Structural Operational Semantics (SOS) \cite{plotkin1981structural}, in which the transitions are generated inductively by syntactical rules.

On the other hand, process calculi, initiated by R. Milner in the early 1980s and tailored to a Calculus of
Communicating Systems (CCS) \cite{Milner1989}, provide an algebraic approach to investigate distributed/concurrent systems.
These calculi enjoy the property of compositionality, but used an interleaving semantics in early days,
which simulates concurrency with nondeterminism and sequentiality,
as expressed by Milner's expansion law \cite{Milner1989}, e.g. $(a\mid b) = a.b + b.a$.
Since the mid-1980s, researchers have begun to explore non-sequential semantics of process calculi.
These semantics are usually obtained,
on one hand,
by enriching the labelled transition systems for CCS through exploring causality \cite{boudol1988non}, or adding locations \cite{boudol1994theory,castellani2001process}, or combining both \cite{degano1990partial,kiehn1994comparing};
on the other hand, by representing concurrent processes by event-based non-sequential models,
e.g. Petri nets and Event Structures. And graph-based process calculi can also be found in \cite{ferrari2006synchronised}.
However, as far as we know, applications of these non-sequential models to programming languages are not sufficiently investigated in the literature.

Recently, a new concurrent theory, called CCS for trees (CCTS) \cite{Ehrhard2013ccts}, has been proposed.
One of the motivations of CCTS is to extend, in a uniform framework, both CCS and top-down tree automata with interacting capacity through parallel composition. The parallel composition in CCTS is parameterized by a graph whose vertices are the locations of the subprocesses and edges specify possible communications between the corresponding subprocesses. It is proved, among others, that CCTS is indeed a conservative extension of both CCS and top-down tree automata. CCTS has a distributed structure supporting concurrency naturally, and the semantics of CCTS is non-sequential, though it is still in an interleaving style in the sense that only one action can occur at each step.

Comparing with CCS, CCTS carries more information about distributions and
interactions of processes, as well as the evolution history of processes.
For instance, a prefixed process $f\cdot(P_1,\ldots,P_n)$ in CCTS,
 where $f$ is a symbol of action with $n$ arity, can evolve to $n$ subprocesses $P_i$,
running concurrently without any communication capacities between each other.
In shared memory multi-threaded programs,
threads can be created dynamically, and these threads cannot communicate with
each other directly, but through shared variables.
From this point of view,
CCTS could be an appropriate model for multi-threaded programming languages.

One of the main contributions of this paper consists of extending CCTS to
value-passing CCTS (VCCTS), whose symbols have the capacity for receiving
and sending data values.
Just like CCS, adding explicit value passing to CCTS
does not increase the expressiveness,
but improves readability. However, different from CCTS,
a new non-sequential semantics for VCCTS is developed in SOS style, allowing
unrelated actions to occur simultaneously.
For instance, the program (1) displayed previously is represented in VCCTS by
$$[\![{\sf x:=1}]\!]\oplus [\![{\sf y:= 2}]\!],$$
meaning that processes $[\![{\sf x:=1}]\!]$ and $[\![{\sf y:= 2}]\!]$ execute independently and concurrently,
while the representation of program (2) in VCCTS can only execute sequentially (see Section \ref{VCCTS_prog}).

Process equivalence is a central idea of process calculi.
As usual, behavioural equivalence of VCCTS is expressed by a concept of weak barbed congruence \cite{Milner1992barbed}, which relates processes with the same behaviours
during evolutions in all contexts. 
Based on the localized transition system in Section \ref{VCCTS_semantics}, the localized early weak
bisimilarity for VCCTS is defined, which is proved to be sufficient for proving weak barbed congruence.
The proofs follow the lines of CCTS, but are a little more
complicated, because, here, one has to deal with multisets of actions instead of
single actions.

Much like CCTS,
VCCTS carries the information about the distribution and the interaction of processes,
as well as the history of process evolution.
A prefixed process in VCCTS at location $p$ can fork $n$ subprocesses, running concurrently without any communication capacities between each other (see Section \ref{VCCTS_semantics}):
$$f(x)\cdot(P_1,\ldots,P_n)\xrightarrow[\lambda]{p:fv\cdot({\vec L})}P_1\{v/x\}\oplus\cdots\oplus P_n\{v/x\}$$
By prefix and $\oplus$ operators, thread creations and communication capacities can be encoded naturally and concisely in VCCTS (see Section \ref{VCCTS_prog}).
Indeed, VCCTS seems to be an appropriate model for multi-threaded programming languages.

Another contribution of this paper concerns the potential applications of
VCCTS, which are illustrated by giving a semantics based on VCCTS to a toy multi-threaded programming language
featuring a core of C/C++ concurrency (by transforming the language into VCCTS). The correctness of this translation is proved.

Formalization of programming languages and of relaxed (or weak) memory models are two major problems in application of concurrent theories. For example, programming languages were modeled in CCS \cite{Milner1989}, Petri Nets \cite{hayman2006independence} etc., while relaxed memory models \cite{Adve1996shared} were formalized in various frameworks usually specifically designed for them. For example, C++ \cite{Batty2011mathematizing,Boehm2008foundations} and Java \cite{Manson2005JMM}, as memory models, are studied in happen-before models \cite{Lamport1978hp}, and Total Store Ordering (TSO) memory model \cite{owens2009better} in abstract machines. 
The expressiveness of VCCTS allows us to discuss both problems in a same framework. The interested reader is referred to Appendix \ref{VCCTS_RMM} for two case studies of relaxed memory models in VCCTS, i.e. compiler reorderings \cite{lea_reordering} and TSO memory model.

The rest of this paper is organized as follows.
In Section \ref{VCCTS_syntax}, the syntax of VCCTS is introduced.
In Section \ref{VCCTS_semantics}, a non-sequential semantics for VCCTS is developed and two kinds of behavioural equivalences are discussed.
Section \ref{VCCTS_prog} focusses on a non-sequential operational semantics, based on VCCTS,
for a toy multi-threaded programming language.
Related work and conclusions
are in Section \ref{Related_work} and Section \ref{Conclusions}, respectively.

For the lack of space, all the proofs are omitted, but they can be found in Appendices.

\section{Value-passing CCTS}\label{VCCTS_syntax}
\subsection{Syntax of Value-passing CCTS}
Let ${\sf Loc}$ be a countable set of locations ranged over by $p,q,\ldots$.
A finite graph $G=(|G|,\frown_G)$ is composed of a finite set of locations $|G|\subseteq{\sf Loc}$ and a set of edges $\frown_G$ which is a symmetric and irreflexive binary relation on $|G|$. 
Given disjoint sets $E$ and $F$ with $p\in E$, let $E[F/p]=(E\setminus \{p\})\cup F$.
Let $G$ and $H$ be graphs with $|G|\cap |H|=\emptyset$ and $p\in |G|$.
A substitution $G[H/p]$ is a graph composed of locations $|G[H/p]|=|G|[|H|/p]$ and  edges $q\frown_{G[H/p]}r$ if $q\frown_G r$, or $q\frown_H r$, or $q\frown_G p$ and $r\in|H|$, or $r\frown_G p$ and $q\in |H|$.

Let ${\bf Var}$ be the set of data variables ranged over by $x$, and let ${\bf Val}$ be the set of data values ranged over  by $v$. The set of arithmetic expressions ${\bf Exp}$ at least includes ${\bf Var}$ and ${\bf Val}$, ranged over by $e$. The set of boolean expressions ${\bf BExp}$ at least includes $\{\it false,true\}$, ranged over by $b$. The substitutions for expressions are defined as usual denoted by $e\{(e^{\prime})/x\}$ and $b\{(b^{\prime})/x\}$.
Let ${\sf fv}(e)$ and ${\sf fv}(b)$ be the set of data variables appearing in $e$ and $b$, respectively. We say that $e$ is closed if ${\sf fv}(e) = \emptyset$, and similarly for $b$.

Let $\mathcal{V}$ be a countable set of process variables ranged over by $X,Y,\ldots$. 
Let $\Sigma = (\Sigma_n)_{n\in {\bf N}}$ be a signature. For each symbol $f\in\Sigma_n$, there is a co-symbol $\overline{f}$. Let $\overline{\Sigma}_n = \Sigma_n\cup \{\overline{f}\mid f\in\Sigma_n\}$, $\overline{\Sigma}=(\overline{\Sigma}_n)_{n\in{\bf N}}$, and $\overline{\overline{f}}=f$, for $f\in \Sigma_n$.
There is only one symbol of arity $0$, denoted by $\ast$. Moreover, $\ast$ dose not pass any value and $\overline{\ast}= \ast$.
$\ast()\cdot()$ is written $\ast$ if there is no confusion.

Let ${\bf Pr}$ be the set of all processes in VCCTS, and we define it inductively as follows:
\begin{center}
$P,Q::=$ $\ast\mid{\bf 0}\mid X\mid \mu X\cdot P\mid  P + Q\mid f(x)\cdot (P_1,\ldots,P_n)$\hfill~\\
 \hfill$\mid$  $\overline{f}(e)\cdot (P_1,\ldots,P_n) \mid G\langle\Phi\rangle\mid P\backslash I\mid {\bf if}~b~{\bf then}~P~{\bf else}~Q$ \\
\end{center}
where $X\in \mathcal{V}$, $x\in{\bf Var}$, $e\in{\bf Exp}$, $b\in {\bf BExp}$, $f\in \Sigma_n (n\ge 1)$, $P_1,\ldots,P_n \in {\bf Pr}$, $G$ is a finite graph, $\Phi$ is a function from $|G|$ to ${\bf Pr}$ and $I$ is a finite subset of $\Sigma$.


Recursive operator $\mu$, sum operator $+$ and symbol restriction $\backslash$ have the same meanings as those in CCS. $\ast$ is an idle process different from empty sum ${\bf 0}$.
In $f(x)\cdot (P_1,\ldots,P_n)$, $x$ is bound; data variables appearing in $e$ are free in $\overline{g}(e)\cdot(Q_1,\ldots,Q_m)$. We use ${\sf fv}(P)$ to represent the free data variables appearing in $P$.
A process $P$ is {\it data closed} if all the data variables appearing in $P$ are bound.
The substitution of an expression for data variable in processes is denoted by $P\{(e)/x\}$, which means substituting $e$ for every free occurrence of $x$ in process $P$.

The parallel composition is defined using a graph.  $G\langle\Phi\rangle$ is the parallel composition of processes $\Phi(p)\in{\bf Pr}$ for each $p\in |G|$ with communication capabilities specified by $G$, and processes $\Phi(p)$ are called the components of $G\langle\Phi\rangle$. In $G\langle\Phi\rangle$, $\Phi(p)$ and $\Phi(q)$ cannot communicate unless there is an edge between $p$ and $q$. The process ${\bf if}~b~{\bf then}~P~{\bf else}~Q$, not in CCTS, acts as $P$ is the value of $b$ is $true$, and as $Q$ otherwise.

$\mu$ is a process variable binder.
$Q[P/X]$ represents substituting $P$ for every free occurrence of $X$ in $Q$. 
Given $P=G\langle\Phi\rangle$ and $Q=H\langle\Psi\rangle$ with $p\in |G|$ and $|G|\cap |H|=\emptyset$, $P[Q/p]$ represents the process $G[H/p]\langle\Phi^{\prime}\rangle$ with $\Phi^{\prime}(p^{\prime}) = \Phi(p^{\prime})$ for $p^{\prime}\notin |H|$ and $\Phi^{\prime}(p^{\prime}) = \Psi(p^{\prime})$ for $p^{\prime}\in |H|$. In general, a substitution may require $\alpha$-conversions on data variables, symbols and process variables.

Given a process $P\in{\bf Pr}$, we say that ${\sf Sort}(P)\subseteq \Sigma$ is the sort of $P$, where ${\sf Sort}$ is a function to extract symbols from processes and it is defined as follows: ${\sf Sort}(X)= {\sf Sort}(\ast)= {\sf Sort}({\bf 0})=\emptyset$, ${\sf Sort}(\mu X\cdot P)={\sf Sort}(P)$,
${\sf Sort}(P\backslash I)={\sf Sort}(P)\setminus I $, ${\sf Sort}(G\langle\Phi\rangle)=\bigcup_{p\in |G|} {\sf Sort}(\Phi(p))$, ${\sf Sort}(P+Q)={\sf Sort}(P)\cup {\sf Sort}(Q)$, ${\sf Sort}({\bf if}~b~{\bf then}~P~{\bf else}~Q)={\sf Sort}(P)\cup {\sf Sort}(Q)$, and ${\sf Sort}(f(x)\cdot(P_1,\ldots,P_n))= {\sf Sort}(\overline{f}(e)\cdot(P_1,\ldots,P_n)) = \{f\}\cup\bigcup_{i=1}^n{\sf Sort}(P_i)$.

Finally, we introduce some notations which will be frequently used later.
Given two graphs $G$ and $H$ with disjoint vertices and $D\subseteq |G|\times|H|$, we define a new graph $K= G \oplus_D H$, where $|K|=|G|\cup|H|$ and for any $p,q\in|K|$ if $p\frown_G q$ or $p\frown_H q$ or $(p,q)\in D$
then $p\frown_K q$. If $D=\emptyset$, we write
$G\oplus H = G\oplus_D H$.
Given $P=G\langle\Phi\rangle$ and $Q=H\langle\Psi\rangle$ with $|G|\cap |H| = \emptyset$, and  $D\subseteq |G|\times|H|$, we define process $P\oplus_D Q$ as
$(G\oplus_D H)\langle\Phi\cup\Psi\rangle$. When $D$ is empty, we write it as
$P\oplus Q$ for simplicity.
The process $P\oplus_D Q$ can be written as $P\mid Q$, if $D=|G|\times|H|$.
More generally, $\oplus{\vec P}$ stands for  $P_1\oplus\cdots\oplus P_n$ when vector ${\vec P} = (P_1,\ldots,P_n)$. When we consider $P_1,\ldots,P_n$ at the same time, we always assume that their associated graphs are pairwise disjoint.

\subsection{Canonical Processes}
Informally, a process is canonical if all sums in it are guarded.
More precisely, we define {\it canonical processes} ({\sf CP}), {\it canonical guarded sums} ({\sf CGS}) and {\it recursive canonical guarded sums} ({\sf RCGS}) following  \cite{Ehrhard2013ccts} by mutual induction as follows:
\begin{center}\small
$\infer[]{X\in {\sf CP}}{X\in \mathcal{V}}$ $\infer[]{G\langle\Phi\rangle\in{\sf CP}}{\Phi: |G|\rightarrow {\sf RCGS}}$ $\infer[]{P\backslash I\in {\sf CP}}{P\in {\sf CP}\quad I\subseteq \Sigma}$ $\infer[]{{\bf 0}\in {\sf CGS}}{}$ \\
$\infer[]{\ast\in {\sf CGS}}{}$ $\infer[]{f(x)\cdot(P_1,\ldots,P_n)\in {\sf CGS}}{x\in {\bf Var}\quad f\in\Sigma_n\quad P_1,\ldots,P_n\in{\sf CP}}$\\
 $\infer[]{\overline{f}(e)\cdot(Q_1,\ldots,Q_n)\in {\sf CGS}}{e\in {\bf Exp}\quad f\in\Sigma_n\quad Q_1,\ldots,Q_n\in{\sf CP}}$ $\infer[]{S_1+S_2\in{\sf CGS}}{S_1,S_2\in{\sf CGS}}$\\
$\infer[]{{\bf if}~ b ~{\bf then}~ S_1~ {\bf else}~ S_2\in{\sf CGS}}{b\in{\bf BExp}\quad S_1,S_2\in{\sf CGS}}$
$\infer[]{S\in {\sf RCGS}}{S\in {\sf CGS}}$ $\infer[]{\mu X\cdot S \in {\sf RCGS}}{S\in {\sf RCGS}}$
\end{center}
For each recursive canonical guarded sum $S$, there is a canonical guarded sum ${\sf cs}(S)$ defined as follows:\\
$${\sf cs}(S)=
  \left\{
    \begin{array}{ll}
      S & \hbox{if $S$ is a canonical guarded sum } \\
      {\sf cs}(T[S/X]) & \hbox{if } S=\mu X\cdot T\\
    \end{array}
  \right.
$$

If $P= G\langle\Phi\rangle$ is a canonical process, we denote $|P| = |G|$. Meanwhile, for $p\in |G|$, let $P(p)$ stand for $\Phi(p)$ and let $\frown_P$ stand for the edges in $G$.
{\it In the rest of this paper, we only consider canonical processes.}
\begin{lemma}\label{lemma:cs}
If $R$ and $P$ are canonical processes, then $R[P/X]$ is a canonical process.
If $R$ is a (recursive) canonical guarded sum, so is $R[P/X]$.
\end{lemma}

\section{Semantics of VCCTS}\label{VCCTS_semantics}
Let ${\sf Proc}$ represent the set of data-closed and canonical processes in VCCTS.
We assume an evaluation function ${\sf eval}$ for the closed expressions in {\bf Exp} and {\bf BExp}.
There are two ways of dealing with input symbols, $f\in \Sigma$. They are usually referred to as {\it early} semantics and {\it late} semantics, and they vary according to the time when the receiving of a value takes place in an input transition. In this paper, we adopt early semantics.
\subsection{Internal Reduction}
Let $P$, $P^{\prime} \in {\sf Proc}$.
$P$ can reduce to $P^{\prime}$, denoted by $P\xrightarrow[]{}P^{\prime}$, if there are $p,q\in|P|$ and $f\in \Sigma_n$ such that $p\frown_P q$,
\begin{itemize}
  \item ${\sf cs}(P(p))$ is one of the forms $f(x)\cdot(P_1,\ldots,P_n) + S$, ${\bf if}~b~{\bf then}~f(x)\cdot(P_1,\ldots,P_n) + S~{\bf else~}Q$ with ${\sf eval} (b) = true$, or
      ${\bf if}~b~{\bf then}~Q~{\bf else~}f(x)\cdot(P_1,\ldots,P_n) + S$ with ${\sf eval}(b) = false$,
  \item ${\sf cs}(P(q))$ is one of the forms $\overline{f}(e)\cdot(P_1,\ldots,P_n) + S$, ${\bf if}~b~{\bf then}~\overline{f}(e)\cdot(P_1,\ldots,P_n) + S~{\bf else~}Q$ with ${\sf eval} (b) = true$, or
      ${\bf if}~b~{\bf then}~Q~{\bf else~}\overline{f}(e)\cdot(P_1,\ldots,P_n) + S$ with ${\sf eval}(b) = false$,
\end{itemize}
and ${\sf eval}(e)=v$.
The definition of $P^{\prime}$ consists of an associated graph and a function from locations to processes.
For the associated graph,
$|P^{\prime}|=(|P|\setminus \{p,q\})\cup
\bigcup_{i=1}^n|P_i\{v/x\}|\cup
\bigcup_{i=1}^n|Q_i|$ and $\frown_{P^{\prime}}$ is the least symmetric relation
on $|P^{\prime}|$ such that, for any $p^{\prime},q^{\prime}\in|P^{\prime}|$,
$p^{\prime}\frown_{P^{\prime}} q^{\prime}$ if one of the following cases is satisfied:
\begin{itemize}
  \item[(a)] $p^{\prime}\frown_{P_i\{v/x\}} q^{\prime}$ or
         $p^{\prime}\frown_{Q_i} q^{\prime}$ for some $i=1,\ldots,n$
  \item[(b)] $p^{\prime}\in |P_i\{v/x\}|$ and
        $q^{\prime}\in |Q_i|$ for some
        $i=1,\ldots,n$ ({\it the same i on both sides})
  \item[(c)] $\{p^{\prime},q^{\prime}\}\nsubseteq \bigcup_{i=1}^n|P_i\{v/x\}|
         \cup\bigcup_{i=1}^n|Q_i|$ and $\lambda(p^{\prime}) \frown_P
         \lambda(q^{\prime})$
\end{itemize}
where $\lambda:|P^{\prime}|\rightarrow |P|$ is a residual function
defined by
$$\lambda(p^{\prime}) =
 \left\{
  \begin{array}{ll}
    p  & \hbox{if } p^{\prime}\in \bigcup_{i=1}^n|P_i\{v/x\}|  \\
    q & \hbox{if } p^{\prime}\in \bigcup_{i=1}^n|Q_i| \\
    p^{\prime} & \hbox{otherwise}
  \end{array}
\right.$$
For the function, $P^{\prime}(p^{\prime})=P_i\{v/x\}(p^{\prime})$
if $p^{\prime}\in|P_i\{v/x\}|$,
$P^{\prime}(p^{\prime})=Q_i(p^{\prime})$ if $p^{\prime}\in |Q_i|$
and $P^{\prime}(p^{\prime})= P(p^{\prime})$ if
$p^{\prime} \notin \bigcup_{i=1}^n|P_i\{v/x\}|\cup
\bigcup_{i=1}^n|Q_i|$, where $i\in \{1,\ldots,n\}$.

In the reduction, $p\frown_P q$ means that the two processes, $P(p)$ and $P(q)$, can interact.
The interaction drops both prefixes and replaces the vertex $p$ in the graph $G$ of $P$ by the graph $G_1\oplus\cdots\oplus G_n$ (where $G_i$ is the graph of $P_i\{v/x\}$) and the vertex $q$ in the graph $G$ by the graph $H_1\oplus\cdots\oplus H_n$ (where $H_i$ is the graph of $Q_i$).
The connection between $p$ and $q$ in $P$ is inherited by the vertices of $G_i$ and $H_i$ in
$P^{\prime}$, but a process located in $G_i$ (i.e. one of the components of
$P_i\{v/x\}$) cannot communicate with processes located
in $H_j$ if $i\neq j$ (cf. (b)).
The connections between $p$ and other vertices of $P$, distinct from $q$, are
inherited by the vertices of $G_i$, similarly for the vertex $q$ and graph $H_i$, $i\in\{1,\ldots,n\}$ (cf. (c)).
$P\backslash I \xrightarrow[]{} P^{\prime}\backslash I$ if $P\xrightarrow[]{} P^{\prime}$. We denote internal reduction with $\xrightarrow[]{}$, and denote its reflexive and transitive closure with $\xrightarrow[]{}^\ast$.

\subsection{Weak Barbed Congruence}
To endow VCCTS with a non-sequential semantics, it could be able to express more than one action occurring simultaneously. The concept of barb is defined as follows.
\begin{definition}[Barb for {\sf RCGS}]
Let $f\in \overline{\Sigma}$ and $P$ be a recursive canonical guarded sum. We say that $f$ is a barb of $P$, written $P\downarrow_{f}$, if one of the following holds:
\begin{itemize}
  \item $f = g$, $g\in \Sigma$ and ${\sf cs}(P)$ is one of the forms $g(x)\cdot(P_1,\ldots,P_n) + S$, ${\bf if}~b~{\bf then}~g(x)\cdot(P_1,\ldots,P_n) + S~{\bf else~}Q$ with ${\sf eval} (b) = true$, or
      ${\bf if}~b~{\bf then}~Q~{\bf else~}g(x)\cdot(P_1,\ldots,P_n) + S$ with ${\sf eval}(b) = false$;
  \item $f = \overline{g}$, $g\in \Sigma$ and ${\sf cs}(P)$ is one of the forms $\overline{g}(e)\cdot(P_1,\ldots,P_n) + S$, ${\bf if}~b~{\bf then}~\overline{g}(e)\cdot(P_1,\ldots,P_n) + S~{\bf else~}Q$ with ${\sf eval} (b) = true$, or
      ${\bf if}~b~{\bf then}~Q~{\bf else~}\overline{g}(e)\cdot(P_1,\ldots,P_n) + S$ with ${\sf eval}(b) = false$.
\end{itemize}
\end{definition}
\begin{definition}[Barb for {\sf CP}]
We say that a finite subset $B$ of $\overline{\Sigma}$ is a barb of a canonical process $Q$, written $Q\downarrow_B$, if there exist distinct locations $q_i \in |Q|$, such that $Q(q_i)\downarrow_{f_i}$ for each $f_i\in B$ and, moreover, $f_i\notin I$ and $\overline{f_i}\notin I$ if $Q$ is of the form $P\backslash I$.
\end{definition}
\begin{definition}[Weak Barbed Bisimulation]
A binary relation $\mathcal{B}$ on ${\sf Proc}$ is a weak barbed bisimulation if it is symmetric and whenever $(P,Q)\in \mathcal{B}$ the following conditions are satisfied:
\begin{itemize}
  \item for any $P^{\prime} \in {\sf Proc}$, if $P\xrightarrow[]{}^\ast P^{\prime}$, then there exists $Q^{\prime}$ such that $Q\xrightarrow[]{}^{\ast}Q^{\prime}$ and $(P^{\prime},Q^{\prime})\in\mathcal{B}$;
  \item for any $P^{\prime} \in {\sf Proc}$ and any finite set $B\subseteq\overline{\Sigma}$, if $P\xrightarrow[]{}^{\ast}P^{\prime}$ and $P^{\prime}\downarrow_{B}$, then there exists $Q^{\prime}\in {\sf Proc}$ such that $Q\xrightarrow[]{}^{\ast}Q^{\prime}$ and   $Q^{\prime}\downarrow_{B}$.
\end{itemize}
Weak barbed bisimilarity, $\stackrel{\bullet}{\approx}$, is the union of all weak barbed bisimulations.
\end{definition}

\begin{lemma}\label{lemma:barbbisim}
$\stackrel{\bullet}{\approx}$ is an equivalence relation.
\end{lemma}

We investigate an important relation contained in weak barbed bisimilarity, i.e. weak barbed congruence with respect to one-hole contexts.
Given a process variable $Y$, a $Y${\it -context} is a canonical process $R$ containing only one free occurrence of $Y$, and $Y$ does not occur in any subprocess of $R$ of the form $\mu X\cdot R^{\prime}$.
A relation $\mathcal{R}\subseteq {\sf Proc}\times{\sf Proc}$ is a congruence if it is an equivalence and for any $Y${\it -context} $R$, $(P, Q)\in \mathcal{R}$ implies $(R[P/Y],R[Q/Y])\in \mathcal{R}$.
\begin{proposition}\label{prop:congruence}
For any equivalence $\mathcal{R}\subseteq {\sf Proc}\times{\sf Proc}$, there exists a largest congruence $\overline{\mathcal{R}}$ contained in $\mathcal{R}$. This relation is characterized by $(P,Q)\in \overline{\mathcal{R}}$ if and only if for any $Y$-context $R$ one has
$(R[P/Y],R[Q/Y])\in \mathcal{R}$.
\end{proposition}

Processes $P,Q\in {\sf Proc}$ are weakly barbed congruent, denoted by $P \cong Q$, if $R[P/Y]\stackrel{\bullet}{\approx}R[Q/Y]$ for any $Y$-context $R$. $\cong$ is the largest congruence in $\stackrel{\bullet}{\approx}$.
\subsection{Localized Transition Systems}
In this part, we add concurrent information of processes to transitions, obtaining {\it localized transition systems} in which unrelated actions (see Definition \ref{unrel_action}) can happen simultaneously.

The localized transition system is defined over ${\sf Proc}$.
The set of actions is denoted by
$Act=\{fv,\overline{f}v\mid v\in {\bf Val}, f\in \Sigma\}$ and ranged over by
$\alpha,\beta,\ldots$.
Given $\alpha = fv$, let $\overline{\alpha}=\overline{f}v$.
We define a function ${\sf symb}: Act \rightarrow \overline{\Sigma}$, satisfying ${\sf symb}(fv)= f$ for any $fv\in Act$.

\begin{figure}[!tp]\small
$$\infer[(\mbox{Input})]{P\xrightarrow[\lambda]{p:fv\cdot({\vec L})}P^{\prime}}{p\in|P|\quad P(p)\in {\sf RCGS}}$$
\begin{itemize}
        \item ${\sf cs}(P(p))$ is one of the forms $f(x)\cdot(P_1,\ldots,P_n) + S$, ${\bf if}~b~{\bf then}~f(x)\cdot(P_1,\ldots,P_n) + S~{\bf else~}Q$ with ${\sf eval} (b) = true$, or
      ${\bf if}~b~{\bf then}~Q~{\bf else~}f(x)\cdot(P_1,\ldots,P_n) + S$ with ${\sf eval}(b) = false$;
        \item ${\vec L} = (|P_1\{v/x\}|, \ldots, |P_n\{v/x\}|)$;
        \item $P^{\prime} = P[\oplus {\vec P}/p]$ with ${\vec P} = (P_1\{v/x\},\ldots,P_n\{v/x\})$;
        \item $\lambda:|P^{\prime}|\rightarrow |P|$ is the residual function defined by $\lambda(p^{\prime}) = p$ if $p^{\prime}\in \bigcup_{i=1}^n |P_i\{v/x\}|$ and $\lambda(p^{\prime}) = p^{\prime}$ otherwise.
      \end{itemize}
$$\infer[(\mbox{Output})]{P\xrightarrow[\lambda]{p:\overline{f}v\cdot({\vec L})}P^{\prime}}{p\in|P|\quad P(p)\in{\sf RCGS}}$$
\begin{itemize}
  \item  ${\sf cs}(P(p))$ is one of the forms $\overline{f}(e)\cdot(P_1,\ldots,P_n) + S$, ${\bf if}~b~{\bf then}~\overline{f}(e)\cdot(P_1,\ldots,P_n) + S~{\bf else~}Q$ with ${\sf eval} (b) = true$, or
      ${\bf if}~b~{\bf then}~Q~{\bf else~}\overline{f}(e)\cdot(P_1,\ldots,P_n) + S$ with ${\sf eval}(b) = false$, and with ${\sf eval}(e)=v$ in each form;
  \item  ${\vec L}= (|P_1|, \ldots, |P_n|)$;
  \item $P^{\prime} = P[\oplus {\vec P}/p]$ with ${\vec P} = (P_1,\ldots,P_n)$;
  \item $\lambda:|P^{\prime}|\rightarrow |P|$ is the residual function defined by $\lambda(p^{\prime}) = p$ if $p^{\prime}\in \bigcup_{i=1}^n |P_i|$ and $\lambda(p^{\prime}) = p^{\prime}$ otherwise.
\end{itemize}
$$\infer[(\mbox{Com1})]{P\oplus_D Q \xrightarrow[\lambda]{\tau} P^{\prime}\oplus_{D^{\prime}}Q^{\prime}}{P\xrightarrow[\lambda_1]{p:\alpha\cdot({\vec L})}P^{\prime}\quad Q\xrightarrow[\lambda_2]{q:\overline{\alpha}\cdot({\vec H})}Q^{\prime}\quad (p,q)\in D  }$$
\begin{itemize}
  \item  $\lambda:|P^{\prime}|\cup |Q^{\prime}|\rightarrow |P|\cup |Q|$ with $\lambda(p^{\prime})=\lambda_1(p^{\prime})$ for $p^{\prime}\in |P^{\prime}|$ and $\lambda(q^{\prime})=\lambda_2(q^{\prime})$ for $q^{\prime}\in |Q^{\prime}|$;
  \item  $(p^{\prime},q^{\prime})\in D^{\prime}$ if either $p^{\prime}\in L_i$ and $q^{\prime}\in H_i$ with $i\in\{1,\ldots,n\}$, or $\{p^{\prime},q^{\prime}\}\nsubseteq \bigcup_{i=1}^n L_i \cup \bigcup_{i=1}^n H_i$ and $(\lambda(p^{\prime}),\lambda(q^{\prime}))\in D$.
\end{itemize}
$$\infer[(\mbox{Res1})]{P\backslash I \xrightarrow[\lambda]{p:\alpha\cdot({\vec L})}P^{\prime}\backslash I}{P \xrightarrow[\lambda]{p:\alpha\cdot({\vec L})}P^{\prime}\quad {\sf symb}(\alpha)\notin I \mbox{ and } \overline{{\sf symb}(\alpha)}\notin I}$$
$$\infer[(\mbox{Res2})]{P\backslash I \xrightarrow[\lambda]{\tau}P^{\prime}\backslash I}{P \xrightarrow[\lambda]{\tau}P^{\prime}}$$
\caption{\small Single-labelled Transitions of VCCTS}\label{LTS1}
\end{figure}
Single-labelled transitions, defined in Fig. \ref{LTS1}, are of the form $P\xrightarrow[\lambda]{\delta}P^{\prime}$, where $\lambda$ is a residual function to keep the traces of locations during transitions and the single label $\delta$ is
$$\delta::= \tau \mid p:\alpha\cdot(\vec{L})$$
where ${\vec L}$ is a vector of sets of vertices, $p$ is a location and $\alpha$ is an action.

In VCCTS, unrelated actions could happen simultaneously, and we use a multiset $\Delta$ to represent it.
Let $\Delta(\delta)$ represent the occurrence number of $\delta$ in $\Delta$. 
We define ${\sf size}(\Delta) = \sum_{\delta\in \Delta}\Delta(\delta)$ to figure out the size of every given multiset $\Delta$.
Let $\Delta_n^{\tau}$ represent a multiset which only contains $n$ $\tau$s, i.e. ${\sf size}(\Delta_n^{\tau})=\Delta_n^{\tau}(\tau)=n$. The union $\uplus$ and the difference $\backslash\!\!\backslash$ on multisets satisfy: for any multisets $\Delta_1$ and $\Delta_2$, $(\Delta_1\uplus\Delta_2)(\delta) = \Delta_1(\delta)+\Delta(\delta)$ and $(\Delta_1\backslash\!\!\backslash\Delta_2)(\delta)=
\mbox{max}(0,\Delta_1(\delta)-\Delta_2(\delta))$.

\begin{definition}[Unrelated Action]\label{unrel_action}
Actions $\alpha_1$ and $\alpha_2$ are unrelated if ${\sf symb}(\alpha_1)$ $\neq {\sf symb}(\alpha_2)$. A multiset of labels $\Delta$ is pairwise unrelated, denoted by ${\sf PUnrel}(\Delta)$, if for every $(p:\alpha_1\cdot({\vec L_1}), q:\alpha_2\cdot({\vec L_2}))\in \Delta\times \Delta$ with $p\neq q$, actions $\alpha_1$ and $\alpha_2$ are unrelated.
\end{definition}

When we say that a multiset $\Delta$ is pairwise unrelated, we do not take $\tau$ into account. For instance, $\{\tau,\tau\}$ is pairwise unrelated trivially.
Fig. \ref{LTS2} defines the multi-labelled transition rule (Com2) for parallel composed processes.
We can easily extend multi-labelled transitions to canonical processes.
Moreover, in $P\xrightarrow[\lambda]{\Delta}P^{\prime}$ when ${\sf size}(\Delta)= 1$, we just use the unique element to represent the multiset.

\begin{figure}[!tp]\small
  $$\infer[(\mbox{Com2})]{P\oplus_D Q\xrightarrow[\lambda]{\Delta} P^{\prime}\oplus_{D^{\prime}} Q^{\prime}}{
    P\xrightarrow[\lambda_1]{\Delta_1}P^{\prime}\quad Q\xrightarrow[\lambda_2]{\Delta_2}Q^{\prime}\quad \Upsilon }$$
  \begin{itemize}
    \item $\Upsilon$ requires ${\sf PUnrel}(\Delta_1), {\sf PUnrel}(\Delta_2)$ and ${\sf PUnrel}(\Delta_1\uplus \Delta_2)$
    \item $\Delta = \Delta^{\prime} \uplus \Delta_m^{\tau}$ with $\Delta^{\prime} = (\Delta_1\uplus \Delta_2)\backslash\!\!\backslash\Delta_0$, $m = \frac{{\sf size}(\Delta_0)}{2}$ and $\Delta_0 = \{p:\alpha\cdot(\vec{L}),q:\overline{\alpha}\cdot
        (\vec{M})\mid (p,q)\in D \mbox{ and } p:\alpha\cdot(\vec{L})\in \Delta_1
        \mbox{ and } q:\overline{\alpha}\cdot(\vec{M})\in \Delta_2\}$;
    \item $\lambda:|P^{\prime}|\cup |Q^{\prime}|\rightarrow |P|\cup|Q|$ with $\lambda(p)=\lambda_1(p)$ for any $p\in |P^{\prime}|$ and $\lambda(q)=\lambda_2(q)$ for any $q\in |Q^{\prime}|$;
    \item  $(p^{\prime},q^{\prime})\in D^{\prime} \subseteq |P^{\prime}|\times |Q^{\prime}|$, if
        \begin{itemize}
        \item either $p^{\prime}\in  L_{i}$ and $q^{\prime}\in M_{i}$ with $i\in\{1,\ldots,n\}$, $p:\alpha\cdot(\vec{L})\in \Delta_0$, $q:\overline{\alpha}\cdot(\vec{M})\in \Delta_0$ and $(p,q)\in D$. ({\it communication between $\alpha$ and $\overline{\alpha}$})
        \item  or $(\lambda(p^{\prime}),\lambda(q^{\prime}))\in D$ and  $\{p^{\prime},q^{\prime}\}\nsubseteq \bigcup_{i=1}^n L_{i} \cup \bigcup_{i=1}^n M_{i}$ for any $p:\alpha\cdot(\vec{L})\in \Delta_0$, $q:\overline{\alpha}\cdot(\vec{M})\in \Delta_0$ and $(p,q)\in D$. ({\it inheritance})
        \end{itemize}
  \end{itemize}
  \caption{\small Multi-labelled Transitions of VCCTS}\label{LTS2}
\end{figure}

In multi-labelled transitions, unrelated actions could occur consecutively and the order in which they occur does not affect the final process, characterized by the following lemma.
\begin{lemma}[Diamond Property]\label{diamond}
\begin{enumerate}
  \item If $P\xrightarrow[\lambda_1]{\delta_1}P^{\prime}$, $Q\xrightarrow[\lambda_2]{\delta_2}Q^{\prime}$ and $P\oplus_D Q\xrightarrow[\lambda]{\{\delta_1,\delta_2\}}
   P^{\prime}\oplus_{D^{\prime}}Q^{\prime}$ (i.e. $\{\delta_1,\delta_2\}$ is pairwise unrelated), then we have
   $P\oplus_D Q\xrightarrow[\mu_1]{\delta_1}P^{\prime}\oplus_{D_1}Q
   \xrightarrow[\mu_2]{\delta_2}
   P^{\prime}\oplus_{D^{\prime}}Q^{\prime}$, $P\oplus_D Q\xrightarrow[\rho_1]{\delta_2}P\oplus_{D_2}Q^{\prime}
   \xrightarrow[\rho_2]{\delta_1}
   P^{\prime}\oplus_{D^{\prime}}Q^{\prime}$ and
   $\mu_1\circ \mu_2 = \rho_1\circ \rho_2 = \lambda$.
  \item Given a process $P$, if $P\xrightarrow[\lambda]{\Delta}P^{\prime}$ then there exist $P_0 = P$, $P_n = P^{\prime}$, and $P_i \xrightarrow[\lambda_{i+1}]{\delta_{i+1}}P_{i+1}$ such that $P_0\xrightarrow[\lambda_1]{\delta_1}P_1
      \xrightarrow[\lambda_2]{\delta_2}\cdots
    \xrightarrow[\lambda_{n}]{\delta_{n}}P_n$,
   where $n={\sf size}(\Delta)$, $\delta_{i+1}\in \Delta$ with $i\in \{0,\ldots,n-1\}$ and $\lambda= \lambda_1\circ\cdots\circ\lambda_n$.
\end{enumerate}
\end{lemma}

\noindent{\it Notations.~}
We write $P \xrightarrow[\lambda]{\tau^{\ast}}P^{\prime}$ if there exists
$n\geq 1$ such that $P = P_1$, $P^{\prime} = P_n$, $P_1\xrightarrow[\lambda_1]{\tau}P_2\cdots P_{n-1}$ $\xrightarrow[\lambda_{n-1}]{\tau}P_n$ and $\lambda = \lambda_1\circ \cdots \circ \lambda_{n-1}$.
$\widehat{\Delta}$ represents the multiset with all the invisible labels removed from $\Delta$.
By diamond property, if $P\xrightarrow[\lambda]{\Delta}P^{\prime}$, then we can get $P\xrightarrow[\lambda_1]{\tau^{\ast}}P_1
\xrightarrow[\lambda_2]{\widehat{\Delta}}P_2\xrightarrow[\lambda_3]{\tau^{\ast}}P^{\prime}$ and $\lambda=\lambda_1\circ\lambda_2\circ\lambda_3$ for some processes $P_1$ and $P_2$. $P\xLongrightarrow[\lambda,\lambda_1,\lambda^{\prime}]
                 {\widehat{\Delta}}P^{\prime}$ means that there exist
                  processes $P_1$ and $P_1^{\prime}$ such that
                 $P\xrightarrow[\lambda]{\tau^{\ast}}P_1\xrightarrow[\lambda_1]
                 {\widehat{\Delta}}P_1^{\prime}\xrightarrow[\lambda^{\prime}]
                {\tau^{\ast}}P^{\prime}$.
\subsection{Weak Bisimulation}
As usual, it is hard to handle barbed congruence directly, and bisimilarity is a convenient tool for this.
We define an early weak bisimulation on ${\sf Proc}$ through triples $(P,E,Q)$ by taking locations into account, where $E \subseteq |P|\times|Q|$ specifies the pairs of corresponding subprocesses to be considered together.

\begin{definition}[Localized Relation \cite{Ehrhard2013ccts}]
A localized relation on ${\sf Proc}$ is a set $\mathcal{R}\subseteq {\sf Proc}\times\mathcal{P}({\sf Loc}^2)\times{\sf Proc}$ such that, if $(P,E,Q)\in \mathcal{R}$ then $E\subseteq |P|\times|Q|$. $\mathcal{R}$ is symmetric if $(P,E,Q)\in \mathcal{R}$ then $(Q,^t\!\!E,P)\in \mathcal{R}$, where $^t\!E=\{(q,p)\mid (p,q)\in E\}$.
\end{definition}

\begin{definition}[Corresponding Multiset]
Given $\widehat{\Delta}$ (containing observable labels only), a corresponding multiset for $\widehat{\Delta}$, denoted by $\widehat{\Delta}^{c}$, is a multiset of labels such that for each $p:\alpha\cdot(\vec{L})\in \widehat{\Delta}$ there exists a unique label $q:\alpha\cdot(\vec{M})\in \widehat{\Delta}^{c}$ (with the same action $\alpha$), and vice versa.
\end{definition}

\begin{definition}[Localized Early Weak Bisimulation]
A symmetric localized relation $\mathcal{S}$ is a localized early weak bisimulation such that:
\begin{itemize}
  \item if $(P,E,Q)\in\mathcal{S}$ and $P \xrightarrow[\lambda]{\tau}P^{\prime}$ then
   $Q\xrightarrow[\rho]{\tau^{\ast}}Q^{\prime}$ with $(P^{\prime},E^{\prime},Q^{\prime})\in\mathcal{S}$ for some $E^{\prime}\subseteq|P^{\prime}|\times|Q^{\prime}|$ such that, if $(p^{\prime},q^{\prime})\in E^{\prime}$ then $(\lambda(p^{\prime}),\rho(q^{\prime}))\in E$;
  \item if $(P,E,Q)\in \mathcal{S}$ and
  $P\xlongrightarrow[\lambda]{\widehat{\Delta}}P^{\prime}$
   then
  $Q\xLongrightarrow[\rho,\rho_1,\rho^{\prime}]
  {\widehat{\Delta}^{c}}Q^{\prime}$ with the conditions that for any $p:\alpha\cdot({\vec L})\in \widehat{\Delta}$ there exists $q:\alpha\cdot({\vec M})\in \widehat{\Delta}^{c}$ such that $(p,\rho(q))\in E$ and $(P^{\prime},E^{\prime},Q^{\prime})\in \mathcal{S}$ for some $E^{\prime}\subseteq|P^{\prime}|\times|Q^{\prime}|$ such that if $(p^{\prime},q^{\prime})\in E^{\prime}$ then $(\lambda(p^{\prime}),\rho\rho_1\rho^{\prime}(q^{\prime}))\in E$, and moreover, if $n\geq 2$ then for any pair of labels $p:\alpha\cdot({\vec L})\in \widehat{\Delta}$ and $q:\alpha\cdot({\vec M})\in \widehat{\Delta}^c$ either $(p^{\prime},\rho^{\prime}(q^{\prime}))\in \bigcup_{i=1}^n(L_i\times M_i)$ or $p^{\prime}\notin \bigcup_{i=1}^n L_i$ and $\rho^{\prime}(q^{\prime})\notin \bigcup_{i=1}^n M_i$.
\end{itemize}
The localized early weak bisimilarity is the union of all localized early weak bisimulations, denoted by $\approx$.
\end{definition}

\begin{lemma}
$\approx$ is a localized early weak bisimulation.
\end{lemma}

\begin{example}
Compared to Milner's expansion law \cite{Milner1989}, e.g. $(a\mid b) = a.b + b.a$, one can easily check that $\overline{f}(1)\cdot({\bf 0})\mid \overline{g}(2)\cdot({\bf 0})~~ /\!\!\!\!\!\!\approx~\overline{f}(1)\cdot(\overline{g}(2)\cdot({\bf 0}))+ \overline{g}(2)\cdot(\overline{f}(1)\cdot({\bf 0}))$.
\end{example}

The relationship between the two bisimulations is characterized by the following proposition.
\begin{proposition}\label{propsitionbisimulationbarb}
If $P\approx Q$ then $P\stackrel{\bullet}{\approx}Q$.
\end{proposition}


\begin{theorem}\label{bisimilationCongruence}
$\approx$ is a congruence.
\end{theorem}

\begin{theorem}\label{maintheorem}
If $P\approx Q$, then $P\cong Q$.
\end{theorem}
\begin{IEEEproof}
It is straightforward, using Theorem \ref{bisimilationCongruence} and Proposition \ref{propsitionbisimulationbarb}.
\end{IEEEproof}

\section{Process Models for Programming languages}\label{VCCTS_prog}
In this section, we define a semantics for a multi-threaded programming language by translating the language into VCCTS, and prove the correctness of this translation. 

\subsection{The Language and Semantics}
\begin{figure}[!tb]\small
\centering
\begin{tabular}{lrl}
   (Value) &${\sf v}\in $& ${\bf Z}$ \\
  (ThdId)  &${\sf t}\in$ & ${\bf N}$ \\
  (MO)&${\sf mo_1}\in$  & $\{{\sf sc, rel}\}$ \\
            &${\sf mo_2}\in$  & $\{{\sf sc, acq}\}$ \\
  (Reg)& ${\sf r}\in$ & $\{{\sf r_1, r_2,r_3, \ldots}\}$\\
  (NVar) & ${\sf x} \in$ & $\{{\sf x, y, z, \ldots}\}$ \\
  (AVar) & ${\sf a} \in$ & $\{{\sf a_1,a_2, a_3,\ldots}\}$ \\
  (Lock) & ${\sf l}\in$ & $\{{\sf l_1,l_2,l_3,\ldots} \}$\\
  (Exp) & ${\sf e}::=$ & ${\sf v \mid r \mid  f_{op}(e_1,\ldots,e_n)}$\\
    (BExp) & ${\sf b}::=$ & ${\sf true}\mid {\sf false}\mid {\sf not~ b}$\\
   & &$\mid {\sf b_1~ and~ b_2}\mid {\sf e_1=e_2}\mid  \ldots$\\
  (Instr) & ${\sf i}::=$ & ${\sf x:=e\mid  a.{\sf store}(e,mo_1) } $\\
  & &$\mid {\sf r:=x\mid r:= a.load(mo_2) } $\\
          &         &${\sf \mid l.{\sf lock}()\mid l.{\sf unlock}()\mid {\sf print}~e}$\\
  (Comm) & ${\sf C}::=$ & ${\sf skip\mid i\mid thread~~t(C(r),e)}$\\
  & &$\mid {\sf  C_1;C_2}\mid {\sf if~ {\sf b} ~{\sf then}~C_1~{\sf else }~C_2} $\\
  & &$\mid {\sf while}~{\sf b}~{\sf do ~C~} $
 \end{tabular}
\caption{\small The syntax of the programming language }\label{fig:langsyntax}
\end{figure}
Fig. \ref{fig:langsyntax} contains the syntax of a toy programming language, featuring a core of C/C++ concurrency \cite{CPP11,C11}.
A program consists of one or more threads 
running concurrently.

We use ${\sf r}$ for registers, ${\sf x}$, ${\sf y}$ and ${\sf z}$ for non-atomic shared variables, and ${\sf a}$ for atomic shared variables.
Because registers in a thread are inaccessible to other threads, we assume that all registers are distinct.
${\bf Z}$ and ${\bf N}$ are the sets of integers and natural numbers, respectively.
Let ${\sf e}$ and ${\sf b}$ be metavariables over arithmetic expressions and boolean expressions, respectively.

The primitive instructions consist of accesses to non-atomic and atomic variables, operations on locks and output instructions.
${\sf a.{\sf store}(e,mo_1)}$ means writing the value of ${\sf e}$ to an atomic variable ${\sf a}$ with memory order ${\sf mo_1}$, and ${\sf r:=a.{\sf load}(mo_2)}$ means reading an atomic variable ${\sf a}$ with memory order ${\sf mo_2}$. Memory orders are fully described in C/C++ standards \cite{CPP11,C11}, and,  for simplicity, we just use a subset of the set of memory-order modifiers to study relaxed memory models further.
${\sf l.{\sf lock}()}$ and ${\sf l.{\sf unlock}()}$ represent acquiring and releasing a lock ${\sf l}$, respectively.
${\sf print~e}$ outputs the value of ${\sf e}$. 
${\sf thread~t(C(r),e)}$ means creating a new thread with the body ${\sf C}$ by passing the value of ${\sf e}$ to the argument ${\sf r}$. For simplicity, we assume that a thread creation cannot be the last command in a thread and that there is only one argument in a thread creation, but it is easy to extend it to multiple arguments. A command may be a ${\sf skip}$, a primitive instruction, a thread creation or a composition of them.

To understand VCCTS model well and to give a reference for the correctness of the translation (defined in Section \ref{Section_trans}), we define an interleaving semantics for the language in a standard way, following \cite{winskel1993formal}. Fig. \ref{fig:runtime} contains run-time constructs for the program.
\begin{figure}[!tp]\small
\begin{center}
\begin{tabular}{lccl}
  (LState) & $\sigma_l$ & $\in$ & Reg $\rightharpoonup$ Value\\
  (GlobalMem) & $\sigma_g$ & $\in$ & (NVar$\cup$AVar) $\rightharpoonup$ Value \\
  (Threads)  & $T$ & $\subseteq$ & LState $\times$ Comm \\
  (AvaLock)  & $\mathbb{L}_a$  & $\subseteq$  & Lock \\
  (BusyLock)     & $\mathbb{L}_b$  & $\subseteq$  & Lock\\
  (GState)  & $s$ & $::=$ & $(\sigma_g,\mathbb{L}_a,\mathbb{L}_b)$\\
  (SeqLabels) & $\iota_{\sf t} $& $::=$ & $\tau\mid  (\overline{\sf out},{\sf v})$\\
  (Labels) & $\iota $& $::=$ & $\iota_{\sf t}\mid  (\overline{\sf fork},{\sf 0})$\\
\end{tabular}
\end{center}
\caption{\small Run-time constructs}\label{fig:runtime}
\end{figure}
A thread is a pair of a local state and a command, $(\sigma_l,{\sf C})$.
A {\it thread configuration} is $(s,(\sigma_l,{\sf C}))$ and $s=(\sigma_g,\mathbb{L}_a,\mathbb{L}_b)$ is the global state for the thread $(\sigma_l,{\sf C})$, where global memory $\sigma_g$ maps shared variables to values, $\mathbb{L}_b$ contains the busy locks and $\mathbb{L}_a$ contains the locks available. $\sigma_l[{\sf r\mapsto v}]$ means updating the register ${\sf r}$ with ${\sf v}$ and leaves other registers unchanged, similarly for $\sigma_g[{\sf x\mapsto v}]$ and $\sigma_g[{\sf a\mapsto v}]$.

\begin{figure}[!tb]\tiny
  $$\infer[(\textsc{WriteX})]{((\sigma_g,\mathbb{L}_a,\mathbb{L}_b),(\sigma_l,{\sf x:=e}))\xlongrightarrow[]{\tau}_{\sf t} ((\sigma_g^{\prime},\mathbb{L}_a,\mathbb{L}_b),(\sigma_l,{\sf skip}))}
  {[\![{\sf e}]\!]_{\sigma_l}={\sf v}& \sigma_g^{\prime} = \sigma_g[{\sf x\mapsto v}]}$$
  $$\infer[(\textsc{ReadX})]{((\sigma_g,\mathbb{L}_a,\mathbb{L}_b),(\sigma_l,{\sf r:=x})) \xlongrightarrow[]{\tau}_{\sf t} ((\sigma_g,\mathbb{L}_a,\mathbb{L}_b),(\sigma_l^{\prime},{\sf skip}))}
  {\sigma_g({\sf x})={\sf v}& \sigma_l^{\prime} = \sigma_l[{\sf r\mapsto v}]}$$
 $$\infer[(\textsc{Print})]{((\sigma_g,\mathbb{L}_a,\mathbb{L}_b),(\sigma_l,{\sf print~e})) \xlongrightarrow[]{(\overline{\sf out},{\sf v})}_{\sf t} ((\sigma_g,\mathbb{L}_a,\mathbb{L}_b),(\sigma_l,{\sf skip}))}
  {[\![{\sf e}]\!]_{\sigma_l}={\sf v}}$$
  $$\infer[(\textsc{Seq1})]{(s,(\sigma_l,{\sf C_1;C_2}))\xlongrightarrow[]{\iota_{\sf t}}_{\sf t}(s^{\prime},(\sigma_l^{\prime},{\sf C_1^{\prime};C_2}))}{(s,(\sigma_l,{\sf C_1}))\xlongrightarrow[]{\iota_{\sf t}}_{\sf t}(s^{\prime},(\sigma_l^{\prime},{\sf C_1^{\prime}}))}$$
  $$\infer[(\textsc{Seq2})]{(s,(\sigma_l,{\sf skip;C}))\xlongrightarrow[]{\iota_{\sf t}}_{\sf t}(s^{\prime},(\sigma_l^{\prime},{\sf C^{\prime}}))}
  {(s,(\sigma_l,{\sf C}))\xlongrightarrow[]{\iota_{\sf t}}_{\sf t}(s^{\prime},(\sigma_l^{\prime},{\sf C^{\prime}}))}$$
  \caption{\small Selected thread transitions}\label{fig:thread transitions}
\end{figure}

A {\it global configuration} is defined as $(s,T)$ which contains  information of the whole program, where $s$ is the global state and $T$ is the set of threads running concurrently. Let $\Gamma$ range over global configurations.
We define two kinds of labelled transitions: $\xlongrightarrow[]{\iota_{\sf t}}_{\sf t}$ for {\it thread transitions} of thread configurations in Fig. \ref{fig:thread transitions} (see Appendix \ref{fullversion} for the full version), and $\xlongrightarrow[]{\iota}_{T}$ for {\it global transitions} of global
configurations in Fig. \ref{fig:global transitions}.

\begin{figure}[!tb]\tiny
  $$\infer[]{(s,T\cup\{(\sigma_l,{\sf C})\})\xlongrightarrow[]{\iota_{\sf t}}_T (s^{\prime},T\cup\{(\sigma_l^{\prime},{\sf C}^{\prime})\})}
  {(s,(\sigma_l,{\sf C}))\xlongrightarrow[]{\iota_{\sf t}}_{\sf t}  (s^{\prime},(\sigma_l^{\prime},{\sf C}^{\prime}))}$$
  $$\infer[]{(s,T\cup\{(\sigma_l,{\sf thread~ t_i(C_i(r),e);C})\})\xlongrightarrow[]{(\overline{\sf fork},{\sf 0})}_T (s,T\cup\{(\sigma_l,{\sf C})\}\cup \{(\{{\sf r}\mapsto {\sf v}\},{\sf C_i})\})}{[\![{\sf e}]\!]_{\sigma_l}={\sf v}}$$
  \caption{\small Global transitions}\label{fig:global transitions}
\end{figure}

\subsection{Translations}\label{Section_trans}
Inspired by \cite{Milner1989}, we give a non-sequential operational semantics to the language by
translating the language into VCCTS. We denote the translation with $[\![~]\!]$.
To handle the local states of threads, we use a process together with an environment to simulate a data closed process.
An {\it environment} is a partial function from data variables to data values, i.e. $\varrho:{\bf Var}\rightharpoonup {\bf Val}$.
An environment is not changed during process evolutions, and it plays the role of ${\sf eval}$.
$P$ is closed under $\varrho$, if ${\sf fv}(P)\subseteq {\sf dom}(\varrho)$.
Let $(P,\varrho)$ implicitly mean that $P$ is closed under $\varrho$.
$\varrho[x\rightarrow v]$ represents a new environment different from $\varrho$ only in mapping $x$ to $v$.

\begin{figure}[!tb]\tiny
\centering
\begin{spacing}{1.2}
\subfloat[Expressions]
{\begin{tabular}{r||l}
   \hline
   Exp(BExp) & {\bf Exp}({\bf BExp})\\
   \hline
   \hline
   ${\sf v}$ & $v$ \\
   \hline
   ${\sf r}$ & $r$ \\
   \hline
   ${\sf f_{ op}(e_1,\ldots,e_n)}$ & $f_{\it op}([\![{\sf e_1}]\!],\ldots,[\![{\sf e_n}]\!])$\\
   \hline
   ${\sf true}$ & ${\it true}$\\
   \hline
   ${\sf false}$ & ${\it false}$\\
   \hline
   ${\sf not~b}$ & $f_{\it not}([\![{\sf b}]\!]) $\\
   \hline
   $ {\sf b_1~and~b_2}$ & $f_{\it and}([\![{\sf b_1}]\!],[\![{\sf b_2}]\!])$\\
   \hline
 \end{tabular}
}
\subfloat[Instuctions]
{\begin{tabular}{r||l}
   \hline
   Instructions & VCCTS\\
   \hline
   \hline
   ${\sf x:=e}$ & $\overline{{\sf write_x}} ([\![{\sf e}]\!])\cdot (\ast)$ \\
   \hline
   ${\sf r:=x}$ & ${\sf read_x}([\![{\sf r}]\!])\cdot (\ast)$ \\
   \hline
   ${\sf print~e}$ & $\overline{{\sf out}}([\![{\sf e}]\!])\cdot (\ast)$\\
   \hline
   ${\sf a. store(e,mo_1)}$ & $\overline{{\sf write_a^{mo_1}}} ([\![{\sf e}]\!])\cdot (\ast)$\\
   \hline
   ${\sf r:= a.load(mo_2)}$ & ${\sf read_a^{mo_2}}([\![{\sf r}]\!])\cdot (\ast)$\\
   \hline
   ${\sf l.lock}()$ & $\overline{{\sf up_l}}(1)\cdot (\ast) $\\
   \hline
   $ {\sf l.unlock}()$ & $\overline{{\sf down_l}}(0)\cdot (\ast)$\\
   \hline
 \end{tabular}
}\\
\subfloat[Commands]
{\begin{tabular}{r||l}
   \hline
   Commands & VCCTS\\
   \hline
   \hline
   ${\sf skip}$ & $\ast$ \\
   \hline
   ${\sf if~b~then~C_1~else~C_2}$ & ${\bf if}~[\![{\sf b}]\!]~{\bf then}~[\![{\sf C_1}]\!]~{\bf else}~[\![{\sf C_2}]\!]$ \\
   \hline
   ${\sf while~b~ do~C}$ & $\mu X.({\bf if}~[\![{\sf b}]\!]~{\bf then}~[\![{\sf C}]\!]\rhd X ~{\bf else}~[\![{\sf skip}]\!])$\\
   \hline
   ${\sf C_1;C_2}$ &
   $\left\{
  \begin{array}{ll}
    ~\overline{{\sf fork}}(0)\cdot([\![{\sf C_2}]\!],[\![{\sf C}]\!]\{[\![{\sf e}]\!]/[\![{\sf r}]\!]\}), & \mbox{if }{\sf C_1=thread~t(C(r),e)} \\
    ~[\![{\sf C_1}]\!]\rhd [\![{\sf C_2}]\!], & \hbox{otherwise}
  \end{array}
  \right.$\\
   \hline
 \end{tabular}
}\\
\subfloat[Run-time states]
{\begin{tabular}{r||l}
   \hline
   Run-time states & VCCTS\\
   \hline
   \hline
   ${\sf x}$ with value ${\sf v}$ & $X_{\sf x}(v) = \mu X\cdot({\sf write_x}(y)\cdot(X(y))+\overline{{\sf read_x}}(v)\cdot(X(v)))$ with $[\![{\sf v}]\!]=v$ \\
   \hline
   ${\sf a}$ with value ${\sf v}$ & $X_{\sf a}(v) = \mu X\cdot({\sf write^{sc}_a}(x)\cdot(X(x))+{\sf
               write^{rel}_a}(x)\cdot(X(x))+$ \\
    & $\qquad\qquad\qquad\overline{{\sf read^{sc}_a}}(v)\cdot(X(v))+
          {\overline{\sf read^{acq}_a}}(v)\cdot(X(v)))$ with $[\![{\sf v}]\!]=v$\\
   \hline
   a available lock&$L_{\sf l}^a = \mu X\cdot({\sf up_l}(x)\cdot({\sf down_l}(y)\cdot(X)))$\\
   \hline
   a busy lock & $L_{\sf l}^b = {\sf down_l}(y)\cdot(L_{\sf l}^a)$\\
   \hline
   $\sigma_g$ & $X_{{\sf x_1}}([\![\sigma_g({\sf x_1})]\!])\oplus\cdots \oplus X_{{\sf x_n}}([\![\sigma_g({\sf x_n})]\!])\oplus X_{{\sf a_1}}([\![\sigma_g({\sf a_1})]\!])$\\
    & $~~\oplus\cdots\oplus X_{{\sf a_m}}([\![\sigma_g({\sf a_m})]\!])~\mbox{ with } {\sf m+n} = |{\sf dom}(\sigma_g)|$\\
   \hline
 $ \mathbb{L}_a$ & $L_{{\sf l_1}}^a\oplus\cdots\oplus L_{{\sf l_n}}^a~~\mbox{ with } {\sf l_i} \in \mathbb{L}_a, {\sf n} = |\mathbb{L}_a|$\\
   \hline
   $ \mathbb{L}_b$ & $L_{{\sf l_1}}^b\oplus\cdots\oplus L_{{\sf l_m}}^b~~\mbox{ with } {\sf l_j}\in \mathbb{L}_b, {\sf m} = |\mathbb{L}_b|$\\
   \hline
   $(\sigma_g,\mathbb{L}_a,\mathbb{L}_b) $& $[\![\sigma_g]\!]\oplus [\![\mathbb{L}_a]\!]\oplus [\![\mathbb{L}_b]\!] $ \\
   \hline
 \end{tabular}
}
\end{spacing}
\caption{\small The definition for translation}\label{fig:translation}
\end{figure}

We translate expressions into expressions in VCCTS in Fig. \ref{fig:translation}(a), and translate instructions and commands into processes in Fig. \ref{fig:translation}(b) and Fig. \ref{fig:translation}(c), respectively. We translate non-atomic variables, atomic variables and locks into processes in Fig. \ref{fig:translation}(d), while translating local states into environments.
Therefore, processes for commands can interact with processes standing for shared variables in VCCTS, just like threads accessing shared variables in concurrent programs.

Here, we use $r,x,y,\ldots$ for data variables in VCCTS.
We define a new operator $\rhd$ for sequential compositions. Given $P$ and $Q$ translated from instructions, $P\rhd Q$ means that $Q$ cannot execute until $P$ terminates, denoted by $P[Q/\ast]$ which means substituting $Q$ for every occurrence of $\ast$ in $P$.
For sequential commands, in Fig. \ref{fig:translation}(c), we distinguish two cases explicitly. The main difference is the treatment of thread creation ${\sf thread~t(C(r),e)}$, which introduces a parallel composition of the created thread and the original thread.

For non-atomic variables, in Fig. \ref{fig:translation}(d), process $X_{\sf x}(v)$ represents a variable ${\sf x}$ with content ${\sf v}$.
The value contained in $X_{\sf x}(v)$ can be read via $\overline{{\sf read_x}}$, and the content can be updated via ${\sf write_x}$.
The translation for an atomic variable ${\sf a}$ with value ${\sf v}$ is similar, and the symbols take memory orders into account.
Let $\mu X\cdot({\sf up_l}(x)\cdot({\sf down_l}(y)\cdot(X)))$ represent a lock.
$L_{\sf l}^a$ means that lock ${\sf l}$ is available, and $L_{\sf l}^b$ means that ${\sf l}$ is busy.


We translate local states into environments.
The environment $\varrho_l$ translated from $\sigma_l$  satisfies that $[\![\sigma_{l}({\sf r})]\!]=\varrho_l([\![{\sf r}]\!])$ for any ${\sf r}\in {\sf dom}(\sigma_{l})$.
For any global configuration, $\Gamma$, we have
\begin{center}
\begin{tabular}{rl}
  $[\![\Gamma]\!]~=$ &$[\![(s,\{(\sigma_{l_{\sf 1}},{\sf C_1}),\ldots, (\sigma_{l_{\sf n}},{\sf C_n})\})]\!]$  \\
   $=$&$(([\![s]\!]\mid ([\![{\sf C_1}]\!]\oplus\cdots\oplus[\![{\sf C_n}]\!]))\backslash {\sf Sort}([\![s]\!]), \varrho )$  \\
\end{tabular}
\end{center}
where $[\![\sigma_{l_{\sf i}}({\sf r})]\!]=\varrho([\![{\sf r}]\!])$ for any ${\sf r}\in {\sf dom}(\sigma_{l_{\sf i}})$, ${\sf 1\leq i\leq n}$.
For labels $\iota$ and $\delta$, we define
$$
{\sf Act}(\delta) =
\left\{
  \begin{array}{ll}
    \alpha & \hbox{if }\delta = p:\alpha\cdot({\vec L}) \\
    \tau   & \hbox{if }\delta = \tau
  \end{array}
\right.
\mbox{~and~}$$
$$
[\![\iota]\!] =
\left\{
  \begin{array}{ll}
    \tau & \hbox{if } \iota = \tau \\
    \overline{\sf out}v & \hbox{if }\iota = (\overline{\sf out},{\sf v}) \mbox{ and } [\![{\sf v}]\!] = v\\
    \overline{\sf fork}0 & \hbox{if }\iota = (\overline{\sf fork},{\sf 0})\mbox{ and } [\![{\sf 0}]\!] = 0
  \end{array}
\right.
$$

\subsection{Correctness of the Translation}
\begin{figure}[!tb]\tiny
\centering
\subfloat[]
{$
\xymatrix{
  \Gamma \ar[d]_{[\![~]\!]} \ar@{}[r]|-{\xlongrightarrow[]{\iota}_{T}} & \Gamma^{\prime}\ar[dr]^{[\![~]\!]}  &\\
  [\![\Gamma]\!]=(P,\varrho) \ar@{}[r]|-{\xlongrightarrow[\lambda]{\delta}}& (P^{\prime},\varrho)\ar@{}[r]|-{\approx }       &[\![\Gamma^{\prime}]\!]      }
$
}
\subfloat[]
{$
\xymatrix{
  [\![\Gamma]\!]=(P,\varrho) \ar@{}[r]|-{\xlongrightarrow[\lambda]{\delta}} & (P^{\prime},\varrho)\ar@{}[r]|-{\approx }       &[\![\Gamma^{\prime}]\!]  \\
   \Gamma \ar[u]^{[\![~]\!]} \ar@{}[r]|-{\xlongrightarrow[]{\iota}_{T}}& \Gamma^{\prime}\ar[ur]_{[\![~]\!]} &     }
$
}
\caption{\small Diagrams of correctness}\label{fig:diagram}
\end{figure}
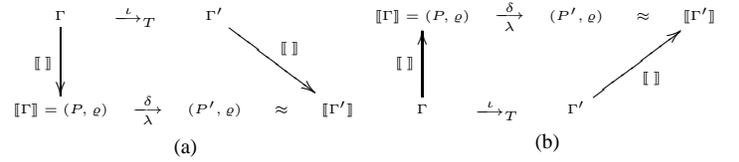
We first prove that
processes translated from the language are canonical.
Since multi-labelled transitions can be serialized (by Lemma \ref{diamond}), we only consider single-labelled transitions here.
Then we prove that a global transition of $\Gamma$ can be simulated by a transition of $[\![\Gamma]\!]$ in VCCTS with respect to weak bisimulation (see Fig. \ref{fig:diagram}(a) and  Theorem \ref{thm-trans}). Conversely, a transition of $[\![\Gamma]\!]$ should reflect a global transition of $\Gamma$ with respect to weak bisimulation (see Fig. \ref{fig:diagram}(b) and Theorem \ref{thm-trans-rev}).
And more examples can be found in Appendix \ref{examples}.
\begin{lemma}\label{transcanonical}
For any $\Gamma$, if $[\![\Gamma]\!] = (P,\varrho)$, then $P$ is canonical.
\end{lemma}
\begin{theorem}\label{thm-trans}
If $\Gamma\xrightarrow[]{\iota}_{T}\Gamma^{\prime}$ and $[\![\Gamma]\!] = (P,\varrho)$, then there exists  $(P,\varrho)\xrightarrow[\lambda]{\delta}(P^{\prime},\varrho)$ such that $(P^{\prime},\varrho)\approx [\![\Gamma^{\prime}]\!]$ and ${\sf Act}(\delta) = [\![\iota]\!]$.
\end{theorem}

\begin{theorem}\label{thm-trans-rev}
Let $[\![\Gamma]\!] = (P,\varrho)$. If $(P,\varrho)\xrightarrow[\lambda]{\delta}(P^{\prime},\varrho)$, then $\Gamma\xrightarrow[]{\iota}_{T}\Gamma^{\prime}$ and $[\![\Gamma^{\prime}]\!] \approx (P^{\prime},\varrho)$ for some $\Gamma^{\prime}$ with ${\sf Act}(\delta) = [\![\iota]\!]$.
\end{theorem}

\section{Related Work}\label{Related_work}
We have no intention of exhausting all the works concerning
concurrent theories and non-sequential semantics,
but only discuss several closely related.
The interested reader is referred to, e.g. \cite{boudol2008twenty}, for a detailed survey of concurrent theories with non-sequential semantics.

The concurrent theory CCTS introduced in \cite{Ehrhard2013ccts} is certainly the most related to the present work,
and the differences between them have been discussed in Section \ref{Intro}.

There are different approaches to provide a non-sequential semantics to CCS:
\begin{itemize}
\item
In \cite{boudol1988non},
a concurrent system with non-sequential semantics,
called the proved transition system, was proposed by G. Boudol and I. Castellani.
This semantics based on a partially ordered multiset labelled transition system,
in which transition labels were identified by proofs, and hence a multiset actions, instead of a simple action.
This semantics preserved the causality and concurrency relations.
In our work, transitions are labelled by multiset of actions and locations,
which endowed a non-sequential semantics for VCCTS which has a richer topological structure inherited from CCTS.
\item
In \cite{boudol1994theory,castellani2001process},
localities were introduced to describe explicitly the distribution of processes,
either from a statical approach where locations are assigned to process statically before processes are executed,
or from a dynamical approach where they are assigned dynamically when executions proceed.
In that setting, a transition carried both an action and a string of locations standing for the accessing path.
However, during the process evolution, the string of locations might be either totally discarded,
or partially recorded.
In our work, we use locations to identify dynamically the distribution of processes,
and use residual functions to keep track of the full information of locations during process evolution.
\item
In \cite{degano1990partial},
Degano et al. proposed an operational semantics for CCS in the SOS style via the partial orderings derivation relation.
The derivation relation was defined on sets of sequential subprocesses of CCS, called grapes,
and described the actions of the sequential subprocesses and the causal dependencies among them.
In our work, canonical guarded sums play the same role as grapes,
but transitions are labelled by multisets which record the actions happening simultaneously at each step and their locations.
 \end{itemize}

As for the concurrent modeling of programming languages,
R. Milner was the first to give an interleaving semantics to a parallel language by translating it into value-passing CCS.
However, the correctness of the translation was discussed informally \cite{Milner1989}.
While, in \cite{hayman2006independence}, Hayman and Winskel studied the semantics of a parallel language in terms of Petri nets, where the semantics of commands were considered as nets,
control flows and program states were treated as conditions in nets.
The correctness of the encoding was also established.
In our work, we provide an operational non-sequential semantics for a toy parallel programming language
based on VCCTS. 
We prove that program executions can be described in term of
VCCTS up-to weak bisimulation.

\section{Conclusions}\label{Conclusions}
The syntax of VCCTS inherits mostly from CCTS. In VCCTS, just like in CCTS, symbols have $n$ arity for connecting with tree automata; graphs, and hence locations, are introduced for defining parallel compositions of processes; locations are therefore needed in semantics and in proofs of some main results. Furthermore, multisets of labels are used for endowing VCCTS with a true concurrency semantics (called non-sequential semantics in the paper). In a sense, VCCTS is more expressive (and more complex) than CCS, much like that tree automata are more expressive (and more complex) than finite automata.

For applications of VCCTS, we give a non-sequential semantics to a multi-threaded programming language by translating it into VCCTS. We propose an algebraical approach for relaxed memory models via transformations, in Appendix \ref{VCCTS_RMM}.

\bibliographystyle{IEEEtran}
\bibliography{IEEEabrv,myref}

\begin{thebibliography}{10}
\providecommand{\url}[1]{#1}
\csname url@samestyle\endcsname
\providecommand{\newblock}{\relax}
\providecommand{\bibinfo}[2]{#2}
\providecommand{\BIBentrySTDinterwordspacing}{\spaceskip=0pt\relax}
\providecommand{\BIBentryALTinterwordstretchfactor}{4}
\providecommand{\BIBentryALTinterwordspacing}{\spaceskip=\fontdimen2\font plus
\BIBentryALTinterwordstretchfactor\fontdimen3\font minus
  \fontdimen4\font\relax}
\providecommand{\BIBforeignlanguage}[2]{{%
\expandafter\ifx\csname l@#1\endcsname\relax
\typeout{** WARNING: IEEEtran.bst: No hyphenation pattern has been}%
\typeout{** loaded for the language `#1'. Using the pattern for}%
\typeout{** the default language instead.}%
\else
\language=\csname l@#1\endcsname
\fi
#2}}
\providecommand{\BIBdecl}{\relax}
\BIBdecl

\bibitem{reisig1985petri}
W.~Reisig, ``Petri nets: an introduction, volume 4 of {EATCS} monographs on
  theoretical computer science,'' 1985.

\bibitem{winskel1989introduction}
G.~Winskel, ``An introduction to event structures,'' \emph{Linear Time,
  Branching Time and Partial Order in Logics and Models for Concurrency}, pp.
  364--397, 1989.

\bibitem{winskel2009events}
------, ``Events, causality and symmetry,'' \emph{The Computer Journal}, p.
  bxp052, 2009.

\bibitem{plotkin1981structural}
G.~D. Plotkin, ``A structural approach to operational semantics,'' 1981.

\bibitem{Milner1989}
R.~Milner, \emph{Communication and Concurrency}.\hskip 1em plus 0.5em minus
  0.4em\relax Upper Saddle River, NJ, USA: Prentice-Hall, Inc., 1989.

\bibitem{boudol1988non}
G.~Boudol and I.~Castellani, ``A non-interleaving semantics for {CCS} based on
  proved transitions,'' \emph{Fundamenta Informaticae}, 1988.

\bibitem{boudol1994theory}
G.~Boudol, I.~Castellani, M.~Hennessy, and A.~Kiehn, ``A theory of processes
  with localities,'' \emph{Formal Aspects of Computing}, vol.~6, no.~2, pp.
  165--200, 1994.

\bibitem{castellani2001process}
I.~Castellani, ``Process algebras with localities,'' \emph{Handbook of Process
  Algebra}, 2001.

\bibitem{degano1990partial}
P.~Degano, R.~De~Nicola, and U.~Montanari, ``A partial ordering semantics for
  {CCS},'' \emph{Theoretical Computer Science}, vol.~75, no.~3, pp. 223--262,
  1990.

\bibitem{kiehn1994comparing}
A.~Kiehn, ``Comparing locality and causality based equivalences,'' \emph{Acta
  Informatica}, vol.~31, no.~8, pp. 697--718, 1994.

\bibitem{ferrari2006synchronised}
G.~L. Ferrari, D.~Hirsch, I.~Lanese, U.~Montanari, and E.~Tuosto,
  ``Synchronised hyperedge replacement as a model for service oriented
  computing,'' in \emph{Formal Methods for Components and Objects}.\hskip 1em
  plus 0.5em minus 0.4em\relax Springer, 2006, pp. 22--43.

\bibitem{Ehrhard2013ccts}
T.~Ehrhard and Y.~Jiang, ``{CCS} for trees,'' 2013,
  \url{http://arxiv.org/abs/1306.1714}.

\bibitem{Milner1992barbed}
R.~Milner and D.~Sangiorgi, ``Barbed bisimulation,'' in \emph{Automata,
  Languages and Programming}.\hskip 1em plus 0.5em minus 0.4em\relax Springer,
  1992, pp. 685--695.

\bibitem{hayman2006independence}
J.~Hayman and G.~Winskel, ``Independence and concurrent separation logic,'' in
  \emph{Logic in Computer Science}, 2006, pp. 147--156.

\bibitem{Adve1996shared}
S.~V. Adve and K.~Gharachorloo, ``Shared memory consistency models: A
  tutorial,'' \emph{Computer}, vol.~29, no.~12, pp. 66--76, 1996.

\bibitem{Batty2011mathematizing}
M.~Batty, S.~Owens, S.~Sarkar, P.~Sewell, and T.~Weber, ``Mathematizing {C++}
  concurrency,'' in \emph{POPL}.\hskip 1em plus 0.5em minus 0.4em\relax ACM,
  2011, pp. 55--66.

\bibitem{Boehm2008foundations}
H.-J. Boehm and S.~V. Adve, ``Foundations of the {C++} concurrency memory
  model,'' in \emph{PLDI}.\hskip 1em plus 0.5em minus 0.4em\relax ACM, 2008,
  pp. 68--78.

\bibitem{Manson2005JMM}
J.~Manson, W.~Pugh, and S.~V. Adve, ``The {Java} memory model,'' in
  \emph{POPL}.\hskip 1em plus 0.5em minus 0.4em\relax New York, NY, USA: ACM,
  2005, pp. 378--391.

\bibitem{Lamport1978hp}
L.~Lamport, ``Time, clocks, and the ordering of events in a distributed
  system,'' \emph{Communications of the ACM}, vol.~21, no.~7, pp. 558--565,
  1978.

\bibitem{owens2009better}
S.~Owens, S.~Sarkar, and P.~Sewell, ``A better x86 memory model: x86-tso,'' in
  \emph{Theorem Proving in Higher Order Logics}.\hskip 1em plus 0.5em minus
  0.4em\relax Springer, 2009, pp. 391--407.

\bibitem{lea_reordering}
D.~Lea, ``The jsr-133 cookbook for compiler writers,'' 2011,
  \url{http://g.oswego.edu/dl/jmm/cookbook.html}.

\bibitem{CPP11}
``{IOS/IEC 14882:2011}. programming language {C++}.''

\bibitem{C11}
``{IOS/IEC 9899:2011}. programming language {C}.''

\bibitem{winskel1993formal}
G.~Winskel, \emph{The formal semantics of programming languages: an
  introduction}.\hskip 1em plus 0.5em minus 0.4em\relax MIT press, 1993.

\bibitem{boudol2008twenty}
G.~Boudol, I.~Castellani, M.~Hennessy, M.~Nielsen, and G.~Winskel, ``Twenty
  years on: Reflections on the {CEDISYS} project. {C}ombining true concurrency
  with process algebra,'' in \emph{Concurrency, Graphs and Models}.\hskip 1em
  plus 0.5em minus 0.4em\relax Springer, 2008, pp. 757--777.

\end{thebibliography}

\newpage
\appendix
\renewcommand{\appendixname}{Appendix~\Alph{section}}

\subsection{Examples For Translations in Section \ref{VCCTS_prog}}\label{examples}
\begin{example}[Concurrency vs. Sequentiality]
Recall the concurrent program ${\sf x:=1}$ $\parallel{\sf y:= 2}$ (in Section \ref{Intro}), which is translated into the following process in VCCTS, denoted by $P$,
$$\overline{{\sf write_x}} ([\![{\sf 1}]\!])\cdot (\ast)\oplus \overline{{\sf write_y}} ([\![{\sf 2}]\!])\cdot (\ast)$$
where $|P| = \{1,2\}$, ${\sf cs}(P(1))= \overline{{\sf write_x}} (1)\cdot (\ast)$ and ${\sf cs}(P(2))= \overline{{\sf write_y}} (2)\cdot (\ast)$.
Because $\overline{{\sf write_x}}$ and $\overline{{\sf write_y}}$ are distinct, they can occur simultaneously in $P$:
$$\overline{{\sf write_x}} (1)\cdot (\ast)\oplus \overline{{\sf write_y}} (2)\cdot (\ast)\xrightarrow[\mathrm{Id}]{\{1:\overline{{\sf write_x}}1\cdot(\{1\}),2:\overline{{\sf write_y}}2\cdot(\{2\})\}}\ast\oplus \ast$$
The program $({\sf x:=1;y:=2}) + ({\sf y:=2;x:=1})$ is translated into
\begin{center}
$(\overline{{\sf write_x}} ([\![{\sf 1}]\!])\cdot (\ast) \rhd \overline{{\sf write_y}} ([\![{\sf 2}]\!])\cdot (\ast)) + \hfill~$\\
$\hfill(\overline{{\sf write_y}} ([\![{\sf 2}]\!])\cdot (\ast) \rhd \overline{{\sf write_x}} ([\![{\sf 1}]\!])\cdot (\ast))$
\end{center}
which can only perform symbols $\overline{{\sf write_x}}$ and  $\overline{{\sf write_y}}$ one by one.
\end{example}

\begin{example}[Thread Creation]\\
The program ${\sf thread~ t(x:=r~(r),1);y:=2}$, written in our toy language, spawns a thread that writes ${\sf 1}$ to ${\sf x}$ and concurrently writes ${\sf 2}$ to ${\sf y}$ in the original thread.
The process $P = [\![{\sf thread~ t(x:=r~(r),1);}$ ${\sf y:=2}]\!] = \overline{\sf fork}(0)\cdot([\![{\sf y:= 2}]\!],[\![{\sf x:=r}]\!]\{1/r\})=\overline{\sf fork}(0)\cdot(\overline{{\sf write_y}} (2)\cdot (\ast),\overline{{\sf write_x}} (1)\cdot (\ast))$ can perform symbol $\overline{\sf fork}$, and then processes $\overline{{\sf write_y}} (2)\cdot (\ast)$ and $\overline{{\sf write_x}} (1)\cdot (\ast)$ execute concurrently without any communications, i.e.
\begin{center}
$ \overline{\sf fork}(0)\cdot(\overline{{\sf write_y}} (2)\cdot (\ast),\overline{{\sf write_x}} (1)\cdot (\ast))\xrightarrow[\lambda]{1:\overline{\sf fork}(0)\cdot(\{2\},\{1\})}\hfill~$\\
$\hfill \overline{{\sf write_y}} (2)\cdot (\ast)\oplus \overline{{\sf write_x}} (1)\cdot (\ast) $
\end{center}
where $|P|=\{1\}$, $|\overline{{\sf write_x}} (1)\cdot (\ast)|=\{1\}$, $|\overline{{\sf write_y}} (2)\cdot (\ast)| = \{2\}$, $\lambda(1) = 1$ and $\lambda(2) = 1$.
Here, locations $1$ and $2$ play the role of thread identifiers.
\end{example}

\subsection{Relaxed Memory Models Based on VCCTS}\label{VCCTS_RMM}

So far we have shown that VCCTS can model multi-threaded programs in a non-sequential approach. In this section, we formalize memory models in VCCTS.

Before doing this, we first recall the concept of data race in programming community, then we define similar concepts for the processes translated from programming languages (in Definition \ref{conflict} and Definition \ref{conflictfree}). In concurrent programs, there exists a data race when different threads simultaneously access the same normal variables and at least one of the accesses is a write:
\begin{definition}[Data Race Freedom]
A configuration $\Gamma = (s,T)$ involves a data race, if there exist $(\sigma_{l_1},{\sf C_1})$, $(\sigma_{l_2},{\sf C_2})\in T$ such that both of them access the same normal variable ${\sf x}$, that is an instruction ${\sf x}:={\sf e}$ or ${\sf r}:= {\sf x}$, and at least one instruction is a write (i.e. ${\sf x}:={\sf e}$). A configuration $\Gamma$ is data race free, if from $\Gamma$ one cannot reach a configuration involving a data race.
\end{definition}

The common way to ensure data race freedom is protecting every normal shared variable with a lock, or using atomic variables (e.g. in  C++ \cite{Boehm2008foundations}) or violate variables (e.g. in Java \cite{Manson2005JMM}).

Compared to concurrent threads, which can concurrently access the same normal variables, processes (corresponding to threads) in different locations can communicate with the same processes corresponding to normal variables.
\begin{definition}[Conflict Relation]\label{conflict}
The conflict relation {\bf CFL} is a binary relation on symbols of processes translated from instructions accessing to normal variables, defined by
$${\bf CFL} = \bigcup_{{\sf x} \in \mbox{NVar}}\{({\sf read_x,\overline{write_x}}),({\sf \overline{write_x},\overline{write_x}}),({\sf \overline{write_x},read_x})\}$$
where $\mbox{NVar}$ is the set of normal variables.
\end{definition}

The conflict relation is symmetric. Intuitively, conflict symbols should not be enabled at different locations simultaneously.
\begin{definition}[Conflict Free Process]\label{conflictfree}
A process $(P,\varrho)$ involves a conflict, if there are distinct locations $p_1,p_2\in|P|$, such that a pair of conflict symbols are enabled at $p_1$ and $p_2$, respectively. A process $(P,\varrho)$ is conflict free, if from it one cannot reach a process which involves a conflict.
\end{definition}
\begin{proposition}
For any configuration $\Gamma$, if $[\![\Gamma]\!]$ is conflict free then $\Gamma$ is data-race free, and vice versa.
\end{proposition}

The transformations, which a multi-threaded system can apply to programs, can be formalized in terms of VCCTS.
The behaviours of a process in a concrete concurrent system can be defined by the semantics of VCCTS.
This procedure is described as follows.
Let $P$ and $Q$ be processes, and $P\hookrightarrow^{\ast} Q$ if $Q$ is obtained from $P$ by applying zero or more times of transformation $\hookrightarrow$.
Given a process $P$, we denote its closure under transformation $\hookrightarrow$ by
${\sf closure}(P) = \{Q\mid P\hookrightarrow^{\ast} Q\}$.
The behaviors of $P$, in a concrete system, are the union of the behaviors of each process in ${\sf closure}(P)$. For the correctness, we have to show that if $Q\in {\sf closure}(P)$ and $P$ is conflict free, then $P\approx Q$ in VCCTS. In other words, the transformation does not introduce any new behaviours.
In the subsequent parts, we will study two instantiations of the transformation, and prove the correctness of each instantiation.

\subsubsection{Compiler Reorderings}
Reordering transformation is a typical technique for compiler optimizations. It allows changing the execution order of memory operations. Of course, not all operations are reorderable without introducing new behaviors. In Doug Lea's cookbook \cite{lea_reordering}, reordering constraints for instructions were discussed in detail.
Here we investigate a class of reorderings appearing in the literature, e.g. \cite{lea_reordering}, in terms of VCCTS.

We use the following terminology to distinguish the classes of processes corresponding to primitive instructions in the programming language. A process is a {\it release process} if it is of the form $\overline{\sf down_l}(0)\cdot(\ast)$, $\overline{\sf write_a^{rel}}(e)\cdot(\ast)$ or $\overline{\sf write_a^{sc}}(e)\cdot(\ast)$, and
a process is an {\it acquire process} if it is of the form $\overline{\sf up_l}(0)\cdot(\ast)$, ${\sf read_a^{acq}}(e)\cdot(\ast)$ or ${\sf read_a^{sc}}(e)\cdot(\ast)$, where ${\sf l}$ is a lock and ${\sf a}$ is an atomic variable. A process is a {\it normal process} if it is of the form $\overline{\sf write_x}(e)\cdot(\ast)$ or
${\sf read_x}(r)\cdot(\ast)$, where ${\sf x}$ is a normal variable.
The basic reordering rules for  processes (corresponding to primitive instructions) are defined in Fig. \ref{fig:reording1}. The rules {\bf WR}, {\bf WW}, {\bf RR} and {\bf RW} describe the reorderings of independent normal processes. We use rules {\bf UW1}, {\bf UW2}, {\bf UR1} and {\bf UR2} to describe the reorderings of a release process and its following normal process.
Rules {\bf WL1}, {\bf WL2}, {\bf RL1} and {\bf RL2} describe the reorderings of an acquire process and its preceding normal process. The reordering template for processes is given in Fig. \ref{fig:reording2}.
\begin{figure}[!tp]
  \begin{description}\tiny
  \begin{spacing}{1.3}
  \item[(WR)] $\overline{\sf write_x}(e)\cdot(\ast)\rhd {\sf read_y}(r)\cdot(\ast)\hookrightarrow{\sf read_y}(r)\cdot(\ast)\rhd\overline{\sf write_x}(e)\cdot(\ast)$, if $r\notin {\sf fv}(e),~{\sf x}\neq {\sf y}$
  \item[(WW)] $ \overline{\sf write_x}(e_1)\cdot(\ast)\rhd \overline{\sf write_y}(e_2)\cdot(\ast)\hookrightarrow\overline{\sf write_y}(e_2)\cdot(\ast)\rhd\overline{\sf write_x}(e_1)\cdot(\ast)$, if ${\sf x}\neq {\sf y}$
  \item[(RR)] ${\sf read_x}(r_1)\cdot(\ast)\rhd {\sf read_y}(r_2)\cdot(\ast)\hookrightarrow{\sf read_y}(r_2)\cdot(\ast)
    \rhd{\sf read_x}(r_1)\cdot(\ast)$, if $r_1\neq r_2$
  \item[(RW)] ${\sf read_x}(r)\cdot(\ast)\rhd\overline{\sf write_y}(e)\cdot(\ast)\hookrightarrow\overline{\sf write_y}(e)\cdot(\ast)\rhd {\sf read_x}(r)\cdot(\ast)$, if $r\notin {\sf fv}(e),~{\sf x}\neq {\sf y}$
  \item[(UW1)] $\overline{\sf down_l}(0)\cdot(\ast)\rhd\overline{\sf write_x}(e)\cdot(\ast)\hookrightarrow\overline{\sf write_x}(e)\cdot(\ast)\rhd\overline{\sf down_l}(0)\cdot(\ast)$, always
  \item[(UW2)]$\overline{\sf write_a^{rel}}(e_1)\cdot(\ast)\rhd\overline{\sf write_x}(e_2)\cdot(\ast)\hookrightarrow\overline{\sf write_x}(e_2)\cdot(\ast)\rhd\overline{\sf write_a^{rel}}(e_1)\cdot(\ast) $, always
  \item[(UR1)] $\overline{\sf down_l}(0)\cdot(\ast)\rhd{\sf read_x}(r)\cdot(\ast)\hookrightarrow{\sf read_x}(r)\cdot(\ast)\rhd\overline{\sf down_l}(0)\cdot(\ast) $, always
  \item[(UR2)] $\overline{\sf write_a^{rel}}(e)\cdot(\ast)\rhd{\sf read_x}(r)\cdot(\ast)\hookrightarrow{\sf read_x}(r)\cdot(\ast)\rhd\overline{\sf write_a^{rel}}(e)\cdot(\ast)$, if $ r\notin {\sf fv}(e)$
  \item[(WL1)] $\overline{\sf write_x}(e)\cdot(\ast)\rhd\overline{\sf up_l}(1)\cdot(\ast)\hookrightarrow\overline{\sf up_l}(1)\cdot(\ast)\rhd\overline{\sf write_x}(e)\cdot(\ast)$, always
  \item[(WL2)] $\overline{\sf write_x}(e_1)\cdot(\ast)\rhd\overline{\sf write_a}(e_2)\cdot(\ast)\hookrightarrow\overline{\sf write_a}(e_2)\cdot(\ast)\rhd\overline{\sf write_x}(e_1)\cdot(\ast)$, always
  \item[(RL1)] ${\sf read_x}(r)\cdot(\ast)\rhd\overline{\sf up_l}(1)\cdot(\ast)\hookrightarrow\overline{\sf up_l}(1)\cdot(\ast)\rhd
    {\sf read_x}(r)\cdot(\ast)$, always
  \item[(RL2)] ${\sf read_x}(r)\cdot(\ast)\rhd\overline{\sf write_a}(e)\cdot(\ast)\hookrightarrow\overline{\sf write_a}(e)\cdot(\ast)\rhd{\sf read_x}(r)\cdot(\ast)$, if $r\notin {\sf fv}(e)$
  \end{spacing}
  \end{description}
\caption{\small Syntactic reordering rules for processes (corresponding to instructions)}\label{fig:reording1}
\end{figure}

\begin{figure}[!tp]\tiny
\setlength{\belowcaptionskip}{-10pt} 
\centering
  $\infer[\mbox{(ID)}]{P\hookrightarrow P}{}$\qquad
  $\infer[\mbox{(RES)}]{P\backslash I \hookrightarrow P^{\prime}\backslash I}{P\hookrightarrow P^{\prime}}$\qquad
  $\infer[\mbox{(SEQ)}]{P\rhd Q \hookrightarrow P^{\prime} \rhd Q^{\prime}}
  {P\hookrightarrow P^{\prime}, Q\hookrightarrow Q^{\prime}}$\\
  ~\\
  $\infer[\mbox{(REC)}]{P[\mu X\cdot P/ X ]\hookrightarrow P^{\prime}[\mu X\cdot P/ X]}{P\hookrightarrow P^{\prime}}$\qquad
  $\infer[\mbox{(SUM)}]{P + Q \hookrightarrow P^{\prime} + Q^{\prime}}
  {P\hookrightarrow P^{\prime}, Q\hookrightarrow Q^{\prime}}$
  $$\infer[\mbox{(IF)}]{{\bf if}~b~{\bf then}~P~{\bf else}~Q\hookrightarrow {\bf if}~b~{\bf then}~P^{\prime}~{\bf else}~Q^{\prime} }
  {P\hookrightarrow P^{\prime}, Q\hookrightarrow Q^{\prime}}$$
  $$\infer[\mbox{(PAR)}]{G\langle\Phi\rangle\hookrightarrow G\langle\Phi^{\prime}\rangle}{\forall p\in|G|.\Phi(p)=P\hookrightarrow P^{\prime}=\Phi^{\prime}(p), {\sf dom}(\Phi)={\sf dom}(\Phi^{\prime})}$$

\caption{The reordering template for processes}\label{fig:reording2}
\end{figure}

\begin{figure}[!tp]\tiny
\setlength{\belowcaptionskip}{-10pt} 
  \centering
  $$\infer[\mbox{if }r\notin {\sf fv}(e)~\mbox{(R-WR)}]{\overline{\sf write_x}(e)\cdot(\ast)\rhd {\sf read_y}(r)\cdot(\ast)\hookrightarrow_{\sf tso} {\sf read_y}(r)\cdot(\ast)\rhd\overline{\sf write_x}(e)\cdot(\ast) }{}$$
  $$\infer[\mbox{(A-WR)}]{\overline{\sf write_x}(e)\cdot(\ast)\rhd {\sf read_x}(r)\cdot(\ast)\rhd P\hookrightarrow_{\sf tso} \overline{\sf write_x}(e)\cdot(\ast)\rhd P\{(e)/r\}}{}$$
  $$\infer[\mbox{(ID)}]{P\hookrightarrow_{\sf tso}P}{} \qquad \infer[\mbox{(RES)}]{P\backslash I \hookrightarrow_{\sf tso} P^{\prime}\backslash I}{P\hookrightarrow_{\sf tso} P^{\prime}}\qquad \infer[\mbox{(SUM)}]{P + Q \hookrightarrow_{\sf tso} P^{\prime} + Q^{\prime}}
  {P\hookrightarrow_{\sf tso} P^{\prime}& Q\hookrightarrow_{\sf tso} Q^{\prime}}$$
  $$\infer[\mbox{(ID-IF)}]{({\bf if}~b~{\bf then}~P_1~{\bf else}~P_2)\rhd Q \hookrightarrow_{\sf tso}{\bf if}~b~{\bf then}~P_1\rhd Q~{\bf else}~P_2\rhd Q}{}$$
  $$\infer[\mbox{(IF)}]
  {\begin{tabular}{l}
    $\overline{\sf write_x}(e)\cdot(\ast)\rhd {\bf if}~b~{\bf then}~P~{\bf else}~Q\hookrightarrow_{\sf tso}$ \\
    $\qquad \qquad \qquad\qquad {\bf if}~b~{\bf then}~\overline{\sf write_x}(e)\cdot(\ast)\rhd P~{\bf else}~\overline{\sf write_x}(e)\cdot(\ast)\rhd Q$ \\
   \end{tabular}
  }{}$$
  $$\infer[\mbox{(REC)}]{\overline{\sf write_x}(e)\cdot(\ast)\rhd\mu X\cdot P \hookrightarrow_{\sf tso} \overline{\sf write_x}(e)\cdot(\ast)\rhd P[(\mu X\cdot P)/X]}{}$$
  $$\infer[\mbox{(PAR)}]{G\langle\Phi\rangle\hookrightarrow_{\sf tso} G\langle\Phi^{\prime}\rangle}{\forall p\in|G|.\Phi(p)=P\hookrightarrow_{\sf tso} P^{\prime}=\Phi^{\prime}(p), {\sf dom}(\Phi)={\sf dom}(\Phi^{\prime})}$$
  \caption{The transformation rules for TSO}\label{fig:tso}
\end{figure}

For the correctness of compiler reorderings, we need to show that the transformation should have the following properties: (i) the transformation preserves conflict freedom, and (ii) every transformed process has the same behaviours as the original process, provided that the original process is conflict free.

\begin{theorem}\label{thm_cr}
For any configuration $\Gamma$ and $[\![\Gamma]\!] = (P,\varrho)$, if $(P,\varrho)$ is conflict free and $P\hookrightarrow^{\ast} Q$, then $(Q,\varrho)$ is also conflict free.
\end{theorem}


\begin{theorem}\label{thm_cr_def}
For any configuration $\Gamma$ and $[\![\Gamma]\!] = (P,\varrho)$, if $(P,\varrho)$ is conflict free and $P\hookrightarrow^{\ast} Q$, then $(P,\varrho)\approx (Q,\varrho)$.
\end{theorem}
\subsubsection{Total Store Ordering}
The Total Store Ordering (TSO) memory model \cite{owens2009better} is a non-trivial model weaker than sequentially consistent model. TSO supports write-to-read reordering in the same threads, and allows a thread to read its own writes earlier. In operational approaches, write buffers are usually taken as a part of the model for TSO.

In this paper, we use transformation rules, given in Fig. \ref{fig:tso}, to reflect write buffers in TSO from a view of the memory. The rule {\bf R-WR} is to describe the reordering of a write with a following read when there is no dependency. The rule {\bf A-WR} allows a thread to read its own write earlier, by eliminating the read process and substituting $e$ for every free occurrence of $r$ in the subsequent process.
The other rules are used to deliver {\bf R-WR} and {\bf A-WR} to the whole processes.
We also establish the correctness of the transformation rules in TSO following the lines in compiler reorderings.
\begin{theorem}\label{thm_tso}
For any configuration $\Gamma$ and $[\![\Gamma]\!] = (P,\varrho)$, if $(P,\varrho)$ is conflict free and $P\hookrightarrow_{\sf tso}^{\ast} Q$, then $(Q,\varrho)$ is also conflict free.
\end{theorem}

\begin{theorem}\label{thm_tso_drf}
For any configuration $\Gamma$ and $[\![\Gamma]\!] = (P,\varrho)$, if $(P,\varrho)$ is conflict free and $P\hookrightarrow_{\sf tso}^{\ast} Q$, then $(P,\varrho)\approx (Q,\varrho)$.
\end{theorem}

\subsection{Basic Proofs in VCCTS}
The proof for Lemma \ref{lemma:cs}.
\begin{IEEEproof}
It is easy by induction on $R$.
\end{IEEEproof}
The proof for Lemma \ref{lemma:barbbisim}.
\begin{IEEEproof}
We need to prove that $\stackrel{\bullet}{\approx}$ is reflexive, symmetric and transitive. It is straightforward from the definition.
\end{IEEEproof}
The proof for Proposition \ref{prop:congruence}.
\begin{IEEEproof}For the first statement,
from the definition of congruence, it is obvious that the identity relation contained in $\mathcal{R}$ is a congruence. And congruences are closed under arbitrary unions and contexts.

For the second statement, let $\mathcal{E}$ be a congruence defined by  $(P,Q)\in \mathcal{E}$ if and only if for any $Y$-context $R$ one has $(R[P/Y],R[Q/Y])\in \mathcal{R}$. Therefore, $\mathcal{E}$ is a congruence contained in $\mathcal{R}$ (because we can take $R = Y$) and hence $\mathcal{E}\subseteq \overline{\mathcal{R}}$. Conversely, let $(P,Q)\in \overline{\mathcal{R}}$ and $R$ be a $Y$-context. Because $\overline{\mathcal{R}}$ is a congruence, we have $(R[P/Y],R[Q/Y])\in \overline{\mathcal{R}}$. We have $(R[P/Y],R[Q/Y])\in \mathcal{R}$ from $\overline{\mathcal{R}}\subseteq \mathcal{R}$ by definition of $\overline{\mathcal{R}}$ and hence $(P,Q)\in \mathcal{E}$.
\end{IEEEproof}

The proof for Lemma \ref{diamond}.
\begin{IEEEproof} For the case (1),
from $P\oplus_D Q\xrightarrow[\lambda]{\{\delta_1,\delta_2\}}
P^{\prime}\oplus_{D^{\prime}}Q^{\prime}$, we have that $\{\delta_1,\delta_2\}$ is unrelated.

From $P\xrightarrow[\lambda_1]{\delta_1}P^{\prime}$ and $Q\xrightarrow[\lambda_2]{\delta_2}Q^{\prime}$, we know $\lambda_1:|P^{\prime}|\rightarrow |P|$ and $\lambda_2:|Q^{\prime}|\rightarrow|Q|$, respectively. From $P\oplus_D Q\xrightarrow[\lambda]{\{\delta_1,\delta_2\}}
P^{\prime}\oplus_{D^{\prime}}Q^{\prime}$, we know $\lambda:|P^{\prime}|\cup |Q^{\prime}|\rightarrow|P|\cup |Q|$ which is consistent with $\lambda_1$ and $\lambda_2$, and $(p^{\prime},q^{\prime})\in D^{\prime}$ if $(\lambda(p^{\prime}),\lambda(q^{\prime}))\in D$.

For $P\oplus_D Q \xrightarrow[\mu_1]{\delta_1}P^{\prime}\oplus_{D_1}Q
\xrightarrow[\mu_2]{\delta_2}
P^{\prime}\oplus_{D_1^{\prime}}Q^{\prime}$, we have $\mu_1:|P^{\prime}|\cup |Q|\rightarrow |P|\cup |Q|$ and $\forall p^{\prime}\in |P^{\prime}|, q\in |Q|, \mu_1(p^{\prime}) = \lambda_1(p^{\prime})$ and $\mu_1(q)=q$. $(p^{\prime},q)\in D_1$ if $(\mu_1(p^{\prime}),\mu_1(q))\in D$, that is $(\lambda_1(p^{\prime}),q)\in D$. We also have $\mu_2:|P^{\prime}|\cup |Q^{\prime}|\rightarrow |P^{\prime}|\cup |Q|$ and $\forall p^{\prime}\in |P^{\prime}|, q^{\prime}\in |Q^{\prime}|, \mu_2(p^{\prime})=p^{\prime}$ and $\mu_2(q^{\prime})=\lambda_2(q^{\prime})$.
$(p^{\prime},q^{\prime})\in D_1^{\prime}$
if $(\mu_2(p^{\prime}),\mu_2(q^{\prime}))\in D_1$, that is $(p^{\prime},\lambda_2(q^{\prime}))\in D_1$. So $(p^{\prime},q^{\prime})\in D_1^{\prime}$, if $(\mu_1\circ \mu_2(p^{\prime}), \mu_1\circ \mu_2(q^{\prime}))\in D$ that is $(\mu_1(p^{\prime}),\mu_2(q^{\prime}))\in D$, i.e. $ (\lambda_1(p^{\prime}),\lambda_2(q^{\prime}))
=(\lambda(p^{\prime}),\lambda(q^{\prime}))\in D$ from the definitions of $\mu_1$ and $\mu_2$.

Therefore, $D_1^{\prime}=D^{\prime}$, and we have
$P\oplus_D Q\xrightarrow[\mu_1]{\delta_1}P^{\prime}\oplus_{D_1}Q
\xrightarrow[\mu_2]{\delta_2}
P^{\prime}\oplus_{D^{\prime}}Q^{\prime}$ with $\mu_1\circ\mu_2 = \lambda$.

For the case $P\oplus_D Q\xrightarrow[\rho_1]{\delta_2}
P\oplus_{D_2}Q^{\prime}\xrightarrow[\rho_2]{\delta_1}
P^{\prime}\oplus_{D_2^{\prime}}Q^{\prime}$, it is similar.

For the case (2), induction on the size of $\Delta$ with the assumption that $\Delta$ is pairwise unrelated.
\end{IEEEproof}
\subsection{Localized Early Weak Bisimulation Is an Equivalence}\label{app1}
\begin{lemma}\label{taus}
Let $\mathcal{R}$ be a localized early weak bisimulation. If $(P,E,Q)\in \mathcal{R}$ and $P\xrightarrow[\lambda]{\tau^{\ast}}P^{\prime}$, then $Q\xrightarrow[\rho]{\tau^{\ast}}Q^{\prime}$ with $(P^{\prime},E^{\prime},Q^{\prime})\in \mathcal{R}$ for some $E^{\prime}\subseteq |P^{\prime}|\times|Q^{\prime}|$ such that if $(p^{\prime},q^{\prime})\in E^{\prime}$ then $(\lambda(p^{\prime}),\rho(q^{\prime}))\in E$.
\end{lemma}
\begin{IEEEproof}
Induction on the length of the derivation of $P\xrightarrow[\lambda]{\tau^{\ast}}P^{\prime}$.
\end{IEEEproof}
\begin{lemma}\label{generals}
If $P\xrightarrow[\lambda]{\tau^{\ast}}P_1$,
$P_1\xLongrightarrow[\lambda_1,\lambda_2,\lambda_3]
{\widehat{\Delta}}P_1^{\prime}$ and $P_1^{\prime}\xrightarrow[\lambda^{\prime}]{\tau^{\ast}}
P^{\prime}$, then
$P\xLongrightarrow[\lambda\lambda_1,\lambda_2,\lambda_3\lambda^{\prime}]
{\widehat{\Delta}}P^{\prime}$.
\end{lemma}
\begin{IEEEproof}
Straightforward.
\end{IEEEproof}
\begin{lemma}\label{lammebisimulation}
A symmetric localized relation $\mathcal{R}\subseteq {\sf Proc}\times\mathcal{P}({\sf Loc}^2)\times{\sf Proc}$ is a localized early weak bisimulation if and only if the following properties hold:
\begin{itemize}
  \item If $(P,E,Q)\in \mathcal{R}$ and $P\xLongrightarrow[\lambda,\lambda_1,\lambda^{\prime}]
      {\widehat{\Delta}}P^{\prime}$, then $Q\xLongrightarrow[\rho,\rho_1,\rho^{\prime}]
      {\widehat{\Delta}^c}Q^{\prime}$ with for any pair of labels $p:\alpha\cdot
      ({\vec L})\in \widehat{\Delta}$ and $q:\alpha\cdot({\vec M})\in \widehat{\Delta}^c$, we have $(\lambda(p),\rho(q))\in E$ and $(P^{\prime},E^{\prime},Q^{\prime})\in \mathcal{R}$ for some $E^{\prime}\subseteq |P^{\prime}|\times|Q^{\prime}|$ such that if $(p^{\prime},q^{\prime})\in E^{\prime}$ then $(\lambda\lambda_1\lambda^{\prime}(p^{\prime}),
       \rho\rho_1\rho^{\prime}(q^{\prime}))\in E$, and moreover, if $n \geq 2$, then  for any pair of labels $p:\alpha\cdot
      ({\vec L})\in \widehat{\Delta}$ and $q:\alpha\cdot({\vec M})\in \widehat{\Delta}^c$ either $(\lambda^{\prime}(p^{\prime}),\rho^{\prime}(q^{\prime}))\in \bigcup_{i=1}^n (L_i\times M_i)$ or $\lambda^{\prime}(p^{\prime})\notin \bigcup_{i=1}^n L_i$ and $\rho^{\prime}(q^{\prime})\notin \bigcup_{i=1}^n M_i$.
  \item If $(P,E,Q)\in \mathcal{R}$ and $P\xrightarrow[\lambda]{\tau^{\ast}}P^{\prime}$, then
      $Q\xrightarrow[\rho]{\tau^{\ast}}Q^{\prime}$ with $(P^{\prime},E^{\prime},Q^{\prime})\in \mathcal{R}$ for some $E^{\prime}\in |P^{\prime}|\times|Q^{\prime}|$ such that if $(p^{\prime},q^{\prime})\in E^{\prime}$ then $(\lambda(p^{\prime}),\rho(q^{\prime}))\in E$.
\end{itemize}
\end{lemma}
\begin{IEEEproof}
($\Leftarrow$) Because $\xrightarrow[\lambda]{\tau}$ and $\xrightarrow[\lambda]{\widehat{\Delta}}$ are special cases of
$\xrightarrow[\lambda]{\tau^{\ast}}$ and $\xLongrightarrow[\lambda,\lambda_1,\lambda^{\prime}]
{\widehat{\Delta}}$ respectively, this direction is obvious.

($\Rightarrow$) For the first statement, assume that $(P,E,Q)\in \mathcal{R}$ and $P\xLongrightarrow[\lambda,\lambda_1,\lambda^{\prime}]
{\widehat{\Delta}}P^{\prime}$ which is $P\xrightarrow[\lambda]{\tau^{\ast}}P_1
\xrightarrow[\lambda_1]{\widehat{\Delta}}
P_1^{\prime}\xrightarrow[\lambda^{\prime}]{\tau^{\ast}}P^{\prime}$, by Lemma \ref{taus} we can get $Q\xrightarrow[\rho]{\tau^{\ast}}Q_1$ with $(P_1,E_1,Q_1)\in \mathcal{R}$ where $E_1$ satisfies the property that if $(p_1,q_1)\in E_1$ then $(\lambda(p_1),\rho(q_1))\in E$.

From $P_1\xrightarrow[\lambda_1]{\widehat{\Delta}}P_1^{\prime}$ and $(P_1,E_1,Q_1)\in \mathcal{R}$, we can get $Q_1\xLongrightarrow[\rho_1,\rho_2,\rho_1^{\prime}]
{\widehat{\Delta}^c}Q_1^{\prime}$ with the condition that for any pair of labels $p:\alpha\cdot({\vec L})\in \widehat{\Delta}$ and $q:\alpha\cdot({\vec M})\in \widehat{\Delta}^c$ we have $(p,\rho_1(q)) \in E_1$ and $(P_1^{\prime},E_1^{\prime},Q_1^{\prime})\in \mathcal{R}$, where $E_1^{\prime}$ satisfies that if $(p_1^{\prime},q_1^{\prime})\in E_1^{\prime}$ then $(\lambda_1(p_1^{\prime}),\rho_1\rho_2\rho_1^{\prime}(q_1^{\prime}))\in E_1$, and, moreover, if $n \geq 2$ then for any pair of labels $p:\alpha\cdot({\vec L})\in \widehat{\Delta}$ and $q:\alpha\cdot({\vec M})\in \widehat{\Delta}^c$ either $(p_1^{\prime},\rho_1^{\prime}(q_1^{\prime}))\in \bigcup_{i=1}^n(L_i\times M_i)$, or $p_1^{\prime}\notin \bigcup_{i=1}^n L_i$ and $\rho_1^{\prime}(q_1^{\prime})\notin \bigcup_{i=1}^n M_i$.

Since $P_1^{\prime}\xrightarrow[\lambda^{\prime}]{\tau^{\ast}}
P^{\prime}$ and $(P_1^{\prime}, E_1^{\prime},Q_1^{\prime})\in \mathcal{R}$,
by Lemma \ref{taus}, we can have $Q_1^{\prime}\xrightarrow[\rho^{\prime}]{\tau^{\ast}}Q^{\prime}$ with $(P^{\prime},E^{\prime},Q^{\prime})\in \mathcal{R}$ where $E^{\prime}$ satisfies that if $(p^{\prime},q^{\prime})\in E^{\prime}$ then $(\lambda^{\prime}(p^{\prime}),\rho^{\prime}(q^{\prime}))\in E_1^{\prime}$.

With $Q\xrightarrow[\rho]{\tau^{\ast}}Q_1$, $Q_1\xLongrightarrow[\rho_1,\rho_2,\rho_1^{\prime}]
{\widehat{\Delta}^c}Q_1^{\prime}$  and $Q_1^{\prime}\xrightarrow[\rho^{\prime}]{\tau^{\ast}}Q^{\prime}$, by Lemma \ref{generals} we can get
$Q\xLongrightarrow[\rho\rho_1,\rho_2,\rho_1^{\prime}\rho^{\prime}]
{\widehat{\Delta}^c}Q^{\prime}$. Meanwhile, we have $(P^{\prime},E^{\prime},Q^{\prime})\in \mathcal{R}$. The conditions on residual functions are satisfied obviously.

For the second statement, it is straightforward from Lemma \ref{taus}.
\end{IEEEproof}
\begin{lemma}[Reflexivity]\label{reflexivity}
Let $\mathcal{I}$ be the localized relation defined by $(P,E,Q)\in \mathcal{I}$ if $P=Q$ and $E={\rm Id}_{|P|}$. Then $\mathcal{I}$ is a localized early weak bisimulation.
\end{lemma}
\begin{IEEEproof}
Straightforward.
\end{IEEEproof}

Let $\mathcal{R}$ and $\mathcal{S}$ be localized relations. We define a localized relation $\mathcal{S}\circ\mathcal{R}$ for the composition of $\mathcal{R}$ and $\mathcal{S}$. $(P,H,R)\in \mathcal{S}\circ\mathcal{R}$ if $H\subseteq |P|\times|R|$ and there exist $Q$, $E$ and $F$ such that $(P,E,Q)\in \mathcal{R}$, $(Q,F,R)\in \mathcal{S}$ and $F\circ E\subseteq H$.
\begin{lemma}[Transitivity]\label{transitivity}
If $\mathcal{R}$ and $\mathcal{S}$ are localized early weak bisimulations, then $\mathcal{S}\circ\mathcal{R}$ is also a localized early weak bisimulation.
\end{lemma}
\begin{IEEEproof}
Obviously, $\mathcal{S}\circ\mathcal{R}$ is symmetric.
Then the proof just follows the definition of the localized early weak bisimulation using the Lemma \ref{lammebisimulation}.

From the hypothesis, let $(P,E,Q)\in \mathcal{R}$, $(Q,F,R)\in \mathcal{S}$ and $(P,H,R)\in\mathcal{S}\circ\mathcal{R}$ with $F\circ E \subseteq H$.

(1) If $P\xLongrightarrow[\lambda,\lambda_1,\lambda^{\prime}]
{\widehat{\Delta}}P^{\prime}$, then $Q\xLongrightarrow[\rho,\rho_1,\rho^{\prime}]
{\widehat{\Delta}^c}Q^{\prime}$ and for any pair of labels $p:\alpha\cdot({\vec L})\in \widehat{\Delta}$ and $q:\alpha\cdot({\vec M})\in \widehat{\Delta}^c$ we have
$(\lambda(p),\rho(q))\in E$ and $(P^{\prime},E^{\prime},Q^{\prime})$ $\in \mathcal{R}$ with $E^{\prime}$ such that if $(p^{\prime}, q^{\prime})\in E^{\prime}$ then $(\lambda\lambda_1\lambda^{\prime}(p^{\prime}),\rho\rho_1
\rho^{\prime}(q^{\prime}))\in E$ and if $n\geq 2$ then either $(\lambda^{\prime}(p^{\prime}),\rho^{\prime}(q^{\prime}))\in \bigcup_{i=1}^n(L_i\times M_i)$ or $\lambda^{\prime}(p^{\prime})\notin \bigcup_{i=1}^n L_i$ and $\rho^{\prime}(q^{\prime})\notin \bigcup_{i=1}^n M_i$.
From $(Q,F,R)\in \mathcal{S}$ and $Q\xLongrightarrow[\rho,\rho_1,\rho^{\prime}]
{\widehat{\Delta}^c}Q^{\prime}$, we have $R\xLongrightarrow[\sigma,\sigma_1,\sigma^{\prime}]
{(\widehat{\Delta}^c)^c}R^{\prime}$ and for any pair of labels $q:\alpha\cdot({\vec M})\in \widehat{\Delta}^c$ and $r:\alpha\cdot({\vec N})\in(\widehat{\Delta}^c)^c$ with $(\rho(q),\sigma(r))\in F$ and $(Q^{\prime},F^{\prime},R^{\prime})\in \mathcal{S}$ with $F^{\prime}$ such that if $(q^{\prime},r^{\prime})\in F^{\prime}$ then $(\rho\rho_1\rho^{\prime}(q^{\prime}),\sigma\sigma_1\sigma^{\prime}(r^{\prime}))\in F$
and if $n\geq 2$ we have either $(\rho^{\prime}(q^{\prime}),\sigma^{\prime}(r^{\prime}))\in \bigcup_{i=1}^n(M_i\times N_i)$ or $\rho^{\prime}(q^{\prime})\notin \bigcup_{i=1}^n M_i$ and $\sigma^{\prime}(r^{\prime})\notin \bigcup_{i=1}^n N_i$.

Therefore, for any pair of labels $p:\alpha\cdot({\vec L})\in \widehat{\Delta}$ and $q:\alpha\cdot({\vec M})\in \widehat{\Delta}^c$ and pair of labels $q:\alpha\cdot({\vec M})\in \widehat{\Delta}^c$ and $r:\alpha\cdot({\vec N})\in(\widehat{\Delta}^c)^c$, we have $(\lambda(p),\sigma(r))$ $\in F\circ E\subseteq H$.
Let $H^{\prime}=\{(p^{\prime},r^{\prime})\in |P^{\prime}|\times|R^{\prime}|\mid (\lambda\lambda_1\lambda^{\prime}(p^{\prime}),
 \sigma\sigma_1\sigma^{\prime}(r^{\prime}))\in H \mbox{ and if }n\geq 2, \mbox{ then either } (\lambda^{\prime}(p^{\prime}),\sigma^{\prime}(r^{\prime}))
 \in \bigcup_{i=1}^n(L_i\times N_i)\mbox{ or } \lambda^{\prime}(p^{\prime})\notin \bigcup_{i=1}^n L_i$ $ \mbox{ and }
 \sigma^{\prime}(r^{\prime})\notin \bigcup_{i=1}^n N_i \}$.
It is obvious that $(P^{\prime},H^{\prime},R^{\prime})$ satisfies the residual conditions from the definition of $H^{\prime}$.
Next, we have to prove $F^{\prime}\circ E^{\prime}\subseteq H^{\prime}$, then show that $(P^{\prime}, H^{\prime}, R^{\prime})\in \mathcal{S}\circ \mathcal{R}$.
If $(p^{\prime},r^{\prime})\in F^{\prime}\circ E^{\prime}$, then there exist $(p^{\prime},q^{\prime})\in E^{\prime}$ and $(q^{\prime},r^{\prime})\in F^{\prime}$. We have to prove $(p^{\prime},r^{\prime})\in H^{\prime}$.

Since $(\lambda\lambda_1\lambda^{\prime}(p^{\prime}),\rho\rho_1\rho^{\prime}
(q^{\prime}))\in E$ and $(\rho\rho_1\rho^{\prime}(q^{\prime}),
\sigma\sigma_1\sigma^{\prime}(r^{\prime}))\in F$, we can get
$(\lambda\lambda_1\lambda^{\prime}(p^{\prime}),
\sigma\sigma_1\sigma^{\prime}(r^{\prime}))\in F\circ E \subseteq H$.
For the case $n\geq 2$, we have to prove that if $\lambda^{\prime}(p^{\prime})\in \bigcup_{i=1}^{n}L_i$ or $\sigma^{\prime}(r^{\prime})\in \bigcup_{i=1}^n N_i$ then $(\lambda^{\prime}(p^{\prime}),\sigma^{\prime}(r^{\prime}))\in L_i\times N_i$ for some $i$.
Without loss of generality, we assume $\lambda^{\prime}(p^{\prime})\in \bigcup_{i=1}^nL_i$ here. So we have $(\lambda^{\prime}(p^{\prime}),\rho^{\prime}(q^{\prime}))\in L_i\times M_i$ for some $i\in \{1,\ldots,n\}$. It implies $\rho^{\prime}(q^{\prime})\in \bigcup_{i=1}^n M_i$.
Then we have $(\rho^{\prime}(q^{\prime}),\sigma^{\prime}(r^{\prime}))\in M_i\times N_i$ for some $i\in \{1,\ldots,n\}$.
So we get $(\lambda^{\prime}(p^{\prime}),\sigma^{\prime}(r^{\prime}))\in \bigcup_{i=1}^n L_i\times N_i$ as required, i.e. $(p^{\prime},r^{\prime})\in H^{\prime}$.

(2) From $(P,E,Q)\in \mathcal{R}$, if $P\xrightarrow[\lambda]{\tau^{\ast}}P^{\prime}$, then we have $Q\xrightarrow[\rho]{\tau^{\ast}}Q^{\prime}$ and $(P^{\prime},E^{\prime},Q^{\prime})\in \mathcal{R}$ with some $E^{\prime}$ such that if $(p^{\prime},q^{\prime})\in E^{\prime}$ then $(\lambda(p^{\prime}),\rho(q^{\prime}))\in E$.
Since $(Q,F,R)\in \mathcal{S}$ and $Q\xrightarrow[\rho]{\tau^{\ast}}Q^{\prime}$, we have
$R\xrightarrow[\sigma]{\tau^{\ast}}R^{\prime}$ and $(Q^{\prime},F^{\prime},R^{\prime})\in \mathcal{S}$ with some $F^{\prime}$ such that if $(q^{\prime},r^{\prime})$ then $(\rho(q^{\prime}),\sigma(r^{\prime}))\in F$. We just get $(P^{\prime},F^{\prime}\circ E^{\prime}, R^{\prime})\in \mathcal{S}\circ \mathcal{R}$ as required. And it is obvious that $F^{\prime}\circ E^{\prime}$ satisfies the residual condition.
\end{IEEEproof}
\begin{proposition}
$\approx$ is an equivalence relation.
\end{proposition}
\begin{IEEEproof}
Because $\approx$ is reflexive by Lemma \ref{reflexivity} , symmetric from the definition and transitive by Lemma \ref{transitivity}.
\end{IEEEproof}

The proof for Proposition \ref{propsitionbisimulationbarb}.
\begin{IEEEproof}
Let $\mathcal{R}$ be a localized early weak bisimulation. Let $\mathcal{B}$ be a binary relation on processes defined by: $(P,Q)\in \mathcal{B}$ if there exists some $E\subseteq |P|\times|Q|$ such that $(P,E,Q)\in \mathcal{R}$. Then we have to prove that $\mathcal{B}$ is a weak bared bisimulation. First, we know that $\mathcal{B}$ is symmetric, because $\mathcal{R}$ is symmetric.\\
(1) Let $(P,Q)\in \mathcal{B}$. If $P\xrightarrow[]{}^{\ast}P^{\prime}$ which is $P\xrightarrow[\lambda]{\tau^{\ast}}P^{\prime}$ for some residual function $\lambda$. Because $\mathcal{R}$ is a localized early weak bisimulation, let $(P,E,Q)\in \mathcal{R}$ with some $E\subseteq |P|\times|Q|$.
We have $Q\xrightarrow[\rho]{\tau^{\ast}}Q^{\prime}$ (i.e. $Q\xrightarrow[]{}^{\ast}Q^{\prime}$) and $(P^{\prime},E^{\prime},Q^{\prime})\in \mathcal{R}$ from Lemma \ref{lammebisimulation}.
Meanwhile, for any $(p^{\prime},q^{\prime})\in E^{\prime}$, the residual function $\rho$ satisfies $(\lambda(p^{\prime}),\rho(q^{\prime}))\in E$ . So we have $(P^{\prime}, Q^{\prime})\in \mathcal{B}$.\\
(2) Let $(P,Q)\in \mathcal{B}$. If $P\xrightarrow[]{}^{\ast}P^{\prime}$ and $P^{\prime}\downarrow_{B}$,
then there exists a transition $P^{\prime}\xrightarrow[\lambda_1^{\prime}]{\widehat{\Delta}}P_1$, where $\widehat{\Delta}$ is a pairwise unrelated multiset of labels of the form $p^{\prime}:fv\cdot({\vec L})$ for each $f\in B$. Since $\mathcal{R}$ is a localized early weak bisimulation, let $(P,E,Q)\in \mathcal{R}$ for some $E\subseteq |P|\times|Q|$.
Then we have $Q\xrightarrow[\rho]{\tau^{\ast}}Q^{\prime}$ (i.e. $Q\xrightarrow[]{}^{\ast}Q^{\prime}$) for some residual function $\rho$, and $E^{\prime}\subseteq |P^{\prime}|\times|Q^{\prime}|$ such that $(P^{\prime},E^{\prime},Q^{\prime})\in \mathcal{R}$.
Because $\mathcal{R}$ is a localized early weak bisimulation and $P^{\prime}\xrightarrow[\lambda_1^{\prime}]{\widehat{\Delta}}P_1$, we have $Q^{\prime}\xLongrightarrow[\rho^{\prime},\rho_1,\rho_2]
{\widehat{\Delta}^c}Q_1$ which means $Q^{\prime}\xrightarrow[\rho^{\prime}]{\tau^{\ast}}Q_1^{\prime}$ (i.e. $Q^{\prime}\xrightarrow[]{}^{\ast}Q_1^{\prime}$) with $Q^{\prime}_1\downarrow_{B}$.
From $P^{\prime}\downarrow_B$ and the transition $P^{\prime}\xrightarrow[\lambda_1^{\prime}]{\widehat{\Delta}}P_1$ satisfying the constraints between $B$ and $\widehat{\Delta}$, we get that
$Q\rightarrow^{\ast}Q_1^{\prime}$ with $Q^{\prime}_1\downarrow_{B}$ as required.
\end{IEEEproof}
\subsection{Localized Early Weak Bisimulation Is a Congruence}\label{app2}
In this part, we intend to prove that localized early weak bisimilarity implies weak barbed congruence. We have proved that localized early weak bisimilarity is an equivalence relation.
To prove an equivalence relation is a congruence, we have to prove that the equivalence relation is preserved by the operators in the structure. Here, the main challenge is to extend the localized relation $\mathcal{R}$ to another localized relation $\mathcal{R}^{\prime}$ to embrace the parallel composition in VCCTS, following the lines in \cite{Ehrhard2013ccts}. In this paper, we extend the syntax and propose a new transition system. Therefore, we have to handle more cases from both canonical guarded sums and multiset transitions. In particular, when consider the cases for single label transitions, e.g. $\tau$-transitions, we just follow the way in \cite{Ehrhard2013ccts} with some modifications to accommodate VCCTS and the new transition rules.

In CCS \cite{Milner1989}, if $\mathcal{R}$ is a weak bisimulation and $P~\mathcal{R}~Q$, then we can prove that $S\mid P$ and $S\mid Q$ are weak bisimilar just by proving that a new relation $\mathcal{R}^{\prime}$ extending $\mathcal{R}$, such that $(S\mid P)~\mathcal{R}^{\prime}~(S\mid Q)$, is a weak bisimulation. However, we cannot simply do this in VCCTS. Moreover we have to record the locations of the subprocesses and the edges of locations which represent the possible communications between subprocesses.
To overcome this obstacle in VCCTS, we use $S\oplus_C P$ to specify the parallel composition of $S$ and $P$ with some $C\subseteq|S|\times|P|$. Similarly, we say that $S\oplus_D Q$ with some relation $D\subseteq |S|\times|Q|$ is a parallel composition of $S$ and $Q$. The relations $C$ and $D$ should satisfy some constraints.
\begin{definition}[Adapted Triple of Relations \cite{Ehrhard2013ccts}]
We say that a triple of relations $(D,D^{\prime},E)$ with $D\subseteq A\times B$, $D^{\prime} \subseteq A\times B^{\prime}$ and $E\subseteq B\times B^{\prime}$ is {\it adapted}, if for any $(a,b,b^{\prime})\in A\times B\times B^{\prime}$ with $(b,b^{\prime})\in E$, $(a,b)\in D$  iff $(a,b^{\prime})\in D^{\prime}$.
\end{definition}

Let $\mathcal{R}$ be a localized relation on processes. We define a new localized relation on processes $\mathcal{R}^{\prime}$, by ensuring that $(U,F,V)\in \mathcal{R}^{\prime}$ and the following conditions are satisfied:
\begin{itemize}
  \item there exist a process $S$, a triple $(P,E,Q)\in \mathcal{R}$, $C\subseteq |S|\times|P|$ and $D\subseteq|S|\times|Q|$ such that $U=S\oplus_C P$ and $V= S\oplus_D Q$,
  \item $(C,D,E)$ is adapted,
  \item $F$ is the relation $(\mbox{Id}_{|S|}\cup E) \subseteq |U|\times|V|$.
\end{itemize}
We call that the relation $\mathcal{R}^{\prime}$ is a {\it parallel extension} of $\mathcal{R}$.
\begin{lemma}[\cite{Ehrhard2013ccts}]\label{lammeextendRsymmtric}
If $R$ is symmetric, then its parallel extension $\mathcal{R}^{\prime}$ is also symmetric.
\end{lemma}
\begin{proposition}\label{propositionparallel}
If $\mathcal{R}$ is a localized early weak bisimulation, then its parallel extension $\mathcal{R}^{\prime}$ is also a localized early weak bisimulation.
\end{proposition}
\begin{IEEEproof}
We can get that $\mathcal{R}^{\prime}$ is symmetric from Lemma \ref{lammeextendRsymmtric}.

Let $(U,F,V)\in \mathcal{R}^{\prime}$ with $(P,E,Q)\in \mathcal{R}$, $U= S\oplus_C P$, $V= S\oplus_D Q$, $(C,D,E)$ is adapted and $F=\mbox{Id}_{|S|}\cup E$.\\
\textit{\textbf{Case of a $\tau$-transition.}} Given $U\xrightarrow[\lambda]{\tau}U^{\prime}$, we have to show $V\xrightarrow[\rho]{\tau^{\ast}}V^{\prime}$ with $(U^{\prime},F^{\prime},V^{\prime})\in \mathcal{R}^{\prime}$ such that for any $(u^{\prime},v^{\prime})\in F^{\prime}$ implies $(\lambda(u^{\prime}),\rho(v^{\prime}))\in F$. There are three cases for a $\tau${\it -transition} for $U=S\oplus_C P$. Meanwhile, we only focus on canonical processes and the canonical guarded sum ${\sf cs}(P)$ for a recursive canonical guarded sum $P$ may have three forms, i.e.
$$\tiny
{\sf cs}(P)
   =\left\{
  \begin{array}{ll}
    pre\cdot(Q_1,\ldots,Q_n)+T, \\
    {\bf if}~b~{\bf then}~pre\cdot(Q_1,\ldots,Q_n)+T~{\bf else}~T_1, & \hbox{with } {\sf eval}(b) = {\it true}, \mbox{or} \\
    {\bf if}~b~{\bf then}~T_1~{\bf else}~pre\cdot(Q_1,\ldots,Q_n)+T, & \hbox{with }{\sf eval}(b) ={\it false}
  \end{array}
\right.$$
where $pre$ is a prefix, $T$ and $T_1$ are canonical guarded sums and $Q_1,\ldots,Q_n$ are canonical processes. Without loss of generality, we only consider the case ${\sf cs}(P)=pre\cdot(Q_1,\ldots,Q_n)+T$ and the other cases are similar referring to localized transition rules.
\\
(1) The two locations are in $S$, and $S\oplus_C P\xrightarrow[\lambda]{\tau}S^{\prime}\oplus_{C^{\prime}}P$. If $s,t\in |S|$ with $s\frown_{S}t$ such that ${\sf cs}(S(s))= f(x)\cdot{\vec S} + \widetilde{S}$ and ${\sf cs}(S(t))=\overline{f}(e)\cdot{\vec T} + \widetilde{T}$ with ${\sf eval}(e)=v$, where $\widetilde{S}$ and $\widetilde{T}$ are canonical guarded sums. So we have $S\xrightarrow[\mu]{\tau}S^{\prime}$ with
\begin{itemize}
  \item $|S^{\prime}|=(|S|\setminus\{s,t\})\cup \bigcup_{i=1}^n|S_i\{v/x\}|\cup \bigcup_{i=1}^n|T_i|$
  \item and $\frown_{S^{\prime}}$ is the least symmetric relation on $|S^{\prime}|$ such that $s^{\prime}\frown_{S^{\prime}}t^{\prime}$ if $s^{\prime}\frown_{S_i\{v/x\}}t^{\prime}$, or $s^{\prime}\frown_{T_i}t^{\prime}$, or $(s^{\prime},t^{\prime})\in |S_i\{v/x\}|\times|T_i|$ for some $i\in \{1,\ldots,n\}$, or $\{s^{\prime},t^{\prime}\}\nsubseteq \bigcup_{i=1}^n|S_i\{v/x\}|\cup\bigcup_{i=1}^n|T_i|$ and $\mu(s^{\prime})\frown_S\mu(t^{\prime})$
\end{itemize}
where, $n$ is the arity of $f$, ${\vec S} = (S_1,\ldots,S_n)$ and ${\vec T} = (T_1,\ldots,T_n)$.
Note that $\mu$ is a residual function which is defined in the internal reduction definition as:
$\mu(s^{\prime})= s$ if $s^{\prime}\in \bigcup_{i=1}^n|S_i|$, $\mu(s^{\prime})= t$ if $s^{\prime}\in \bigcup_{i=1}^n|T_i|$, and $\mu(s^{\prime}) = s^{\prime}$ otherwise.
Then we have $U^{\prime}=S^{\prime}\oplus_{C^{\prime}} P$, where $C^{\prime}=\{(s^{\prime},p)\in |S^{\prime}|\times|P|\mid (\mu(s^{\prime}),p)\in C\}$ and $\lambda = \mu \cup \mbox{Id}_{|P|}$.

Similarly, for $V=S\oplus_D Q$ we have $V\xrightarrow[\rho]{\tau}V^{\prime}=S^{\prime}\oplus_{D^{\prime}}Q$ with $\rho = \mu \cup \mbox{Id}_{|Q|}$, and $D^{\prime}=\{(s^{\prime},q)\in |S^{\prime}|\times|Q|\mid (\mu(s^{\prime}),q)\in D\}$.

Then we have to show that the triple $(C^{\prime},D^{\prime},E)$ is adapted.
Let $s^{\prime}\in |S^{\prime}|$, $p\in |P|$ and $q\in |Q|$ such that $(p,q)\in E$. If $(s^{\prime},p)\in C^{\prime}$ then $(\mu(s^{\prime}),p)\in C$. Since $(C,D,E)$ is adapted, we have $(\mu(s^{\prime}),q)\in D$. So $(s^{\prime},q)\in D^{\prime}$. The converse is similar and we omit it here.

So we have $(U^{\prime},F^{\prime},V^{\prime})\in \mathcal{R}^{\prime}$ where $F^{\prime} = \mbox{Id}_{|S^{\prime}|}\cup E$. Then we check the residual condition. Given $(u^{\prime},v^{\prime})\in F^{\prime}$, either if $u^{\prime}=v^{\prime}\in |S^{\prime}|$ then $\lambda(u^{\prime})=\rho(v^{\prime})\in |S|$, or if $(u^{\prime},v^{\prime})\in E$ then $(\lambda(u^{\prime}),\rho(v^{\prime}))\in E$. So, in both cases we have $(\lambda(u^{\prime}),\rho(v^{\prime}))\in F$.

The symmetric case is similar, where we have $s,t\in |S|$ with $s\frown_{S}t$ such that ${\sf cs}(S(s))= \overline{f}(e)\cdot{\vec S} + \widetilde{S}$ with ${\sf eval}(e)=v$ and ${\sf cs}(S(t))=f(x)\cdot{\vec T} + \widetilde{T}$ , where $\widetilde{S}$ and $\widetilde{T}$ are canonical guarded sums. \\
~\\
(2) The two locations are in $P$ and $S\oplus_C P\xrightarrow[\lambda]
{\tau} S\oplus_{C^{\prime}}P^{\prime}$. Let $p,r\in |P|$ with $p\frown_{P}r$ such that ${\sf cs}(P(p))=f(x)\cdot{\vec P}+ \widetilde{P}$ and ${\sf cs}(P(r))= \overline{f}(e)\cdot{\vec R} + \widetilde{R}$ with ${\sf eval}(e) = v$, where $\widetilde{P}$ and $\widetilde{R}$ are canonical guarded sums. So we have $P\xrightarrow[\mu]{\tau}P^{\prime}$ with
\begin{itemize}
  \item $|P^{\prime}|=(|P|\setminus\{p,r\})\cup \bigcup_{i=1}^n|P_i\{v/x\}|\cup \bigcup_{i=1}^n|R_i|$
  \item and $\frown_{P^{\prime}}$ is the least symmetric relation on $|P^{\prime}|$ such that $p^{\prime}\frown_{P^{\prime}}r^{\prime}$ if $p^{\prime}\frown_{P_i\{v/x\}}r^{\prime}$, or $p^{\prime}\frown_{R_i}r^{\prime}$, or $(p^{\prime},r^{\prime})\in |P_i\{v/x\}|\times|R_i|$ for some $i\in \{1,\ldots,n\}$, or $\{p^{\prime},r^{\prime}\}\nsubseteq \bigcup_{i=1}^n|P_i\{v/x\}|\cup\bigcup_{i=1}^n|R_i|$ and $\mu(p^{\prime})\frown_P\mu(r^{\prime})$
\end{itemize}
where, $n$ is the arity of $f$, ${\vec P} = (P_1,\ldots, P_2)$ and ${\vec R} = (R_1,\ldots,R_n)$.
$\mu$ is a residual function defined as:
$\mu(p^{\prime}) = p$ if $p^{\prime}\in \bigcup_{i=1}^n|P_i|$, $\mu(p^{\prime}) = r$ if $p^{\prime}\in \bigcup_{i=1}^n|R_i|$, and $\mu(p^{\prime}) = p^{\prime}$ otherwise.
So we have $U^{\prime} = S\oplus_{C^{\prime}} P^{\prime}$ where $C^{\prime}= \{(s,p^{\prime})\in |S|\times|P^{\prime}|\mid (s,\mu(p^{\prime}))\in C\}$ and the residual function $\lambda = \mbox{Id}_{|S|}\cup \mu$.

Since $(P,E,Q)\in \mathcal{R}$, from $P\xrightarrow[\mu]{\tau}P^{\prime}$, we have $Q\xrightarrow[\nu]{\tau^{\ast}}Q^{\prime}$ with $(P^{\prime}, E^{\prime},Q^{\prime})\in \mathcal{R}$ where $E^{\prime}\subseteq|P^{\prime}|\times|Q^{\prime}|$ such that $(p^{\prime},q^{\prime})\in E^{\prime}$ implies $(\mu(p^{\prime}),\nu(q^{\prime}))\in E$. Let $D^{\prime}= \{(s,q^{\prime})\in |S|\times|Q^{\prime}|\mid (s,\nu(q^{\prime}))\in D\}$. From $V^{\prime} = S\oplus_{D^{\prime}}Q^{\prime}$, we have $V\xrightarrow[\rho]{\tau^{\ast}}V^{\prime}$ with $\rho= \mbox{Id}_{|S|}\cup \nu$.

Then we show that the triple $(C^{\prime},D^{\prime},E^{\prime})$ is adapted.
Let $(p^{\prime},q^{\prime})\in E^{\prime}$ and $s\in |S|$. If $(s,p^{\prime})\in C^{\prime}$, then we have $(s,\mu(p^{\prime}))\in C$. Since $(\mu(p^{\prime}),\nu(q^{\prime}))\in E$ and $(C,D,E)$ is adapted, we have $(s,\nu(q^{\prime}))\in D$. So we can get $ (s,q^{\prime})\in D^{\prime}$ from the definition of $D^{\prime}$. The other direction is similar.

So we have $(U^{\prime},F^{\prime}, V^{\prime})\in \mathcal{R}^{\prime}$ where $F^{\prime}=\mbox{Id}_{|S|}\cup E^{\prime}\subseteq |U^{\prime}|\times|V^{\prime}|$. Then we have to check the residual condition. Given $(u^{\prime},v^{\prime})\in F^{\prime}$, either $u^{\prime}=v^{\prime}\in |S|$ and then $\lambda(u^{\prime})=\rho(v^{\prime}) = u^{\prime}$, or $u^{\prime}\in |P^{\prime}|$, $v^{\prime}\in |Q^{\prime}|$ and $(u^{\prime},v^{\prime})\in E^{\prime}$ and then $(\lambda(u^{\prime}),\rho(v^{\prime}))=(\mu(u^{\prime}),\nu(v^{\prime}))\in E$. So we get $(\lambda(u^{\prime}),\rho(v^{\prime}))\in F$.

The symmetric case is similar, where $p,r\in |P|$ with $p\frown_{P}r$ such that ${\sf cs}(P(p))=\overline{f}(e)\cdot{\vec P}+ \widetilde{P}$  with ${\sf eval}(e) = v$ and ${\sf cs}(P(r))= f(x)\cdot{\vec R} + \widetilde{R}$, where $\widetilde{P}$ and $\widetilde{R}$ are canonical guarded sums.\\
~\\
(3) One of the locations from $S$ and the other from $P$, i.e. $S\oplus_C P \xrightarrow[\lambda]{\tau}S^{\prime}\oplus_{C^{\prime}}P^{\prime}$. Let $p\in |P|$ and $s\in |S|$ with $(s,p)\in C$. And we have ${\sf cs}(P(p))= f(x)\cdot{\vec P}+ \widetilde{P}$ and ${\sf cs}(S(s))= \overline{f}(e)\cdot{\vec S} + \widetilde{S}$ with ${\sf eval}(e) = v$, where $\widetilde{P}$ and $\widetilde{S}$ are canonical guarded sums.
Then we have $U^{\prime}= S^{\prime}\oplus_{C^{\prime}}P^{\prime}$ with $S^{\prime} = S[\oplus{\vec S}/s]$ and $P^{\prime} = P[\oplus{\vec P}\{v/x\}/p]$, where $n$ is the arity of $f$, ${\vec S} = (S_1,\ldots,S_n)$ and ${\vec P}\{v/x\} = (P_1\{v/x\},\ldots,P_n\{v/x\})$. Let $C^{\prime}\subseteq|S^{\prime}|\times|P^{\prime}|$, and $(s^{\prime},p^{\prime})\in C^{\prime}$ if
\begin{itemize}
  \item $(s^{\prime},p^{\prime})\in |S_i|\times |P_i\{v/x\}|$ for some $i\in\{1,\ldots,n\}$
  \item or, $(s^{\prime},p^{\prime})\nsubseteq (\bigcup_{i=1}^n |S_i|)\times(\bigcup_{i=1}^{n} |P_i\{v/x\}|)$ and $(\lambda(s^{\prime}),\lambda(p^{\prime}))\in C$
\end{itemize}
where, residual function $\lambda:|U^{\prime}|=|S^{\prime}|\cup|P^{\prime}|\rightarrow |U|= |S|\cup |P|$, and it is defined as follows: $\lambda(s^{\prime})= s$ if $s^{\prime}\in\bigcup_{i=1}^n |S_i|$, $\lambda(p^{\prime}) = p$ if $p^{\prime}\in \bigcup_{i=1}^n |P_i\{v/x\}|$ and $\lambda(u^{\prime})= u^{\prime}$ if $u^{\prime} \in (|S^{\prime}|\setminus \bigcup_{i=1}^n |S_i|) \cup (|P^{\prime}|\setminus \bigcup_{i=1}^n|P_i\{v/x\}|)$.

We have $P\xrightarrow[\lambda]{p:\alpha\cdot({\vec L})}P^{\prime}$ with $\alpha = fv$  such that $v$ is just the value received from $S$, and $L_i = |P_i\{v/x\}|$ for $i\in\{1,\ldots,n\}$. By $(P,E,Q)\in \mathcal{R}$, we have $Q\xLongrightarrow[\rho,\rho_1,\rho^{\prime}]{q:\alpha\cdot({\vec M})}Q^{\prime}$ with $(p,\rho(q))\in E$ and $(P^{\prime},E^{\prime},Q^{\prime})\in \mathcal{R}$ with $E^{\prime}$ such that $(p^{\prime},q^{\prime})\in E^{\prime}$ implies $(\lambda(p^{\prime}),\rho\rho_1\rho^{\prime}(q^{\prime}))\in E$, and
moreover, if $n\geq 2$, then $(p^{\prime},\rho^{\prime}(q^{\prime}))\in L_i\times M_i$ for some $i\in \{1,\ldots,n\}$, or $p^{\prime}\notin \bigcup_{i=1}^n L_i$ and $\rho^{\prime}(q^{\prime})\notin \bigcup_{i=1}^n M_i$.

We can decompose $Q\xLongrightarrow[\rho,\rho_1,\rho^{\prime}]{q:\alpha\cdot({\vec M})}Q^{\prime}$ as
$$ Q\xrightarrow[\rho]{\tau^{\ast}}Q_1\xrightarrow[\rho_1]
{q:\alpha\cdot({\vec M})}Q_1^{\prime}\xrightarrow[\rho^{\prime}]
{\tau^{\ast}}Q^{\prime}.$$
We have $V\xrightarrow[\mu]{\tau^{\ast}}V_1$ with $V_1=S\oplus_{D_1}Q_1$, $D_1=\{(s,q_1)\in |S|\times |Q_1|\mid (s,\rho(q_1))\in D\}$ and $\mu= \mbox{Id}_{|S|}\cup \rho$.

Since $(p,\rho(q))\in E$, $(s,p)\in C$ and $(C,D,E)$ is adapted, we have $(s,\rho(q))\in D$. So $(s,q)\in D_1$ from the definition of $D_1$.
We have $q\in Q_1$ with ${\sf cs}(Q_1(q))= f(x)\cdot{\vec R}+ \widetilde{R}$ and ${\sf cs}(S(s))= \overline{f}(e)\cdot{\vec S}+ \widetilde{S}$ with ${\sf eval}(e)=v$ where $v$ is the same value as the part of derivation for $S(s)$ in $U\xrightarrow[\lambda]{\tau}U^{\prime}$. Then we have $M_i = |R_i\{v/x\}|$ for $i\in \{1,\dots,n\}$.
We can get  $V_1\xrightarrow[\theta]{\tau}V_1^{\prime}=S^{\prime}\oplus_{D_1^{\prime}}
Q_1^{\prime}$ where $D_1^{\prime}\subseteq|S^{\prime}|\times|Q_1^{\prime}|$ which is defined as follows: given $(s^{\prime},q_1^{\prime})\in |S^{\prime}|\times|Q_1^{\prime}|$, we have $(s^{\prime},q_1^{\prime})\in D_1^{\prime}$
\begin{itemize}
  \item if $s^{\prime}\in |S_i|$ and $q_1^{\prime}\in |R_i\{v/x\}|$ for some $i\in\{1,\ldots,n\}$
  \item or, $s^{\prime}\notin \bigcup_{i=1}^n|S_i|$ or $q_1^{\prime} \notin \bigcup_{i=1}^n|R_i\{v/x\}|$ and $(\theta(s^{\prime}),\theta(q_1^{\prime}))\in D_1$,
\end{itemize}
and the residual function $\theta$ is defined by $\theta(v_1^{\prime})=v_1^{\prime}$ if $v_1^{\prime}\in (|S|\setminus\bigcup_{i=1}^n|S_i|)\cup
(|Q_1|\setminus\bigcup_{i=1}^n|R_i\{v/x\}|)$, $\theta(s^{\prime})=s$ if $s^{\prime}\in \bigcup_{i=1}^n|S_i|$ and $\theta(q_1^{\prime})= q_1$ if $q_1^{\prime}\in \bigcup_{i=1}^n|R_i\{v/x\}|$.

We also have $\theta(q_1^{\prime})=\rho_1(q_1^{\prime})$ for any $q_1^{\prime}\in |Q_1^{\prime}|$.

From $Q_1^{\prime}\xrightarrow[\rho^{\prime}]{\tau^{\ast}}Q^{\prime}$, we have $V_1^{\prime}=S^{\prime}\oplus_{D_1^{\prime}}Q_1^{\prime}
\xrightarrow[\mu^{\prime}]{\tau^{\ast}}V^{\prime}= S^{\prime}\oplus_{D^{\prime}}Q^{\prime}$ where $\mu^{\prime}=\mbox{Id}_{|S^{\prime}|}\cup \rho^{\prime}$ and
$D^{\prime}=\{(s^{\prime},q^{\prime})\in |S^{\prime}|\times |Q^{\prime}|\mid (s^{\prime},\rho^{\prime}(q^{\prime}))\in D_1^{\prime}\}$. So, we have $V\xrightarrow[\mu\theta\mu^{\prime}]{\tau^{\ast}}V^{\prime}$.
Let $F^{\prime}\subseteq |U^{\prime}|\times|V^{\prime}|$ be defined by $F^{\prime}= \mbox{Id}_{|S^{\prime}|}\cup E^{\prime}$. It is clear that
$(u^{\prime},v^{\prime})\in F^{\prime}$ implies $(\lambda(u^{\prime}),\mu\theta\mu^{\prime}(v^{\prime}))\in F$, since $(p^{\prime},q^{\prime})\in E^{\prime}$ implies $(\lambda(p^{\prime}),\rho\rho_1\rho^{\prime}(q^{\prime}))\in E$ and $\theta$ and $\rho_1$ coincide on $|Q_1^{\prime}|$.

Then we have to prove $(U^{\prime},F^{\prime},V^{\prime})\in \mathcal{R}^{\prime}$. To prove it, we can just show that the triple $(C^{\prime},D^{\prime},E^{\prime})$ is adapted. Let $s^{\prime}\in |S^{\prime}|$, $p^{\prime}\in |P^{\prime}|$ and $q^{\prime}\in |Q^{\prime}|$ with $(p^{\prime},q^{\prime})\in E^{\prime}$ (i.e. particularly $(\lambda(p^{\prime}),\rho\theta\rho^{\prime}(q^{\prime}))\in E$).

If $(s^{\prime},p^{\prime})\in C^{\prime}$, then we have to show that $(s^{\prime},q^{\prime})\in D^{\prime}$ which is $(s^{\prime},\rho^{\prime}(q^{\prime}))\in D_1^{\prime}$. Referring to the definition of $C^{\prime}$, we analyse it in three cases:
\begin{itemize}
  \item First case: $(s^{\prime},p^{\prime})\in |S_i|\times|P_i\{v/x\}|$ for some $i$. We distinguish two cases as the value of $n$ (the arity of $f$). For $n\geq 2$, since $p^{\prime}\in |P_i\{v/x\}|=L_i$, we must have $\rho^{\prime}(q^{\prime})\in M_i = |R_i\{v/x\}|$ because $(p^{\prime},q^{\prime})\in E^{\prime}$. Then we have $(s^{\prime},\rho^{\prime}(q^{\prime}))\in D_1^{\prime}$ as required.
      For $n=1$, if $\rho^{\prime}(q^{\prime})\in M_1$ we can reason as above.  So assume that $\rho^{\prime}(q^{\prime})\notin M_1=\bigcup_{i=1}^n|R_i\{v/x\}|$. Coming back to the definition of $D_1^{\prime}$, it suffices to prove that $(\theta(s^{\prime}),\rho\theta\rho^{\prime}(q^{\prime}))
      =(s,\rho\rho^{\prime}(q^{\prime}))\in D$. Since $(p^{\prime},q^{\prime})\in E^{\prime}$, we have $(\lambda(p^{\prime}),\rho\theta\rho^{\prime}(q^{\prime}))= (p,\rho\rho^{\prime}(q^{\prime}))\in E$. We also have $(s,p)\in C$, and hence $(s,\rho\rho^{\prime}(q^{\prime}))\in D$ as required for $(C,D,E)$ is adapted.
  \item Second case: $s^{\prime}\notin \bigcup_{i=1}^n|S_i|$. In order to prove $(s^{\prime},q^{\prime})\in D^{\prime}$, it suffices to prove that $(\theta(s^{\prime}),\rho\theta\rho^{\prime}(q^{\prime})) = (s^{\prime},\rho\theta\rho^{\prime}(q^{\prime}))\in D$. And we have $(s^{\prime},p^{\prime})\in C^{\prime}$ and $s^{\prime}\notin \bigcup_{i=1}^n |S_i|$, hence $(\lambda(s^{\prime}),\lambda(p^{\prime}))=
      (s^{\prime},\lambda(p^{\prime}))\in C$. Since $(p^{\prime},q^{\prime})\in E^{\prime}$, we have $(\lambda(p^{\prime}),\rho\theta\rho^{\prime}(q^{\prime}))\in E$. Thus we have $(s^{\prime},\rho\theta\rho^{\prime}(q^{\prime}))\in D$ since $(C,D,E)$ is adapted.
  \item Third case: $s^{\prime}\in \bigcup_{i=1}^n|S_i|$ and $p^{\prime}\notin \bigcup_{i=1}^n|P_i\{v/x\}|$, so we have $(\lambda(s^{\prime}),\lambda(p^{\prime}))=(s,p^{\prime})\in C$ (by definition of $C^{\prime}$ and $(s^{\prime},p^{\prime})\in C^{\prime}$). Assume $n\geq 2$. Since $(p^{\prime},q^{\prime})\in E^{\prime}$, we must have $\rho^{\prime}(q^{\prime})\notin \bigcup_{i=1}^n M_i$. To prove $(s^{\prime},\rho^{\prime}(q^{\prime}))\in D_1^{\prime}$, it suffices to check that $(\theta(s^{\prime}),\rho\theta\rho^{\prime}(q^{\prime})) = (s,\rho\rho^{\prime}(q^{\prime}))\in D$. It holds since $(C,D,E)$ is adapted, $(s,p^{\prime})\in C$ and $(p^{\prime},\rho\rho^{\prime}(q^{\prime}))\in E$ for $(p^{\prime},q^{\prime})\in E^{\prime}$.
      For $n=1$, if $\rho^{\prime}(q^{\prime})\notin \bigcup_{i=1}^n M_i = M_1$, we can reason as above. If we assume that $\rho^{\prime}(q^{\prime})\in M_1$, then we have $(s^{\prime},\rho^{\prime}(q^{\prime}))\in |S_1|\times M_1$, so $(s^{\prime},\rho^{\prime}(q^{\prime}))\in D_1^{\prime}$.
\end{itemize}

Now we prove the converse. If $(s^{\prime},q^{\prime})\in D^{\prime}$, i.e. $(s^{\prime},\rho^{\prime}(q^{\prime}))\in D_1^{\prime}$, we have to show $(s^{\prime},p^{\prime})\in C^{\prime}$. We also consider three cases.
\begin{itemize}
  \item First case: $s^{\prime}\in |S_i|$ and $\rho^{\prime}(q^{\prime})\in M_i=|R_i\{v/x\}|$ for some $i\in \{1,\ldots,n\}$. If $n\geq 2$ the fact that $(p^{\prime},q^{\prime})\in E^{\prime}$ implies $p^{\prime}\in L_i = |P_i\{v/x\}|$ and hence $(s^{\prime},p^{\prime})\in C^{\prime}$ as required.
      Assume $n=1$. If $p^{\prime}\in L_1$, then $p^{\prime}\in |P_1\{v/x\}|$ and we have $(s^{\prime},p^{\prime})\in |S_1|\times|P_1\{v/x\}|$. We have
      $(s^{\prime},p^{\prime})\in C^{\prime}$. If $p^{\prime}\notin L_1$, then $p^{\prime}\notin |P_1\{v/x\}|$.
      Since $(p^{\prime},q^{\prime})\in E^{\prime}$, we have $(\lambda(p^{\prime}),\rho\theta\rho^{\prime}(q^{\prime}))\in E$, i.e. $(p^{\prime},\rho(q))\in E$. Since we have $(s^{\prime},q^{\prime})\in D^{\prime}$, we have $(\theta(s^{\prime}),\rho\theta\rho^{\prime}(q^{\prime}))\in D$, i.e.
      $(s,\rho(q))\in D$. So we have $(s,p^{\prime})\in C$ as $(C,D,E)$ is adapted. Since $(\lambda(s^{\prime}),\lambda(p^{\prime})) = (s,p^{\prime})\in C$ and $p^{\prime}\notin L_1$, we have $(s^{\prime},p^{\prime})\in C^{\prime}$.

  \item Second case: $s^{\prime}\notin \bigcup_{i=1}^n|S_i|$. For the definition of $C^{\prime}$, it suffices to prove $(\lambda(s^{\prime}),\lambda(p^{\prime}))=
      (s^{\prime},\lambda(p^{\prime}))\in C$. Since $(s^{\prime},q^{\prime})\in D^{\prime}$ and $s^{\prime}\notin \bigcup_{i=1}^n|S_i|$, we have $(\theta(s^{\prime}),\rho\theta\rho^{\prime}(q^{\prime}))=
      (s^{\prime},\rho\theta\rho^{\prime}(q^{\prime}))\in D$. Since $(p^{\prime},q^{\prime})\in E^{\prime}$ we have $(\lambda(p^{\prime}),\rho\theta\rho^{\prime}(q^{\prime}))\in E$, and hence $(s^{\prime},\lambda(p^{\prime}))\in C$ for $(C,D,E)$ is adapted.
  \item Third case: $s^{\prime}\in |S_i|$ for some $i\in\{1,\ldots,n\}$ and $\rho^{\prime}(q^{\prime})\notin\bigcup_{i=1}^n M_i$. If $n\geq 2$, we have $p^{\prime}\notin\bigcup_{i=1}^n L_i$ for $(p^{\prime},q^{\prime})\in E^{\prime}$. To check $(s^{\prime},p^{\prime})\in C^{\prime}$, it suffices to prove
      $(\lambda(s^{\prime}),\lambda(p^{\prime}))=(s,p^{\prime})\in C$.
      We have $(s^{\prime},q^{\prime})\in D^{\prime}$ and hence $(\theta(s^{\prime}),\rho\theta\rho^{\prime}(q^{\prime})) =
      (s,\rho\rho^{\prime}(q^{\prime}))\in D$. Since $(p^{\prime},q^{\prime})\in E^{\prime}$, we have $(\lambda(p^{\prime}),\rho\theta\rho^{\prime}(q^{\prime}))=
      (p^{\prime},\rho\rho^{\prime}(q^{\prime}))\in E$ and hence $(s,p^{\prime})\in C$ for $(C,D,E)$ is adapted. For $n=1$, if $p^{\prime}\in L_1$, we have $(s^{\prime},p^{\prime})\in C^{\prime}$
      since $(s^{\prime},p^{\prime})\in |S_1|\times|P_1\{v/x\}|$. If
      $p^{\prime}\notin L_1$, then $p^{\prime}\notin \bigcup_{i=1}^n|P_i\{v/x\}|$. It suffices to prove that $(\lambda(s^{\prime}),\lambda(p^{\prime}))=(s,p^{\prime})\in C$. We have $(\lambda(p^{\prime}),\rho\theta\rho^{\prime}(q^{\prime}))=
      (p^{\prime},\rho\rho^{\prime}(q^{\prime}))\in E$ because $(p^{\prime},q^{\prime})\in E^{\prime}$ and $(\theta(s^{\prime}),\rho\theta\rho^{\prime}(q^{\prime}))=
      (s,\rho\rho^{\prime}(q^{\prime}))\in D$ for $(s^{\prime},q^{\prime})\in D^{\prime}$ and
      $\rho^{\prime}(q^{\prime})\notin \bigcup_{i=1}^n M_i$. It follows that $(s,p^{\prime})\in C$ as required since $(C,D,E)$ is adapted.
\end{itemize}

The other case is similar, where $p\in |P|$ and $s\in |S|$ such that $(s,p)\in C$, ${\sf cs}(P(p))= \overline{f}(e)\cdot{\vec P}+ \widetilde{P}$ with ${\sf eval}(e) = v$ and ${\sf cs}(S(s))= f(x)\cdot{\vec S}+ \widetilde{S}$, and $\widetilde{P}$ and $\widetilde{S}$ are canonical guarded sums.\\
~\\
\textit{\textbf{Case of $\widehat{\Delta}$ transition.}} Since $(U,F,V)\in \mathcal{R}^{\prime}$ and $(P,E,Q)\in \mathcal{R}$, we have to prove that if $U\xrightarrow[\lambda]{\widehat{\Delta}}U^{\prime}$, then $V\xLongrightarrow[\rho,\rho_1,\rho^{\prime}]
{\widehat{\Delta}^c}V^{\prime}$ and $(U^{\prime},F^{\prime},V^{\prime})\in \mathcal{R}^{\prime}$.

Now, we assume that $S\oplus_C P\xrightarrow[\lambda]{\widehat{\Delta}}
S^{\prime}\oplus_{C^{\prime}}P^{\prime}$. Because we have considered the communications between $S$ and $P$ in the first part above. Here, we only consider the observable transitions from $S$ and $P$ without any communication. So, we have the following two transitions
$S\xrightarrow[\lambda_1]{\widehat{\Delta}_1}S^{\prime}$ and $P\xrightarrow[\lambda_2]{\widehat{\Delta}_2}P^{\prime}$ for $S$ and $P$, respectively, where $\widehat{\Delta}_1\uplus\widehat{\Delta}_2=\widehat{\Delta}$. For the residual functions, we have
$\lambda:|S^{\prime}|\cup |P^{\prime}|\rightarrow |S|\cup |P|$, $\lambda_1:|S^{\prime}|\rightarrow|S|$ and  $\lambda_2:|P^{\prime}|\rightarrow|P|$ with  $\lambda(s^{\prime})=\lambda_1(s^{\prime})$ for any $s^{\prime}\in S^{\prime}$ and $\lambda(p^{\prime})=\lambda_2(p^{\prime})$ for any $p^{\prime}\in P^{\prime}$.

Since $(P,E,Q)\in \mathcal{R}$  and $P\xrightarrow[\lambda_2]{\widehat{\Delta}_2}P^{\prime}$
we have $Q\xLongrightarrow[\rho,\rho_2,\rho^{\prime}]
{\widehat{\Delta}_2^c}Q^{\prime}$
for any $p:\alpha\cdot({\vec L})\in \widehat{\Delta}_2$ there exists $q:\alpha\cdot({\vec M})\in \widehat{\Delta}_2^{c}$ such that $(p,\rho(q))\in E$ and $(P^{\prime},E^{\prime},Q^{\prime})\in \mathcal{R}$ for some $E^{\prime}\subseteq|P^{\prime}|\times|Q^{\prime}|$
such that if $(p^{\prime},q^{\prime})\in E^{\prime}$ then $(\lambda_2(p^{\prime}),\rho\rho_2\rho^{\prime}(q^{\prime}))\in E$, and, moreover, if $n\geq 2$ then for any pair of labels $p:\alpha\cdot({\vec L})\in \widehat{\Delta}_2$ and $q:\alpha\cdot({\vec M})\in \widehat{\Delta}_2^c$ either $(p^{\prime},\rho^{\prime}(q^{\prime}))\in \bigcup_{i=1}^n(L_i\times M_i)$ or $p^{\prime}\notin \bigcup_{i=1}^n L_i$ and $\rho^{\prime}(q^{\prime})\notin \bigcup_{i=1}^n M_i$.

Therefore we have $V\xLongrightarrow[\nu,\nu_1,\nu^{\prime}]
{\widehat{\Delta}^c}V^{\prime}$
where $\widehat{\Delta}^c = \widehat{\Delta}_1 \uplus \widehat{\Delta}_2^c$, $V^{\prime}=S^{\prime}\oplus_{D^{\prime}}Q^{\prime}$ with $D^{\prime}=\{(s^{\prime},q^{\prime})\in |S^{\prime}|\times|Q^{\prime}|\mid (\nu_1(s^{\prime}),\rho\nu_1\rho^{\prime}(q^{\prime}))\in D\}$.
$V\xLongrightarrow[\nu,\nu_1,\nu^{\prime}]
{\widehat{\Delta}^c}V^{\prime}$ can be decomposed as
$$S\oplus_D Q \xrightarrow[\nu]{\tau^{\ast}}S\oplus_{D_1}Q_1 \xrightarrow[\nu_1]{\widehat{\Delta}^c}
S^{\prime}\oplus_{D_1^{\prime}}Q_1^{\prime}
\xrightarrow[\nu^{\prime}]{\tau^{\ast}}
S^{\prime}\oplus_{D^{\prime}}Q^{\prime}$$
with $\nu=\mbox{Id}_{|S|}\cup \rho$ and $\nu^{\prime}=\mbox{Id}_{|S^{\prime}|}\cup \rho^{\prime}$. $\nu_1:|S^{\prime}|\cup |Q_1^{\prime}|\rightarrow |S|\cup |Q_1|$ with $\nu_1(s^{\prime})=\lambda_1(s^{\prime})$ for any $s^{\prime}\in |S^{\prime}|$ and $\nu_1(q_1^{\prime})= \rho_2(q_1^{\prime})$ for any $q_1^{\prime}\in |Q_1^{\prime}|$.

Let $F^{\prime}\subseteq |U^{\prime}|\times |V^{\prime}|$ be defined as $F^{\prime}= \mbox{Id}_{|S^{\prime}|}\cup E^{\prime}$.
For  $(u^{\prime},v^{\prime})\in F^{\prime}$, if $u^{\prime}\in |S^{\prime}|$ or $v^{\prime}\in |S^{\prime}|$, we must have $u^{\prime}=v^{\prime}$. If $u^{\prime}\notin |S^{\prime}|$ and $v^{\prime}\notin |S^{\prime}|$ then we have $(u^{\prime},v^{\prime})\in E^{\prime}$. Hence $(\lambda(u^{\prime}),\nu\nu_1\nu^{\prime}(v^{\prime}))=
(\lambda_2(u^{\prime}),\rho\rho_2\rho^{\prime}(v^{\prime}))\in E$, and for any pair of labels $p:\alpha\cdot({\vec L})\in \widehat{\Delta}_2$ and $q:\alpha\cdot({\vec M})\in \widehat{\Delta}_2^c$, if $n
\geq 2$, either there exists $i$ such that $u^{\prime}\in L_i$ and $\nu^{\prime}(v^{\prime})=\rho^{\prime}(v^{\prime})\in M_i$, or $u^{\prime}\notin \bigcup_{i=1}^n L_i$ and $\nu^{\prime}(v^{\prime})=\rho^{\prime}(v^{\prime})\notin \bigcup_{i=1}^n M_i$.

Moreover, the triple $(C^{\prime},D^{\prime},E^{\prime})$ is adapted: let $(p^{\prime},q^{\prime})\in E^{\prime}$ and $s^{\prime}\in |S^{\prime}|$. We have $(\lambda_2(p^{\prime}),\rho\rho_2\rho^{\prime}(q^{\prime}))\in E$. We have $(s^{\prime},p^{\prime})\in C^{\prime}$ iff $(\lambda(s^{\prime}),\lambda(p^{\prime}))\in C$ iff $(\lambda_1(s^{\prime}),\lambda_2(p^{\prime}))\in C$ iff $(\lambda_1(s^{\prime}),\rho\rho_2\rho^{\prime}(q^{\prime}))\in D$ iff
$(\nu\nu_1\nu^{\prime}(s^{\prime}),\nu\nu_1\nu^{\prime}(q^{\prime}))\in D $ iff $(s^{\prime},q^{\prime})\in D^{\prime}$.
\end{IEEEproof}

The proof for Theorem \ref{bisimilationCongruence}.
\begin{IEEEproof}
Let $\mathcal{R}$ be a localized early weak bisimulation. Let $R$ be a $Y${ -context}. We define a new localized relation denoted by $R[\mathcal{R}/Y]$:
\begin{itemize}
  \item if $Y = R$ then $R[\mathcal{R}/Y] = \mathcal{R}$
  \item if $Y\neq R$ then we make $(P^{\prime},E^{\prime},Q^{\prime})\in R[\mathcal{R}/Y]$ if there exist $(P,E,Q)\in \mathcal{R}$, $E^{\prime}= \mbox{Id}_{|R|}$, $P^{\prime}=R[P/Y]$ and $Q^{\prime}= R[Q/Y]$. Since $R\neq Y$, it is obvious that $|P^{\prime}|=|Q^{\prime}|= |R|$.
\end{itemize}
We define a localized relation $\mathcal{R}^+$ as the union of $\mathcal{I}$ (the set of all triples $(U,E,U)$ where $U\in {\sf Proc}$ and $E = \mbox{Id}_{|U|}$), the parallel extension $\mathcal{R}^{\prime}$ of $\mathcal{R}$ and all the relations of the shape $R[\mathcal{R}/Y]$ for all $Y$-context $R$. Then what we have to do is to prove that $\mathcal{R}^+$ is a localized early weak bisimulation. It is easy to check that $\mathcal{R}^+$ is symmetric.

Let $(U,F,V)\in \mathcal{R}^+$ and we have to analyse the two following situations:
\begin{itemize}
  \item[(1)] $U\xrightarrow[\mu]{\tau}U^{\prime}$
  \item[(2)] or $U\xrightarrow[\mu]{\widehat{\Delta}}U^{\prime}$
\end{itemize}
In each case, we analyse all the possible transitions from the challenger, and then we show that there are corresponding transitions of the defender to respond to the challenger. We consider all the possible relations from $\mathcal{R}^+$. We analyse the two cases in details.
\begin{itemize}
  \item For case (1) we must show that $V\xrightarrow[\nu]{\tau^{\ast}}V^{\prime}$ with $(U^{\prime},F^{\prime}, V^{\prime})\in \mathcal{R}^+$ for some $F^{\prime}\subseteq |U^{\prime}|\times|V^{\prime}|$ such that for any $(u^{\prime},v^{\prime})\in F^{\prime}$, we have $(\mu(u^{\prime}),\nu(v^{\prime}))\in F$.
  \item For case (2) we must show that $V\xLongrightarrow[\nu,\nu_1,\nu^{\prime}]
      {\widehat{\Delta}^c}V^{\prime}$ with $(U^{\prime},F^{\prime},V^{\prime})\in \mathcal{R}^+$ and for any pair of labels $p:\alpha\cdot({\vec L})\in \widehat{\Delta}$ and $q:\alpha\cdot({\vec M})\in \widehat{\Delta}^c$, $(p,\nu(q))\in F$.
      And for some $F^{\prime}\subseteq|U^{\prime}|\times|V^{\prime}|$ such that for any $(u^{\prime},v^{\prime})\in F^{\prime}$, we have $(\mu(u^{\prime}),\nu\nu_1\nu^{\prime}(v^{\prime}))\in F$ and for any pair of labels $p:\alpha\cdot({\vec L})\in \widehat{\Delta}$ and $q:\alpha\cdot({\vec M})\in \widehat{\Delta}^c$, if $n\geq 2$, then we have either $(u^{\prime},\nu^{\prime}(v^{\prime}))\in \bigcup_{i=1}^n (L_i\times M_i)$ or $u^{\prime}\notin \bigcup_{i=1}^n L_i$ and $\nu^{\prime}(v^{\prime})\notin \bigcup_{i=1}^nM_i$.
\end{itemize}
Now, we analyse the possible relations in $\mathcal{R}^+$.

The case where $(U,F,V)\in \mathcal{I}$ is trivial.

If $(U,F,V)\in \mathcal{R}^{\prime}$, we can directly apply  Proposition \ref{propositionparallel} to the both cases.

Assume that $(U,F,V)\in R[\mathcal{R}/Y]$ for some $Y$-context $R$, so that $U=R[P/Y]$ and $V= R[Q/Y]$ with $(P,E,Q)\in \mathcal{R}$ such that $F=E$ if $R = Y$ and $F=\mbox{Id}_{|R|}$ otherwise. If $ R = Y$, we can directly use the fact that $\mathcal{R}$ is a localized weak bisimulation to show that $V^{\prime}$ and $F^{\prime}$ satisfy the required conditions.

At last we consider $R\neq Y$, so we have $F = \mbox{Id}_{|R|}$.
In this paper, we only focus on canonical processes and the canonical guarded sum ${\sf cs}(P)$ for a recursive canonical guarded sum $P$ may have three forms, i.e.
$$\tiny
{\sf cs}(P)
   =\left\{
  \begin{array}{ll}
    pre\cdot(Q_1,\ldots,Q_n)+T, \\
    {\bf if}~b~{\bf then}~pre\cdot(Q_1,\ldots,Q_n)+T~{\bf else}~T_1, & \hbox{with } {\sf eval}(b) = {\it true}, \mbox{or} \\
    {\bf if}~b~{\bf then}~T_1~{\bf else}~pre\cdot(Q_1,\ldots,Q_n)+T, & \hbox{with }{\sf eval}(b) ={\it false}
  \end{array}
\right.$$
where $pre$ is a prefix, $T$ and $T_1$ are canonical guarded sums and $Q_1,\ldots,Q_n$ are canonical processes. Without loss of generality, we only consider the case ${\sf cs}(P)=pre\cdot(Q_1,\ldots,Q_n)+T$ and the other cases are similar referring to transition rules.

By the definition of the $Y$-context, there is exactly one $r\in |R|$ such that $Y$ occurs free in $R(r)$. And ${\sf cs}(R(r))= f(x)\cdot{\vec R} + \widetilde{R}$ and $Y$ does not occur free in $\widetilde{R}$ and occurs exactly in one of the processes $R_1,\ldots,R_n$. Without loss of generality we can assume that $R_1$ is a $Y$-context and $Y$ does not occur free in $R_2,\ldots,R_n$.\\
~\\

We assume that $R_1\neq Y$. In both cases (1) and (2), we have $U^{\prime}= R^{\prime}[P/Y]$ with $R\xrightarrow[\mu]{\tau}R^{\prime}$ (case (1)) or $R\xrightarrow[\mu]{\widehat{\Delta}}R^{\prime}$ (case (2)). Let $V^{\prime}=R^{\prime}[Q/Y]$.
In case (1), we have $V\xrightarrow[\mu]{\tau}V^{\prime}$ and in case (2) we have $V\xrightarrow[\mu]{\widehat{\Delta}^c}V^{\prime}$. Since $Y\neq R^{\prime}$, we have $(U^{\prime},\mbox{Id}_{|R^{\prime}|}, V^{\prime})\in \mathcal{R}^{+}$ for $(P,E,Q)\in \mathcal{R}$. The residual condition is obviously satisfied in both cases.

At last we assume that $R_1=Y$.

\textit{\textbf{For case (1)}}. There are two cases to consider the locations $s,t\in |U|$ involved in the transition $U\xrightarrow[\mu]{\tau}U^{\prime}$. The case $s\neq r$ and $t\neq r$ is similar to the case above where $R_1 \neq Y$. The other two cases are  the case $s=r$ (hence $t\neq r$) and the symmetric case $t=r$ (hence $s\neq r$). We just consider the case $s=r$.

So $U(t)=R(t)=\overline{f}(e)\cdot{\vec T}+\widetilde{T}$ with ${\sf eval}(e)=v$ and the guarded sum $R(r)$ has an unique summand involved in the transition $U\xrightarrow[\mu]{\tau}U^{\prime}$ and this summand is of the form $f(x)\cdot{\vec S}$ (called active summand in the text that follows).

If the active summand is $f(x)\cdot{\vec R}$ then we have $U(r)= f(x)\cdot(P,R_2,\ldots,R_n) + \widetilde{S}$. $U^{\prime}$ can be written as $U^{\prime}=R^{\prime}\oplus_C P\{v/x\}$ for some process $R^{\prime}$ which can be defined using only $R$ and $C\subseteq |R^{\prime}|\times |P\{v/x\}|$. $R^{\prime}$ is defined as follows:
\begin{itemize}
  \item $|R^{\prime}|= (|R|\setminus\{t,r\})\cup \bigcup_{i=2}^n |R_i\{v/x\}|\cup \bigcup_{i=1}^n |T_i|$
  \item and $\frown_{R^{\prime}}$ is the least symmetric relation on $|R^{\prime}|$ such that $r^{\prime}\frown_{R^{\prime}} t^{\prime}$
      if $r^{\prime}\frown_{R_i\{v/x\}}t^{\prime}$ for some $i\in \{2,\ldots,n\}$,
      or $r^{\prime}\frown_{T_i}t^{\prime}$ for some $i\in\{1,\ldots,n\}$,
      or $(r^{\prime},t^{\prime})\in |R_i\{v/x\}|\times|T_i|$ for some $i\in \{2,\ldots,n\}$,
      or $\mu(r^{\prime})\frown_R\mu(t^{\prime})$ with $r^{\prime}\notin \bigcup_{i=2}^n |R_i\{v/x\}|$ or $t^{\prime}\notin \bigcup_{i=1}^n|T_i|$. 

\end{itemize}
where the residual function $\mu:|U^{\prime}|\rightarrow |U|$ is given by $\mu(r^{\prime})= r$ if $r^{\prime}\in |P\{v/x\}|\cup\bigcup_{i=2}^n|R_i\{v/x\}|$,
$\mu(r^{\prime})= t$ if $r^{\prime}\in \bigcup_{i=1}^n|T_i|$, and $\mu(r^{\prime})=r^{\prime}$ otherwise.

The relation $C$ is defined as follows: given $(r^{\prime},p)\in |R^{\prime}|\times |P\{v/x\}|$,
one has $(r^{\prime},p)\in C$ if $r^{\prime}\in |T_1|$, or $r^{\prime}\notin \bigcup_{i=2}^n|R_i\{v/x\}|\cup \bigcup_{i=1}^n|T_i|$ and $r^{\prime}\frown_R r$.

Let $V^{\prime} = R^{\prime}\oplus_D Q\{v/x\}$, where $ D\subseteq |R^{\prime}|\times |Q\{v/x\}|$ is defined similarly in the way for $C$ by replacing $P\{v/x\}$ by $Q\{v/x\}$. From $(p,q)\in E$ and the definitions of $C$ and $D$, we have $(r^{\prime},p)\in C$ iff $(r^{\prime},q)\in D$. So $(C,D,E)$ is adapted. We can make the reduction on $V$, such that $V\xrightarrow[\nu]{\tau}V^{\prime}$
for the residual function $\nu$ which is defined like $\mu$ by replacing $P\{v/x\}$ by $Q\{v/x\}$. We have $(U^{\prime},F^{\prime},V^{\prime})\in \mathcal{R}^{\prime}\subseteq \mathcal{R}^{+}$ where $F^{\prime}= \mbox{Id}_{|R^{\prime}|}\cup E$. If $(u^{\prime},v^{\prime})\in F^{\prime}$, then we have $\mu(u^{\prime})=\nu(v^{\prime})$, that is $(\mu(u^{\prime}),\mu(v^{\prime}))\in F$ so that the condition on residuals holds.

If the active summand is not $f(x)\cdot{\vec R}$, then we have $V\xrightarrow[\mu]{\tau}U^{\prime}$ (both $P$ and $Q$ are discarded in the corresponding reductions, respectively). We just finish the proof because of $(U^{\prime},\mbox{Id}_{|U^{\prime}|},U^{\prime})\in \mathcal{I}\subseteq\mathcal{R}^{\prime}$.

\textit{\textbf{For case (2)}}. In the transition $U\xrightarrow[\mu]{\widehat{\Delta}}U^{\prime}$, if $r$ is not mentioned in $\widehat{\Delta}$, then we have $R[P/Y]=U\xrightarrow[\mu]{\widehat{\Delta}}
U^{\prime}=R^{\prime}[P/Y]$. We also have $R[Q/Y]=V\xrightarrow[\mu]{\widehat{\Delta}}V^{\prime}=R^{\prime}[Q/Y]$ so $(U^{\prime},\mbox{Id}_{|R^{\prime}|},V^{\prime})\in R^{\prime}[\mathcal{R}/Y]\subseteq \mathcal{R}^{+}$ and the residual condition is satisfied.

If $r:\alpha\cdot(\vec{L})$ is mentioned in $\widehat{\Delta}$, then there exists exactly one of the summands of the guarded sum $R(r)$ being the prefixed process preforming action $\alpha$ in $U\xrightarrow[\mu]{\widehat{\Delta}}U^{\prime}$.

The case where the active summand is not $f(x)\cdot(P,R_2,\ldots,R_n)$ is similar to the previous case, because $P$ is discarded in the transition.

For $R[P/Y]$, we can rewrite it as $R_1\oplus_{C_1} R(r)$, where $R_1(s)=R(s)$ for $s\in (|R|\setminus\{r\})$, and $(s,r)\in C_1$ if $s\frown_R r$.

If the active summand is $f(x)\cdot(P,R_2,\ldots, R_n)$, then
$U=R[P/Y]= R_1\oplus_{C_1} R(r) \xrightarrow[\mu]{\widehat{\Delta}}U^{\prime}=R_1^{\prime}\oplus_{C_1^{\prime}}
(P\{v/x\}\oplus R_2\{v/x\}\ldots\oplus R_n\{v/x\})$ for $v\in {\bf Val}$.

We rewrite $U^{\prime}$ as $R^{\prime}\oplus_C P\{v/x\}$ for $v\in {\bf Val}$, where $R^{\prime}$ is defined by
\begin{itemize}
  \item $|R^{\prime}|= |R_1^{\prime}|\cup \bigcup_{i=2}^n|R_i\{v/x\}|$
       and $\frown_{R^{\prime}}$ is the least symmetric relation on $|R^{\prime}|$ such that $r^{\prime}\frown_{R^{\prime}}t^{\prime}$ if $r^{\prime}\frown_{R_i\{v/x\}}t^{\prime}$ for some $i\in \{2,\ldots, n\}$ or $\mu(r^{\prime})\frown_R \mu(t^{\prime})$.
\end{itemize}

The relation $C\subseteq |R^{\prime}|\times |P\{v/x\}|$ is defined by $(r^{\prime},q)\in C$ if $r^{\prime}\notin \bigcup_{i=2}^n|R_i|$ and $\mu(r^{\prime})\frown_R r$.

Then we have
$V=R[Q/Y]= R_1\oplus_{C_1} R(r) \xrightarrow[\mu]{\widehat{\Delta}^c}V^{\prime}=R_1^{\prime}\oplus_{C_1^{\prime\prime}}
(Q\{v/x\}\oplus R_2\{v/x\}\ldots\oplus R_n\{v/x\})$ for $v\in {\bf Val}$.

We rewrite $V^{\prime}$ as $R^{\prime}\oplus_D Q\{v/x\}$ where $R^{\prime}$ is defined as above and  $D$ is defined like $C$ by replacing $P\{v/x\}$ by $Q\{v/x\}$.
Then we have $(U^{\prime},F^{\prime},V^{\prime})\in \mathcal{R}^{\prime}\subseteq\mathcal{R}^{+}$ where $F^{\prime}=\mbox{Id}_{|R^{\prime}|}\cup E$
since $(C,D,E)$ is adapted. Moreover the condition on residuals is obviously satisfied.

The symmetric case that ${\sf cs}(R(r))= \overline{f}(e)\cdot{\vec R} + \widetilde{R}$ with ${\sf eval}(e)=v$ and $Y$ does not occur free in $\widetilde{R}$ and occurs exactly in one of the processes $R_1,\ldots,R_n$, is similar. So, we show the fact that $\mathcal{R}^{+}$ is a localized early weak bisimulation.

We can now prove that $\approx$ is a congruence. Assume that $P\approx Q$ and let $R$ be a $Y$-context. Let $E\subseteq|P|\times|Q|$ and let $\mathcal{R}$ be a localized early weak bisimulation such that $(P,E,Q)\in \mathcal{R}$.
Then we have $(R[P/Y],\mbox{Id}_{|R|}, R[Q/Y])\in R[\mathcal{R}/Y]\subseteq \mathcal{R}^{+}$ and hence $R[P/Y]\approx R[Q/Y]$ since $\mathcal{R}^{+}$ is a localized early weak bisimulation.
\end{IEEEproof}

\subsection{Full Version of Thread transitions}\label{fullversion}
The full version of thread transitions is given in Fig. \ref{fig:full thread transitions}.
\begin{figure}[!htb]\tiny
  $$\infer[(\textsc{WriteX})]{((\sigma_g,\mathbb{L}_a,\mathbb{L}_b),(\sigma_l,{\sf x:=e}))\xlongrightarrow[]{\tau}_{\sf t} ((\sigma_g^{\prime},\mathbb{L}_a,\mathbb{L}_b),(\sigma_l,{\sf skip}))}
  {[\![{\sf e}]\!]_{\sigma_l}={\sf v}& \sigma_g^{\prime} = \sigma_g[{\sf x\mapsto v}]}$$
  $$\infer[(\textsc{ReadX})]{((\sigma_g,\mathbb{L}_a,\mathbb{L}_b),(\sigma_l,{\sf r:=x})) \xlongrightarrow[]{\tau}_{\sf t} ((\sigma_g,\mathbb{L}_a,\mathbb{L}_b),(\sigma_l^{\prime},{\sf skip}))}
  {\sigma_g({\sf x})={\sf v}& \sigma_l^{\prime} = \sigma_l[{\sf r\mapsto v}]}$$
  $$\infer[(\textsc{ReadA})]{((\sigma_g,\mathbb{L}_a,\mathbb{L}_b),(\sigma_l,{\sf r:=a.load(mo_2)})) \xlongrightarrow[]{\tau}_{\sf t} ((\sigma_g,\mathbb{L}_a,\mathbb{L}_b),(\sigma_l^{\prime},{\sf skip}))}
  {\sigma_g({\sf a})={\sf v}& \sigma_l^{\prime} = \sigma_l[{\sf r\mapsto v}]}$$
  $$\infer[(\textsc{WriteA})]{((\sigma_g,\mathbb{L}_a,\mathbb{L}_b),(\sigma_l,{\sf a.store(e,mo_1)})) \xlongrightarrow[]{\tau}_{\sf t} ((\sigma_g^{\prime},\mathbb{L}_a,\mathbb{L}_b),(\sigma_l,{\sf skip}))}
  {[\![{\sf e}]\!]_{\sigma_l}={\sf v}& \sigma_g^{\prime} = \sigma_g[{\sf a\mapsto v}]}$$
  $$\infer[(\textsc{Lock})]{((\sigma_g,\mathbb{L}_a,\mathbb{L}_b),(\sigma_l,{\sf l.lock()})) \xlongrightarrow[]{\tau}_{\sf t} ((\sigma_g,\mathbb{L}_a^{\prime},\mathbb{L}_b^{\prime}),(\sigma_l,{\sf skip}))}
  {{\sf l} \in \mathbb{L}_a& \mathbb{L}_a^{\prime}=\mathbb{L}_a\setminus\{{\sf l}\} & \mathbb{L}_b^{\prime} = \mathbb{L}_b\cup \{{\sf l}\}}$$
  $$\infer[(\textsc{Unlock})]{((\sigma_g,\mathbb{L}_a,\mathbb{L}_b),(\sigma_l,{\sf l.unlock()})) \xlongrightarrow[]{\tau}_{\sf t} ((\sigma_g,\mathbb{L}_a^{\prime},\mathbb{L}_b^{\prime}),(\sigma_l,{\sf skip}))}
  {{\sf l} \in \mathbb{L}_b& \mathbb{L}^{\prime}_b = \mathbb{L}_b\setminus \{{\sf l}\}& \mathbb{L}_a^{\prime} = \mathbb{L}_a\cup\{{\sf l}\}}$$
  $$\infer[(\textsc{Print})]{((\sigma_g,\mathbb{L}_a,\mathbb{L}_b),(\sigma_l,{\sf print~e})) \xlongrightarrow[]{(\overline{\sf out},{\sf v})}_{\sf t} ((\sigma_g,\mathbb{L}_a,\mathbb{L}_b),(\sigma_l,{\sf skip}))}
  {[\![{\sf e}]\!]_{\sigma_l}={\sf v}}$$
  $$\infer[(\textsc{Seq1})]{(s,(\sigma_l,{\sf C_1;C_2}))\xlongrightarrow[]{\iota_{\sf t}}_{\sf t}(s^{\prime},(\sigma_l^{\prime},{\sf C_1^{\prime};C_2}))}{(s,(\sigma_l,{\sf C_1}))\xlongrightarrow[]{\iota_{\sf t}}_{\sf t}(s^{\prime},(\sigma_l^{\prime},{\sf C_1^{\prime}}))}$$
  $$\infer[(\textsc{Seq2})]{(s,(\sigma_l,{\sf skip;C}))\xlongrightarrow[]{\iota_{\sf t}}_{\sf t}(s^{\prime},(\sigma_l^{\prime},{\sf C^{\prime}}))}
  {(s,(\sigma_l,{\sf C}))\xlongrightarrow[]{\iota_{\sf t}}_{\sf t}(s^{\prime},(\sigma_l^{\prime},{\sf C^{\prime}}))}$$
  $$\infer[(\textsc{If-t})]{(s,(\sigma_l,{\sf if ~b~then~C_1~else~C_2}))\xlongrightarrow[]{\iota_{\sf t}}_{\sf t} (s^{\prime},(\sigma_l^{\prime},{\sf C_1^{\prime}}))}{[\![{\sf b}]\!]_{\sigma_l}= {\sf true}& (s,(\sigma_l,{\sf C_1}))\xlongrightarrow[]{\iota_{\sf t}}_{\sf t}(s^{\prime},(\sigma_l^{\prime},{\sf C_1^{\prime}}))}$$
  $$\infer[(\textsc{If-f})]{(s,(\sigma_l,{\sf if ~b~then~C_1~else~C_2}))\xlongrightarrow[]{\iota_{\sf t}}_{\sf t} (s^{\prime},(\sigma_l^{\prime},{\sf C_2^{\prime}}))}{[\![{\sf b}]\!]_{\sigma_l}= {\sf false}& (s,(\sigma_l,{\sf C_2}))\xlongrightarrow[]{\iota_{\sf t}}_{\sf t}(s^{\prime},(\sigma_l^{\prime},{\sf C_2^{\prime}}))}$$
  $$\infer[(\textsc{While-t})]{(s,(\sigma_l,{\sf while~b~do~C}))\xlongrightarrow[]{\iota_{\sf t}}_{\sf t} (s^{\prime},(\sigma_l^{\prime},{\sf C^{\prime};while~b~do~C}))}{[\![{\sf b}]\!]_{\sigma_l}= {\sf true}& (s,(\sigma_l,{\sf C}))\xlongrightarrow[]{\iota_{\sf t}}_{\sf t}(s^{\prime},(\sigma_l^{\prime},{\sf C^{\prime}}))}$$
  $$\infer[(\textsc{While-f})]{(s,(\sigma_l,{\sf while~b~do~C}))\xlongrightarrow[]{\tau}_{\sf t} (s,(\sigma_l,{\sf skip}))}{[\![{\sf b}]\!]_{\sigma_l}= {\sf false}}$$
  \caption{Full version of thread transitions}\label{fig:full thread transitions}
\end{figure}
\section{Correctness of Translation}\label{appB}
\begin{proposition}\label{exp-env}
For any $v\in {\bf Val}$, $e\in {\bf Exp}$ and environment $\varrho$ such that ${\sf fv}(e)\subseteq {\sf dom}(\varrho)$, $\varrho(e\{v/x\})=\varrho[x\rightarrow v](e)$.
\end{proposition}
\begin{IEEEproof}
Straightforward.
\end{IEEEproof}
\begin{proposition}\label{bexp-env}
For any $v\in {\bf Val}$, $b\in {\bf BExp}$ and environment $\varrho$ such that ${\sf fv}(b)\subseteq {\sf dom}(\varrho)$, $\varrho(b\{v/x\})=\varrho[x\rightarrow v](b)$.
\end{proposition}
\begin{IEEEproof}
Straightforward.
\end{IEEEproof}
\begin{lemma}\label{envbisim}
For any $v\in {\bf Val}$, canonical process $P$ and environment $\varrho$ such that ${\sf fv}(P)\subseteq {\sf dom}(\varrho)$, $(P\{v/x\},\varrho)\approx(P,\varrho[x\rightarrow v])$.
\end{lemma}
\begin{IEEEproof}
We just need to prove that localized relation
\begin{center}
\begin{tabular}{lc}
  $\{~((P\{v/x\},\varrho),\mbox{Id}_{|P|},(P,\varrho[x\rightarrow v])),$ &  \\
   \quad$((P,\varrho[x\rightarrow v]),\mbox{Id}_{|P|},(P\{v/x\},\varrho))$&\\
   \qquad $\mid P \mbox{ is canonical and } {\sf fv}(P)\subseteq {\sf dom}(\varrho)~\} $ & \\
\end{tabular}
\end{center}
is a localized early weak bisimulation. And it is easy to check that the localized relation is symmetric.

First, we prove that for each transition of $(P\{v/x\},\varrho)$, $(P,\varrho[x\rightarrow v])$ can make the same transition. From Proposition \ref{exp-env} and Proposition \ref{bexp-env}, we have that for
any $e$ or $b$ involved during the derivation $\varrho(e\{v/x\})$ $ = \varrho[x\rightarrow v](e)$ and $\varrho(b\{v/x\}) = \varrho[x\rightarrow v](b)$. So for any transition of $(P\{v/x\},\varrho)$, $(P,\varrho[x\rightarrow v])$ can make the same transition as $(P\{v/x\},\varrho)$, selecting the same branch based on conditions of processes by Proposition \ref{bexp-env} and passing the same value by Proposition \ref{exp-env}.

Similarly, we can check that for any transition of $(P,\varrho[x\rightarrow v])$, $(P\{v/x\},\varrho)$ can make the same transition.
\end{IEEEproof}

The proof for Lemma \ref{transcanonical}:
\begin{IEEEproof}
It is easy to check that processes for variables and locks are canonical. For commands, we can easily check them by
induction on the structure of commands.
\end{IEEEproof}
Then we show that, a global transition can be simulated by a transition in VCCTS with respect to weak bisimulation.
The proof for Theorem \ref{thm-trans}:
\begin{IEEEproof}
Analyse the global transitions and all the possible labels one by one. We first consider the global transition rule
$$\infer[]{(s,T\cup\{(\sigma_l,{\sf C})\})\xlongrightarrow[]{\iota}_T (s^{\prime},T\cup\{(\sigma_l^{\prime},{\sf C}^{\prime})\})}
  {(s,(\sigma_l,{\sf C}))\xlongrightarrow[]{\iota}_{\sf t}  (s^{\prime},(\sigma_l^{\prime},{\sf C}^{\prime}))}$$
Therefor, $\iota$ can be either $\tau$ or an output action.

(1) Suppose $\iota = \tau$ and the global transition is of the form
$$\infer[]{(s,T\cup\{(\sigma_l,{\sf C})\})\xlongrightarrow[]{\tau}_T (s^{\prime},T\cup\{(\sigma_l^{\prime},{\sf C}^{\prime})\})}
  {(s,(\sigma_l,{\sf C}))\xlongrightarrow[]{\tau}_{\sf t}  (s^{\prime},(\sigma_l^{\prime},{\sf C}^{\prime}))}$$
Then we have to analyse all possible transitions for $(s,(\sigma_l,{\sf C}))\xlongrightarrow[]{\tau}_{\sf t}  (s^{\prime},(\sigma_l^{\prime},{\sf C}^{\prime}))$.

Case $\textsc{WriteX}$: For $(s,(\sigma_l,{\sf C}))$, the thread modifies the global state by updating a variable ${\sf x}$ and the thread local variable is unchanged. That is $(s,(\sigma_l,{\sf C}))$ $\xlongrightarrow[]{\tau}_{\sf t}$ $(s^{\prime},(\sigma_l,{\sf C}^{\prime}))$.
Thus, there is a communication between the process for  ${\sf x}$ and process $[\![{\sf C}]\!]$. Without loss of generality, we assume ${\sf C \equiv x:=e;C^{\prime}}$, $s=(\sigma_g,(\mathbb{L}_a,\mathbb{L}_b))$, $[\![{\sf e}]\!]_{\sigma_l}={\sf v}$ for some ${\sf v}$ and $\sigma_g^{\prime}=\sigma_g[{\sf x}\mapsto {\sf v}]$. From the translation, we have $Q_1 = [\![{\sf C}]\!]$, $Q_1^{\prime} = [\![{\sf C^{\prime}}]\!]$, $Q_2= X_{\sf x}(v^{\prime})$ for some $v^{\prime} = [\![\sigma_g({\sf x})]\!]$ and $Q_2^{\prime}= X_{\sf x}(v)$ with $v = [\![{\sf v}]\!]$.
From the encoding and the transition of VCCTS, we have $[\![\Gamma]\!]=(P,\varrho)\xrightarrow[\lambda]{\tau}(P^{\prime},\varrho)$, where, $\lambda = \mbox{Id}_{|P|}$ and the communication occurs between $Q_1$ and $Q_2$ (there exists an edge between the locations for them from the translation of global configuration). Therefore, $[\![\Gamma]\!]=(P,\varrho)$, $[\![\Gamma^{\prime}]\!]=(P^{\prime},\varrho)$ and $(P^{\prime},\varrho)\approx (P^{\prime},\varrho)$. And it is easy to check that ${\sf Act}(\delta) = [\![\iota]\!]$, i.e. $\tau = \tau$.
The cases for $\textsc{WriteA,Unlock}$ and $\textsc{Lock}$ are similar.

Case $\textsc{ReadX}$:  Without loss of generality, we assume ${\sf C \equiv r:=x;C^{\prime}}$ and $s=(\sigma_g,(\mathbb{L}_a,\mathbb{L}_b))$. For $(s,(\sigma_l,{\sf C}))$, the thread modifies a local variable ${\sf r}$, $[\![{\sf r}]\!]_{\sigma_l}={\sf v}^{\prime}$ and ${\sf r}$ is updated by the value ${\sf v}=\sigma_g({\sf x})$. That is
$(s,(\sigma_l,{\sf C}))\xlongrightarrow[]{\tau}_{\sf t}  (s,(\sigma_l^{\prime},{\sf C}^{\prime}))$.
From the encoding and the transition rules of VCCTS, we have $(P,\varrho)\xrightarrow[\lambda]{\tau}(P^{\prime}\{v/r\},\varrho)$ and $(P^{\prime}\{v/r\},\varrho)
\approx (P^{\prime},\varrho[r\rightarrow v])$ by Lemma \ref{envbisim}, with $\lambda = \mbox{Id}_{|P|}$, $[\![\Gamma]\!]=(P,\varrho)$, $[\![\Gamma^{\prime}]\!]=(P^{\prime},\varrho^{\prime})$, $[\![{\sf v}]\!] = v$, $[\![{\sf r}]\!] = r$ and $\varrho^{\prime}=\varrho[{ r}\rightarrow v]$. The case for $\textsc{ReadA}$ is similar.

(2) Suppose $\iota = (\overline{\sf out},{\sf v})$ and the global transition is of the form
$$\infer[]{(s,T\cup\{(\sigma_l,{\sf C})\})\xlongrightarrow[]{(\overline{\sf out},{\sf v})}_T (s,T\cup\{(\sigma_l,{\sf C}^{\prime})\})}
  {(s,(\sigma_l,{\sf C}))\xlongrightarrow[]{(\overline{\sf out},{\sf v})}_{\sf t}  (s,(\sigma_l,{\sf C}^{\prime}))}$$
Without loss of generality, we assume ${\sf C \equiv print~e;C^{\prime}}$, $s=(\sigma_g,(\mathbb{L}_a,\mathbb{L}_b))$, and $[\![{\sf e}]\!]_{\sigma_l}={\sf v}$ for some ${\sf v}$.
Then, we have the promise $(s,(\sigma_l,{\sf C}))\xlongrightarrow[]{(\overline{\sf out},{\sf v})}_{\sf t}  (s,(\sigma_l,{\sf C}^{\prime}))$.
From the translation for $\textsc{Print}$, it is trivial. $(P,\varrho)\xrightarrow[\lambda]{p:\overline{\sf out}v\cdot(\{p\})}(P^{\prime},\varrho)$, where $\lambda= \mbox{Id}_{|P|}$, $p$ is the location for $[\![{\sf C}]\!]$ and $v = [\![{\sf v}]\!]$. Therefore $[\![\Gamma^{\prime}]\!] = (P^{\prime},\varrho)$ and $(P^{\prime},\varrho)\approx (P^{\prime},\varrho)$. And it is easy to check that ${\sf Act}(\delta) = [\![\iota]\!]$, i.e. $\overline{\sf out}v = \overline{\sf out}v$.

Then we consider the global transition rule for thread creation,
and the global transition is of the form
$$\tiny
\infer[]{(s,T\cup\{(\sigma_l,{\sf thread~ t_i(C_i(r),e);C})\})\xlongrightarrow[]{ (\overline{\sf fork},{\sf 0})}_T (s,T\cup\{(\sigma_l,{\sf C})\}\cup \{(\{{\sf r}\mapsto {\sf v}\},{\sf C_i})\})}{[\![{\sf e}]\!]_{\sigma_l}={\sf v}}$$
Therefore, the label only has one form, i.e. $\iota = (\overline{\sf fork},{\sf 0})$.

Let $[\![\Gamma]\!]=[\![(s,T\cup\{(\sigma_l,{\sf thread~ t_i(C_i(r),e);C})\})]\!] = (P,\varrho)$. There is a transition from the subprocess $P_1 = [\![{\sf thread~ t_i(C_i(r),e);C}]\!] = \overline{\sf fork}(0)\cdot([\![{\sf C}]\!],[\![{\sf C_i}]\!]\{[\![{\sf e}]\!]/[\![{\sf r}]\!]\})$.
We have $(P,\varrho)\xrightarrow[\lambda]{p:\overline{\sf fork}0\cdot(\{p\},\{q\})}(P^{\prime},\varrho^{\prime})$, where $p$ is the location of $P_1$, $\varrho([\![{\sf e}]\!]) = v$, $q\notin|P|$ and $\lambda:|P^{\prime}|\rightarrow|P|$ with $|P^{\prime}| = |P|\cup \{q\}$, $\lambda(q)=p$ and $\lambda(p^{\prime}) = p^{\prime}$ for any $p^{\prime}\in |P|$. $\varrho^{\prime}$ is a new environment obtained from $\varrho$ by updating the corresponding local variables appearing in ${\sf C_i}$, i.e. $\varrho^{\prime} = \varrho[r\rightarrow \varrho(e)]$ with $r = [\![{\sf r}]\!]$ and $e = [\![{\sf e}]\!]$.
And $[\![\Gamma^{\prime}]\!]=[\![(s,T\cup\{(\sigma_l,{\sf C})\}\cup \{(\{{\sf r}\mapsto [\![{\sf e}]\!]_{\sigma_l}\},{\sf C_i})\})]\!] = (P^{\prime},\varrho^{\prime})$.
We have $(P^{\prime}\{e/r\},\varrho)\approx (P^{\prime},\varrho^{\prime})$.
And it is easy to check that ${\sf Act}(\delta) = [\![\iota]\!]$, i.e. $\overline{\sf fork}0 = \overline{\sf fork}0$.
\end{IEEEproof}
The proof for Theorem \ref{thm-trans-rev}:
\begin{IEEEproof} We analysis all the possible transitions of $(P,\varrho)\xrightarrow[\lambda]{\delta}(P^{\prime},\varrho)$.

From the translation and $[\![\Gamma]\!]=(P,\varrho_1)$, we know that the processes for commands can communicate with the processes via normal variables, atomic variables and locks.
So, we analyze all the possible transitions from $(P,\varrho)$.

If $\delta=\tau$ and $\lambda=\mbox{Id}_{|P|}$, without loss of generality then the communication can happen between a process $Q_1$ for a normal variable ${\sf x}$ and a process $Q_2$ for a thread $(\sigma_l, {\sf C})$.
If $Q_1$ receives a value $v$ and $Q_2$ sends the value $v$, then this corresponds to a write to the normal variable with value $v$. So we have
$((\sigma_g,\mathbb{L}_a,\mathbb{L}_b),(\sigma_l,{\sf x:=e;C_1}))\xrightarrow[]{\tau}_{\sf t} ((\sigma_g^{\prime},\mathbb{L}_a,\mathbb{L}_b),(\sigma_l,{\sf C_1}))$ and $[\![\Gamma^{\prime}]\!] = (P_2,\varrho_2) = (P^{\prime},\varrho)$ with $[\![{\sf e}]\!]_{\sigma_l}={\sf v}$, $[\![{\sf v}]\!] = v$, $ P_2=P^{\prime}$ and $\varrho_2=\varrho$ as required.
If $Q_1$ sends a value $v$ and $Q_2$ receives the value $v$, then this corresponds to read a value ${\sf v}$ with $[\![{\sf v}]\!] = v$ from the normal variable ${\sf x}$ and update the local state $\sigma_l$ for some register ${\sf r}$ with ${\sf v}$. Thus we have $((\sigma_g,\mathbb{L}_a,\mathbb{L}_b),(\sigma_l,{\sf r:=x;C_1}))\xrightarrow[]{\tau}_{\sf t} ((\sigma_g,\mathbb{L}_a,\mathbb{L}_b),(\sigma_l^{\prime},{\sf C_1}))$.
Thus $[\![\Gamma^{\prime}]\!] = (P_2,\varrho_2)$, $P_2\{v/r\} = P^{\prime}$ and $\varrho_2 =\varrho[{ r}\rightarrow v]$.
Since for any process $Q$, data variable $r$, data value $v$ and environment $\varrho$, we have $(Q\{v/r\},\varrho) \approx (Q,\varrho[r\mapsto v])$. Therefore, we get $[\![\Gamma^{\prime}]\!] = (P_2,\varrho_2) \approx (P^{\prime},\varrho)$ as required. And it is easy to check that ${\sf Act}(\delta) = [\![\iota]\!]$, i.e. $\tau =\tau$.

The cases for $\delta=\tau$, $\lambda=\mbox{Id}_{|P|}$, and the communication happens between a process $Q_1$ for an atomic variable (or a lock) and a process $Q_2$ for a thread in $P$, are similar.

If $\delta=p:\overline{\sf out}~v\cdot(\{p\})$ and $\lambda = \mbox{Id}_{|P|}$ for some location $p\in |P|$, then this corresponds to a print in the global transition. We have $((\sigma_g,\mathbb{L}_a,\mathbb{L}_b),(\sigma_l,{\sf print~e;C_1}))$ $\xrightarrow[]{(\overline{\sf out},{\sf v})}_{\sf t}$ $ ((\sigma_g,\mathbb{L}_a,\mathbb{L}_b),(\sigma_l,{\sf C_1}))$ with $v = [\![{\sf v}]\!]$ and $[\![\Gamma^{\prime}]\!] = (P_2,\varrho_2) = (P^{\prime},\varrho)$ with $P_2=P^{\prime}$ and $\varrho_2 = \varrho$.  And it is easy to check that ${\sf Act}(\delta) = [\![\iota]\!]$, i.e. $\overline{\sf out}v = \overline{\sf out}v$.

If $\delta=p:\overline{\sf fork}v\cdot(\{p\},\{q\})$ with $p\in |P|$, $q\notin |P|$, $\lambda(q) = p$ and $\lambda(p^{\prime})=p^{\prime}$ for any $p^{\prime}\in |P|$, then this transition corresponds to a thread fork transition in the global transition. So we have
$$\tiny
\infer[]{(s,T\cup\{(\sigma_l,{\sf thread~ t_i(C_i(r),e);C})\})\xlongrightarrow[]{ (\overline{\sf fork},{\sf v})}_T (s,T\cup\{(\sigma_l,{\sf C})\}\cup \{(\{{\sf r}\mapsto {\sf v}\},{\sf C_i})\})}{[\![{\sf e}]\!]_{\sigma_l}={\sf v}}$$
Thus $[\![\Gamma^{\prime}]\!] = (P_2,\varrho_2)$, $P_2\{v/r\} = P^{\prime}$ and $\varrho_2 =\varrho[{ r}\rightarrow v]$ with $v = [\![{\sf v}]\!]$ and $r = [\![{\sf r}]\!]$.
Since for any process $Q$, data variable $r$, data value $v$ and environment $\varrho$, we have $(Q\{v/r\},\varrho) \approx (Q,\varrho[r\mapsto v])$. Therefore, we get $[\![\Gamma^{\prime}]\!] = (P_2,\varrho_2) \approx (P^{\prime},\varrho)$ as required. And it is easy to check that ${\sf Act}(\delta) = [\![\iota]\!]$, i.e. $\overline{\sf fork}0 = \overline{\sf fork}0$.

\end{IEEEproof}
\subsection{Proofs for Transformations}
The proof for Theorem \ref{thm_cr}.

\begin{IEEEproof} We assume that $(Q,\varrho)$ is not conflict free, i.e. there exists a process $Q^{\prime}$ reached from $Q$ and $p_1,p_2\in |Q^{\prime}|$ such that $p_1\neq p_2$, $Q^{\prime}(p_1)=Q_1$, $Q^{\prime}(q_2)=Q_2$ and there exist two transitions in $Q_1$ and $Q_2$ with the actions $\alpha_1$ and $\alpha_2$, respectively, such that $({\sf symb}(\alpha_1),{\sf symb}(\alpha_2))\in {\bf CFL}$. Then, we have to prove that $(P,\varrho)$ involves a conflict, which contradicts the fact that $(P,\varrho)$ is conflict free.
We prove it by induction on the number of steps from $P$ to $Q$.

If $n = 0$, then $P = Q$. So $Q^{\prime}$ can also be reached from $P$, which contradicts that $(P,\varrho)$ is conflict free.

Induction hypothesis: when $n = k$, we have the fact that $P\hookrightarrow^{\ast} Q$ with $k$ steps and if $(P,\varrho)$ is conflict free, then $(Q,\varrho)$ is also conflict free.

Then we have to show that when $n= k + 1$, the result is also valid.
Let $n = k +1$. Without loss of any generality, we assume that the additional reordering is applied to $Q_1$ or $Q_2$. Here, we just focus on $Q_1$, since the case for $Q_2$ is symmetric.
We first consider reorderings involving only normal variables, i.e. {\bf WR}, {\bf WW}, {\bf RR} and {\bf RW}.
We consider the rule {\bf WR}, and the other three rules are similar.

If there exists a derivation $Q^{\prime}$ of $Q$ involving a conflict between processes $Q_1$ and $Q_2$, then let $Q_1 = Q_{12}\rhd Q_{11}\rhd Q_1^{\prime}$ reordered form $ Q_{11}\rhd Q_{12}\rhd Q_1^{\prime}$, where $Q_{11} = \overline{\sf write_x}(e)\cdot(\ast)$ and $Q_{12} =  {\sf read_y}(r)\cdot(\ast)$ with $r\notin {\sf fv}(e)$ and ${\sf x}\neq {\sf y}$. Since $Q_{12}\rhd Q_{11}\rhd Q_1^{\prime}$ and $Q_2$ involve a conflict, then $Q_{12}\rhd Q_1^{\prime}$ and $Q_2$ involve a conflict which is a derivation of some $Q$ by using $k$ steps of reordering from $P$, contradicting with the induction hypothesis.

For the rules {\bf UW1}, {\bf UW2}, {\bf UR1} and {\bf UR2}, we only consider {\bf UW1}, and the others are similar.
If there exists a derivation $Q^{\prime}$ of $Q$ involving a conflict between processes $Q_1$ and $Q_2$, then let $Q_1 = Q_{12}\rhd Q_{11}\rhd Q_1^{\prime}$ reordered form $ Q_{11}\rhd Q_{12}\rhd Q_1^{\prime}$, where $Q_{11} = \overline{\sf down_l}(0)\cdot(\ast)$ and $Q_{12} =  \overline{\sf write_x}(e)$. Since $Q_{12}\rhd Q_{11}\rhd Q_1^{\prime}$ and $Q_2$ involve a conflict, then $Q_{12}\rhd Q_1^{\prime}$ and $Q_2$ involve a conflict which is a derivation of some $Q$ by using $k$ steps of reordering from $P$. Since $Q_{11}\rhd Q_{12}\rhd Q_1^{\prime}$ can always release a lock that it has owned, i.e. playing $\overline{\sf down_l}(0)\cdot(\ast)$.

For the rules {\bf WL1}, {\bf WL2}, {\bf RL1} and {\bf RL2}, we only consider {\bf WL1}, and the other three are similar.
If there exists a derivation $Q^{\prime}$ of $Q$ involving a conflict between processes $Q_1$ and $Q_2$, then let $Q_1 = Q_{12}\rhd Q_{11}\rhd Q_1^{\prime}$ reordered form $ Q_{11}\rhd Q_{12}\rhd Q_1^{\prime}$, where $Q_{11} = \overline{\sf write_x}(e)\cdot(\ast)$ and $Q_{12} =  \overline{\sf up_l}(0)\cdot(\ast)$. Let  $Q_2$ and $Q_{11}\rhd Q_1^{\prime}$ derived from $Q_{12}\rhd Q_{11}\rhd Q_1^{\prime}$  involve a conflict, then $Q_{11}\rhd Q_{12}\rhd Q_1^{\prime}$ and $Q_2$ will also involve a conflict which is a derivation of some $Q$ by using $k$ steps of reordering from $P$, in the situation that $Q_{11}\rhd Q_{12}\rhd Q_1^{\prime}$ can successfully acquire a lock, i.e. playing $\overline{\sf up_l}(0)\cdot(\ast)$, with the fact that the other processes different from $Q_{11}\rhd Q_{12}\rhd Q_1^{\prime}$  did not play $\overline{\sf up_l}(0)\cdot(\ast)$.

Thus, if $(P,\varrho)$ is conflict free and $P\hookrightarrow^{\ast} Q$, then $(Q,\varrho)$ is also conflict free.
\end{IEEEproof}
The proof for Theorem \ref{thm_cr_def}.

\begin{IEEEproof}
We prove it by induction on the number of steps using $\hookrightarrow$ from $P$ to $Q$.

If $n = 0$, then $P = Q$. It is straightforward that  $(P,\varrho) \approx (P,\varrho)$.

Induction hypothesis: if $(P,\varrho)$ is conflict free, we have $n=k$ steps from $P$ to $Q$ using $\hookrightarrow $, and $(P,\varrho)\approx(Q,\varrho)$ with $E = Id_{|P|}$ for the triples in $((P,\varrho),E,(Q,\varrho))$ (since the reordering does not change the locations of the processes).

We have to establish that if $(P,\varrho)$ is conflict free and $P\hookrightarrow^{\ast} Q$ with $n = k+1$ steps then $(P,\varrho)\approx(Q,\varrho)$.
For $P\hookrightarrow^{\ast} Q$ with $k+1$ step, we have $P\hookrightarrow^{\ast} Q^{\prime}$ with $k$ steps, $Q^{\prime}$ is conflict free by Theorem \ref{thm_cr} and $Q^{\prime}\hookrightarrow^{\ast} Q$ with one step. From the induction hypothesis we have $(P,\varrho)\approx(Q^{\prime},\varrho)$, then we need to show that $(Q^{\prime},\varrho)\approx(Q,\varrho)$. $Q^{\prime}\hookrightarrow^{\ast} Q$ using one step of the reordering rules. We just consider the rule {\bf WR} and the other rules are similar.

Without loss of any generality, we assume that the reordering effects on location $p\in |P|=|Q^{\prime}|=|Q|$.
$\overline{\sf write_x}(e)\cdot(\ast)\rhd {\sf read_y}(r)\cdot(\ast)$
is a segment in $Q^{\prime}(p)$ with $r$ not occurring in $e$ and ${\sf x\neq y}$. After the reordering, $Q(p)$ is the same as $Q^{\prime}(p)$ except that $\overline{\sf write_x}(e)\cdot(\ast)\rhd {\sf read_y}(r)\cdot(\ast)$ in $Q^{\prime}$ is replaced by ${\sf read_y}(r)\cdot(\ast)\rhd\overline{\sf write_x}(e)\cdot(\ast)$.

For any derivation of $Q^{\prime}$, we need to show that $Q$ can match the derivation of $Q^{\prime}$ with the same observations and vice versa.

Assume that $Q^{\prime}$ can reduce to $Q^{\prime}_1$ before referring to $\overline{\sf write_x}(e)\cdot(\ast)\rhd {\sf read_y}(r)\cdot(\ast)$
involved in the {\bf WR} reordering in location $p$. Then $Q$ can make the same derivation as $Q^{\prime}$ to $Q_1$ with $Q^{\prime}_1 = Q_1$. Both $Q^{\prime}_1$ and $Q_1$ are conflict free. And we have $Q^{\prime}_1(q) = Q_1(q)$ for $q\in |Q_1|$ and $q\neq p$.

We analyze the possible derivations involving $\overline{\sf write_x}(e)\cdot(\ast)$ and ${\sf read_y}(r)\cdot(\ast)$. Because we only consider observational behaviors corresponding to the outputs in programs, the observational actions are those involving ${\sf \overline{out}}$. And the behaviors are $\tau$ actions from communications between processes (for states and commands in programs).

We just focus on the transitions involving $\overline{\sf write_x}(e)\cdot(\ast)$ and ${\sf read_y}(r)\cdot(\ast)$, in location $p$ respectively, communicating with ${\sf write_x}(x)$ and $\overline{\sf read_y}(v_2)$ which are prefixes from $X_{\sf x}(v_1) $ and $ X_{\sf y}(v_2)$, respectively, in states with locations $q_1$ and $q_2$.
Then we have to show that $R^{\prime}=((Q_1^{\prime}(q_1)\oplus Q_1^{\prime}(q_2))\mid Q_1^{\prime}(p))\backslash I$ and $R=((Q_1(q_1)\oplus Q_1(q_2))\mid Q_1(p))\backslash I$ are weak bisimilar in the environment $\varrho$, where $I = {\sf Sort}((Q_1^{\prime}(q_1)\oplus Q_1^{\prime}(q_2))={\sf Sort}((Q_1(q_1)\oplus Q_1(q_2))$.

$R^{\prime}\xrightarrow[\mathrm{Id}]{\tau}R^{\prime}_1\xrightarrow[\mathrm{Id}]
{\tau}R^{\prime}_2$ involving $\overline{\sf write_x}(e)\cdot(\ast)$ and ${\sf read_y}(r)\cdot(\ast)$, respectively, in the location $p$ and in the environment $\varrho$.
Then $R$ can make the transition $R\xrightarrow[\mathrm{Id}]{\tau}R_1\xrightarrow[\mathrm{Id}]{\tau}R_2$ involving ${\sf read_y}(r)\cdot(\ast)$ and $\overline{\sf write_x}(e)\cdot(\ast)$, respectively, both in the location $p$ and in the environment $\varrho$ too. Meanwhile, $R^{\prime}_2 = R_2$.
For responses from $R$ respect to $R^{\prime}\xrightarrow[\mathrm{Id}]{\tau}R_1^{\prime}$, according to the definition of localized weak bisimulation, let $R$ be stable. And when $R_1^{\prime}\xrightarrow[\mathrm{Id}]{\tau}R_2^{\prime}$, let $R\xrightarrow[\mathrm{Id}]{\tau}R_1\xrightarrow[\mathrm{Id}]{\tau}R_2$ and $R_2^{\prime}=R_2$ obviously. Conversely, one can show that $R^{\prime}$ can match the derivation of $R$ with the same observations in the environment $\varrho$.

We have prove that, the localized weak bisimilarity is a congruence. So, we can easily obtained that  $(Q^{\prime},\varrho)\approx(Q,\varrho)$.
Since localized weak bisimulations are transitive, from $(P,\varrho)\approx(Q^{\prime},\varrho)$ by induction hypothesis and $(Q^{\prime},\varrho)\approx(Q,\varrho)$, we get $(P,\varrho)\approx(Q,\varrho)$.

The cases for other reordering rules are similar to the rule {\bf WR}.
\end{IEEEproof}
The proof for Theorem \ref{thm_tso}.
\begin{IEEEproof} We assume that $(Q,\varrho)$ is not conflict free, i.e. there exists a process $Q^{\prime}$ reached from $Q$ and $p_1,p_2\in |Q^{\prime}|$ such that $p_1\neq p_2$, $Q(p_1)^{\prime}=Q_1$, $Q^{\prime}(q_2)=Q_2$ and there exist two transitions in $Q_1$ and $Q_2$ with the actions $\alpha_1$ and $\alpha_2$, respectively, such that $({\sf symb}(\alpha_1),{\sf symb}(\alpha_2))\in {\bf CFL}$. Then, we have to prove that $(P,\varrho)$ involves a conflict, which contradicts the fact that $(P,\varrho)$ is conflict free.

We prove it by induction on the number of steps from $P$ to $Q$.

If $n = 0$, then $P = Q$, so $Q^{\prime}$ can also be reached from $P$, which contradicts that $(P,\varrho)$ is conflict free.

Induction hypothesis: when $n = k$, if we have $P\hookrightarrow_{\sf tso}^{\ast} Q$ with $k$ steps and $(P,\varrho)$ is conflict free, then $(Q,\varrho)$ is also conflict free.

When $n = k +1$, without loss of any generality, we assume that the additional reordering works on $Q_1$ or $Q_2$. Here, we just focus on $Q_1$, since the case for $Q_2$ is symmetric.
For the reordering rule {\bf R-WR}, it is similar to the case {\bf WR} in Theorem \ref{thm_cr}.
For the reordering rule {\bf A-WR},
if there exists a derivation $Q^{\prime}$ of $Q$ involving a conflict between processes $Q_1$ and $Q_2$, then let $Q_1 = Q_{11}\rhd Q_1^{\prime\prime}$, which is transformed form $ Q_{11}\rhd Q_{12}\rhd Q_1^{\prime}$, where $Q_{11} = \overline{\sf write_x}(e)\cdot(\ast)$, $Q_{12} = {\sf read_x}(r)\cdot(\ast)$ and $Q_1^{\prime\prime} = Q_1^{\prime}\{(e)/r\}$. Since $Q_{11}\rhd Q_1^{\prime\prime}$ and $Q_2$ involve a conflict, then $Q_{11}\rhd Q_{12} \rhd Q_1^{\prime}$ and $Q_2$ also involve a conflict which is a derivation of some $Q$ by use $k$ steps of reordering from $P$, contradicting with the induction hypothesis.
The other rules are similar.

Therefore, if $(P,\varrho)$ is conflict free and $P\hookrightarrow_{\sf tso}^{\ast} Q$, then $(Q,\varrho)$ is also conflict free.
\end{IEEEproof}
The proof for Theorem \ref{thm_tso_drf}.
\begin{IEEEproof}
We prove it by induction on the number of steps from $P$ to $Q$.

If $n = 0$, then $P = Q$. It is easy to check that $(P,\varrho)\approx (P,\varrho)$.

Induction hypothesis: if $(P,\varrho)$ is conflict free, we have $n=k$ steps from $P$ to $Q$ using $\hookrightarrow_{\sf tso} $, and $(P,\varrho)\approx (Q,\varrho)$ with $E = \mbox{Id}_{|P|}$ in the triple (since the reordering does not change the locations of the processes).

We have to establish that if $(P,\varrho)$ is conflict free and $P\hookrightarrow_{\sf tso}^{\ast} Q$ with $n = k +1$ steps then $(P,\varrho) \approx (Q,\varrho)$. For $P\hookrightarrow_{\sf tso}^{\ast} Q$ with $k+1$ step, we have $P\hookrightarrow_{\sf tso}^{\ast} Q^{\prime}$ with $k$ steps, $Q^{\prime}$ is conflict free by Theorem \ref{thm_tso} and $Q^{\prime}\hookrightarrow_{\sf tso}^{\ast} Q$ with one step. From induction hypothesis we have $(P,\varrho)\approx (Q^{\prime},\varrho)$, then one needs to show that $(Q^{\prime},\varrho)\approx (Q,\varrho)$. Since $Q^{\prime}\hookrightarrow_{\sf tso}^{\ast} Q$ using only one step of the transformation rules, we just consider the rule {\bf A-WR}, and the rule {\bf R-WR} is similar to the case {\bf WR} in Theorem \ref{thm_cr_def}.

Without loss of any generality, we assume that the rule {\bf A-WR} effects on location $p\in |P|=|Q^{\prime}|=|Q|$.
Assume that $Q^{\prime}$ can reduce to $Q^{\prime}_1$ before referring to $\overline{\sf write_x}(e)\cdot(\ast)\rhd {\sf read_x}(r)\cdot(\ast)\rhd P_1$
involved in {\bf A-WR}, in location $p$. Then $Q$ can make the same derivation as $Q^{\prime}$ to $Q_1$ with $Q^{\prime}_1 = Q_1$. Both $Q^{\prime}_1$ and $Q_1$ are conflict free. And we have $Q^{\prime}_1(q) = Q_1(q)$ for $q\in |Q_1|$ and $q\neq p$.

Let $Q^{\prime}(p) = \overline{\sf write_x}(e)\cdot(\ast)\rhd {\sf read_x}(r)\cdot(\ast)\rhd P_1$.
After the transformation, $Q(p)= \overline{\sf write_x}(e)\cdot(\ast)\rhd P_1 \{(e)/r\}$.
We just focus on the transitions involving $\overline{\sf write_x}(e)\cdot(\ast)$ and ${\sf read_x}(r)\cdot(\ast)$, in location $p$ respectively, communicating with ${\sf write_x}(x)$ and $\overline{\sf read_x}(v)$ which are prefixes from $X_{\sf x}(v)$, in states with the location $q\in |Q_1^{\prime}|$.
Then we have to show that $R^{\prime}=(Q_1^{\prime}(q)\mid Q_1^{\prime}(p))\backslash I$ and $R=(Q_1(q)\mid Q_1(p))\backslash I$ are weak bisimilar in the environment $\varrho$, where $I = {\sf Sort}(Q_1^{\prime}(q))= {\sf Sort}(Q_1(q))$.

$R^{\prime}\xrightarrow[\mathrm{Id}]{\tau}R^{\prime}_1\xrightarrow[\mathrm{Id}]
{\tau}R^{\prime}_2$ involves $\overline{\sf write_x}(e)\cdot(\ast)$ and ${\sf read_x}(r)\cdot(\ast)$, respectively, in the locations $p$ and in the environment $\varrho$.
Then $R$ can make the transition $R\xrightarrow[\mathrm{Id}]{\tau}R_1$ involving $\overline{\sf write_x}(e)\cdot(\ast)$, both in the location $p$ and in the environment $\varrho$ too.
Meanwhile, $R_1=R^{\prime}_2\{(e/r)\}$.
For responses from $R$ respect to $R^{\prime}\xrightarrow[\mathrm{Id}]{\tau}R_1^{\prime}$, according to the definition of localized weak bisimulation, let $R$ be stable. And when $R_1^{\prime}\xrightarrow[\mathrm{Id}]{\tau}R_2^{\prime}$, let $R\xrightarrow[\mathrm{Id}]{\tau}R_1$ and $R_1=R^{\prime}_2\{(e/r)\}$ obviously. Conversely, one can show that $R^{\prime}$ can match the derivation of $R$ with the same observations in the environment $\varrho$.

We have prove that, the localized weak bisimilarity is a congruence. So, we can easily obtained that  $(Q^{\prime},\varrho)\approx(Q,\varrho)$.
Since localized weak bisimulations are transitive, from $(P,\varrho)\approx (Q^{\prime},\varrho)$ and $(Q^{\prime},\varrho)\approx (Q,\varrho)$, we get $(P,\varrho)\approx (Q,\varrho)$.

The other transformation rules are similar.

\end{IEEEproof}
\end{document}